\documentclass[aps,prb,reprint,superscriptaddress,twocolumn,longbibliography]{revtex4-2}
\usepackage[utf8]{inputenc}
\usepackage{graphicx}
\usepackage{amsmath}
\usepackage{amsfonts}
\usepackage{amssymb}
\usepackage{bbm}
\usepackage[usenames,dvipsnames,svgnames,table]{xcolor}
\usepackage[USenglish]{babel}
\usepackage{hyperref}
\usepackage{grffile}
\usepackage{xcolor}
\usepackage{bm}
\usepackage{tabularx}
\usepackage{array}
\usepackage{mathtools}
\usepackage{makecell}
\usepackage{booktabs}
\usepackage{blindtext}
\usepackage{csquotes}
\usepackage{lipsum}
\usepackage{soul}
\usepackage{threeparttable}

\usepackage{environ}
\usepackage{siunitx}
\usepackage{tikz}
\usetikzlibrary{positioning}
\usepackage{ifthen}
\usepackage{tikz-3dplot}
\usetikzlibrary{fadings}

\newcommand{\ignore}[1]{}

\newcolumntype{P}[1]{>{\centering\arraybackslash}p{#1}}

\allowdisplaybreaks

\newcommand{\ket}[1]{\left| #1 \right\rangle}

\newcommand{\braket}[2]{\langle #1 | #2 \rangle}

\newcommand{\bbone}{\mathbbm{1}}

\begin{document}
\title{Irreducible momentum-space spin structure of Weyl semimetals\\ and its signatures in Friedel oscillations}

\author{Andy Knoll}
\email{andy.knoll@tu-dresden.de}

\author{Carsten Timm}
\email{carsten.timm@tu-dresden.de}
\affiliation{Institute of Theoretical Physics, Technische Universit\"at Dresden, 01062 Dresden, Germany}
\affiliation{W\"urzburg--Dresden Cluster of Excellence ct.qmat, Technische Universit\"at Dresden, 01062 Dresden, Germany}

\begin{abstract}
Materials that break time-reversal or inversion symmetry possess nondegenerate electronic bands, which can touch at so-called Weyl points. The spinor eigenstates in the vicinity of a Weyl point exhibit a well-defined chirality $\pm 1$. Numerous works have studied the consequences of this chirality, for example in unconventional magnetoelectric transport. However, even a Weyl point with isotropic dispersion is not only characterized by its chirality but also by the momentum dependence of the spinor eigenstates. For a single Weyl point, this momentum-space spin structure can be brought into standard ``hedgehog'' form by a unitary transformation, but for two or more Weyl points, this is not possible. In this work, we show that the relative spin structure of a pair of Weyl points has strong qualitative signatures in the electromagnetic response. Specifically, we investigate the Friedel oscillations in the induced charge density due to a test charge for a centrosymmetric system consisting of two Weyl points with isotropic dispersion. The most pronounced signature is that the amplitude of the Friedel oscillations falls off as $1/r^4$ in directions in which both Weyl points exhibit the same spin structure, while for directions with inverted spin structures, the amplitude of the Friedel oscillations decreases as $1/r^3$.
\end{abstract}

\maketitle

\section{Introduction}

The Dirac equation is one of the fundamental equations of relativistic quantum mechanics \cite{Dirac_1928}. Shortly after its discovery, Hermann Weyl investigated the massless solutions of this equation, which describe what is nowadays called Weyl fermions \cite{Weyl_1929}. These fermions are chiral particles that occur as left-handed and right-handed variants. Mathematically speaking, the chirality of a relativistic spin-$\frac{1}{2}$ particle is determined by its transformation behavior with respect to the left-handed or right-handed representations of the Lorentz group \cite{Peskin_Schroeder_book}. For the massless Weyl fermions, helicity and chirality are identified, which allows for an alternative interpretation of the handedness: If momentum and spin are aligned (anti-aligned), the Weyl fermion is right handed (left handed).

While they have not been found as fundamental particles, Weyl fermions have been observed in condensed-matter systems as low-energy excitations around twofold-degenerate, linearly dispersing band-touching points \cite{Xu_Nasser_Belopolski_2015, Yang_Liu_Sun_2015, Lv_Xu_Weng_2015}, so-called Weyl points \cite{Wan_Turner_2011}. In the vicinity of a Weyl point that is not tilted, the effective Bloch Hamiltonian is written as
\begin{align}
H_{\mathrm{Weyl}}(\bm{k}) &= \hbar\,\sum_{i,j}k_i v_{ij} \sigma_j \nonumber \\
&= \hbar \sum_i k_i \begin{pmatrix}
    v_{iz} & v_{ix} - iv_{iy} \\
    v_{ix} + iv_{iy} & - v_{iz}
  \end{pmatrix} ,
\label{eq:eff_Ham_iso_WP}
\end{align}
where $\bm{\sigma}=(\sigma_x,\sigma_y,\sigma_z)^T$ is the vector of Pauli matrices representing the spin degree of freedom. The real velocity tensor $v \equiv{} (v_{ij})$ (we notate tensors by writing a characteristic component in parentheses) contains effective velocities that express the anisotropy of the Weyl point \cite{Armitage_Mele_2018,Wan_Turner_2011}. The resulting dispersion is
\begin{equation}
E_{\mathrm{Weyl}}(\bm{k}) = \pm \hbar \sqrt{\sum_{i,j,l} k_i v_{ij} v^T_{jl} k_l} .
\label{eq:eff_E_WP}
\end{equation}
The chirality of the Weyl fermions is conveyed to the Weyl points by $v$: If the determinant of $v$ is positive (negative), the chirality of the Weyl point is also positive (negative), i.e., it is right handed (left handed) \cite{Armitage_Mele_2018}. A Weyl point is also a monopole of the Berry curvature and its chirality equals the Chern number obtained by integrating the Berry curvature over a surface enclosing only this Weyl point \cite{Berry_1985,Volovik_helium_drop_2009}. This topological invariant protects the Weyl point against small symmetry-preserving perturbations \cite{Simon_1983,Berry_1984}. According to the fermion-doubling theorem \cite{Nielsen_Ninomiya_1981_a,Nielsen_Ninomiya_1981_b}, lattice realizations containing Weyl fermions, so-called Weyl semimetals \cite{Wan_Turner_2011}, require the presence of pairs of Weyl points of opposite chirality so that the net Chern number vanishes~\cite{Armitage_Mele_2018}.

For a single Weyl point, the presence of a magnetic field splits the Weyl bands into Landau levels \cite{Lv_Qian_Ding_2021}. In the quantum limit, only the zeroth Landau level, which is chiral, is partially occupied. Now, in the presence of an electric field $\bm{E}$ parallel to the magnetic field, the electrons are accelerated according to $\dot{\bm{k}}=-e\bm{E}$ leading to the nonconservation of charge at this Weyl point. A second Weyl point of opposite chirality compensates this imbalance. The nonconservation of charge at a single Weyl point is a manifestation of the Adler--Bell--Jackiw  or chiral anomaly \cite{Adler_1969, Bell_Jackiw_1969, Nielsen_Ninomiya_1983}. In the simplest case, a Weyl semimetal consists of two isotropic Weyl points. The chiral anomaly then leads to a negative longitudinal magnetoresistance (LMR) \cite{Son_Spivak_2013, Kim_Kim_Sasaki_2014, Burkov_chiral_a_2014, Xiong_Kushwaha_Liang_2015, Huang_Zhao_Long_2015, Zhang_Xu_2016, Hirschberger_Kushwaha_Wang_2016, Li_Wang_Li_2017, Dantas_Benitez_Roy_2018}. However, the interplay with the orbital magnetic moment \cite{Xiao_Chang_2010,Sundaram_Niu_1999} and strong intervalley scattering can change the sign of the LMR within the semiclassical framework \cite{KTM20}. Another consequence of the chiral nature of Weyl semimetals is the formation of Fermi arcs \cite{Wan_Turner_2011,Xu_Weng_Wang_2011,Lv_Weng_Fu_2015}. These arcs are formed by surface states that connect the projections of the Weyl points onto the surface Brillouin zone.

The chiral anomaly and the presence of Fermi arcs only rely on the presence of pairs of Weyl points of opposite chirality, not on details such as the anisotropies of the Weyl cones \cite{Zhang_Xu_2016, Lv_Weng_Fu_2015}.
Other physical phenomena and quantities are more sensitive to anisotropies. To name a few examples, good agreement between experimentally observed and theoretically calculated optical conductivities in TaAs, TaP, NbAs, and NbP is only achieved if the anisotropy of the Weyl cones is taken into account \cite{Grassano_Pulci_2018, Grassano_Pulci_Cannuccia_2020}. Furthermore, the dimensionless conductance is strongly affected by the transport direction for anisotropic Weyl cones \cite{Trescher_Sbierski_2015}. In addition, the Fano factor, often regarded as independent of system-specific details, heavily depends on the tilt of the Weyl cones \cite{Trescher_Sbierski_2015}. Finally, anisotropic Weyl cones cause additional terms in the charge conductivity that can change the sign of the LMR \cite{Johansson_Henk_2019}.

Aside from the anisotropy, the velocity tensor $(v_{ij})$ also determines the spin structure of the Weyl point, i.e., it contains information on the eigenstates of the Hamiltonian in Eq.~\eqref{eq:eff_Ham_iso_WP}. A central point of this paper is that this spin structure has important consequences even for isotropic Weyl points. Equation \eqref{eq:eff_E_WP} shows that an isotropic dispersion requires that
\begin{equation}\label{eq:cond_iso_WP}
v v^T = v_F^2 \bbone ,
\end{equation}
where $v_F>0$ is the Fermi velocity. The real velocity tensor $v$ thus has to be of the form $v = v_F R$, where $R$ is an orthogonal matrix. The matrix $R$ describes a proper or improper rotation if $\det R = 1$ or $\det R = -1$, respectively. In particular, both cases correspond to a continuum of eigenstates. If $R$ corresponds to a proper rotation one can find a unitary transformation on the spin-$\frac{1}{2}$ Hilbert space that transforms $R$ into the identity matrix, i.e., there exists a unitary $2\times 2$ matrix $U$ acting on spin space such that
\begin{equation}
U H_{\mathrm{Weyl}}(\bm{k})\, U^\dagger = v_F \hbar \sum_{i,j} k_i R_{ij} U \sigma_j U^\dagger
  = v_F \hbar\, \sum_i k_i \sigma_i .
\label{eq:HWeyl_transf.1}
\end{equation}
This is possible because the Pauli matrices are irreducible tensor operators belonging to the three-dimensional irreducible representation of $\mathrm{SO}(3)$. If $R$ describes an improper rotation one can write it as $-1$ multiplied by an orthogonal matrix with determinant $+1$. The same argument as above then gives
\begin{equation}
U H_{\mathrm{Weyl}}(\bm{k})\, U^\dagger = -v_F\hbar\, \sum_i k_i \sigma_i .
\label{eq:HWeyl_transf.2}
\end{equation}
Hence, for a single isotropic Weyl point we can assume the Weyl Hamiltonian to have the form of either Eq.\ \eqref{eq:HWeyl_transf.1} or \eqref{eq:HWeyl_transf.2}, depending on its chirality, without loss of generality. The spin structure in momentum space than has isotropic hedgehog form. Equations \eqref{eq:HWeyl_transf.1} and \eqref{eq:HWeyl_transf.2} are often referred to as \enquote{standard Weyl Hamiltonians}, which were first proposed by Weyl in the context of relativistic massless quantum mechanics~\cite{Weyl_1929}.

The band structure of any Weyl semimetal contains in general multiple Weyl points. Therefore, a unitary transformation of the entire Hamiltonian can generally only transform the spin structure of one Weyl point to the desired form. The spin structures of the other Weyl points will transform along with the first one. The \emph{relative} spin structure between multiple Weyl points thus remains invariant.
Signatures of the relative spin structure occur due to transitions between different Weyl cones, i.e., intervalley processes. Such processes have frequently been neglected when calculating linear response functions of Weyl semimetals \cite{Panfilov_Burkov_2014, Kapusta_Toimela_1988, Lv_Zhang_2013, Burkov_2014, Zhou_Chang_2015, Ghosh_Timm_2019}, which is a valid approximation in the long-wavelength limit. However, the emergence of spin-structure-dependent characteristics are anticipated for momenta that are approximately equal to the separation of the Weyl points in momentum space.

Closely connected to the charge-density response function, or polarizability, are Friedel oscillations in the induced charge density due to a test charge. In Ref.~\cite{Lv_Zhang_2013}, these oscillations have been investigated in the single-Weyl-point approximation, where the authors made assumptions about the behavior of the Friedel oscillations caused by intervalley processes based on the standard Weyl Hamiltonians in Eqs.~\eqref{eq:HWeyl_transf.1} and \eqref{eq:HWeyl_transf.2}. Energy-resolved Friedel oscillations in the local density of states have previously been studied for three-dimensional Weyl semimetals \cite{Zheng_Wang_Zhong_2016, Diaz_Fernandez_Domingues_Adame_2021,Hosur_2012}, while Friedel oscillations in the induced charge density have been examined for graphene \cite{Wunsch_2006, ChF06, Wang_Raikh_2021}, which can be understood as two degenerate copies of a two-di\-men\-sio\-nal Weyl semimetal if spin-orbit coupling is neglected. However, to the best of our knowledge, there have not been any quantitative studies of the relative spin structure of Weyl points and its signatures in the electromagnetic response or the resulting Friedel oscillations.

In this work, we show that intervalley processes induce strong qualitative signatures of the relative spin structure in the charge-density Friedel oscillations. The paper is organized as follows. In Sec.~\ref{sec:model_systems}, we introduce three low-energy model systems consisting of two isotropic Weyl points, where each model exhibits a distinct relative spin structure. In Sec.~\ref{sec:polarizability}, we calculate analytically the extrinsic polarizability for each model system and investigate singularities of the intrinsic polarizability based on a tight-binding model. Based on the results, we numerically compute the Friedel oscillations for each model system in Sec.~\ref{sec:FO}. In Sec.~\ref{sec:beyond_cWSM}, we discuss extensions beyond the considered minimal models. Finally, we summarize our results in Sec.~\ref{sec:summary_conclusion} and draw conclusions.

\section{Model systems}\label{sec:model_systems}

We introduce three centrosymmetric model systems that break time-reversal symmetry and differ in their spin structures. Each system consists of two isotropic Weyl points of opposite chirality that are separated along the $k_z$ axis. For isotropic Weyl points, the Fermi velocity tensor in Eq.~\eqref{eq:eff_Ham_iso_WP} is written as $(v_{ij})=v_F\, (R^{\alpha\chi}_{ij})$, where $(R^{\alpha\chi}_{ij})$ is the orthogonal spin structure tensor for the Weyl point with chirality $\chi=\pm 1$, and $\alpha$ specifies the specific model as described below. Without loss of generality, we assume the Fermi energy $E_F$ to coincide with the Weyl points or to lie in the conduction band, i.e., $E_F\ge 0$.

\begin{figure}
\includegraphics[scale=1]{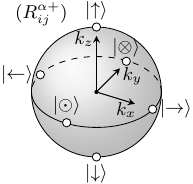}\hspace{1em}%
\includegraphics[scale=1]{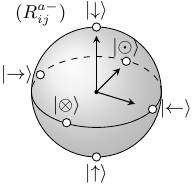}\\[2ex]
\includegraphics[scale=1]{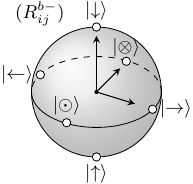}\hspace{1em}%
\includegraphics[scale=1]{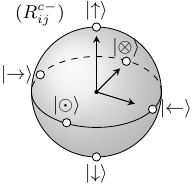}
		\caption{Top left: Spin structure of the Weyl point with positive chirality for all three models. The spin of the Bloch states on a constant-energy surface surrounding the Weyl point are given by $\ket{\rightarrow} = (\ket{\uparrow}+\ket{\downarrow})/\sqrt{2}$, $\ket{\leftarrow} = (\ket{\uparrow}-\ket{\downarrow})/\sqrt{2}$, $\ket{\otimes} = (\ket{\uparrow}+i\ket{\downarrow})/\sqrt{2}$, and $\ket{\odot} = (\ket{\uparrow}-i\ket{\downarrow})/\sqrt{2}$, where $\{\ket{\uparrow},\ket{\downarrow}\}$ is the eigenbasis of the Pauli matrix $\sigma_z$.
		Top right to bottom right: Spin structures of the Weyl point with negative chirality for models $\alpha=a, b, c$. \label{fig_1}}
\end{figure}

The spin structures are different for the three model systems. As we have argued above, the tensor $(R^{\alpha\chi}_{ij})$ of one Weyl point can always be made proportional to the identity matrix. Therefore, without loss of generality, the spin structure tensor of the Weyl point of positive chirality is chosen as
\begin{equation}
	(R^{\alpha +}_{ij}) = \begin{pmatrix}
		+1 & 0 & 0 \\
		0 & +1 & 0 \\
		0 & 0 & +1
	\end{pmatrix} .\label{eq:pseudo_pos}
\end{equation}
As mentioned below Eq.\ \eqref{eq:cond_iso_WP}, the chirality of a Weyl point is associated with a continuum of eigenstates. Out of the continuum of possible spin structures of the Weyl point with negative chirality, we choose three limiting cases. Taking the spin structure in Eq.~\eqref{eq:pseudo_pos} of the Weyl point with positive chirality as the reference, the spin structure of the Weyl point with negative chirality for model system $a$ is inverted along all directions, i.e., the Weyl points are described by the standard forms in Eqs.~\eqref{eq:HWeyl_transf.1} and \eqref{eq:HWeyl_transf.2}. In contrast, for model system $b$, the spin structure of the latter Weyl point is inverted only along the direction parallel to the separation vector of the Weyl points, i.e., the $z$ direction. For model system $c$, the spin structure is inverted along the $x$ direction, which is orthogonal to the separation vector of the Weyl points. For the three model systems, the spin-structure tensors of the Weyl point with negative chirality then read as
\begin{align}
	(R^{a -}_{ij})&= \begin{pmatrix}
		-1 & 0 & 0 \\
		0 & -1 & 0 \\
		0 & 0 & -1
	\end{pmatrix} ,\label{eq:pseudo_a}\\
	(R^{b -}_{ij})&= \begin{pmatrix}
		+1 & 0 & 0 \\
		0 & +1 & 0 \\
		0 & 0 & -1
	\end{pmatrix} ,\label{eq:pseudo_b}\\
	(R^{c -}_{ij})&= \begin{pmatrix}
		-1 & 0 & 0 \\
		0 & +1 & 0 \\
		0 & 0 & +1
	\end{pmatrix} .\label{eq:pseudo_c}
\end{align}
In Fig.~\ref{fig_1}, the spin structures of the Weyl points are illustrated.

In the vicinity of each Weyl point, the Bloch Hamiltonian for model $\alpha=a,b,c$ is given by \begin{equation}\label{eq:model_Hamiltonian}
H^{\alpha\chi}(\bm{k})=v_F\hbar\, \sum_{i,j}
	\left(k_i-\chi\frac{Q_i}{2}\right)
	R^{\alpha\chi}_{ij}\, \sigma_j ,
\end{equation}
where $\bm{Q} = (Q_x,Q_y,Q_z)^T$ is the separation vector between the two Weyl points. Here, we assume $\bm{Q}=Q\hat{\bm{e}}_z$, where $\hat{\bm{e}}_z$ is the unit vector in $z$ direction. The eigenvalue equation for $H^{\alpha\chi}(\bm{k})$ reads as
\begin{equation}
	H^{\alpha\chi}(\bm{k})\ket{u^{\alpha\chi}_s(\bm{k})}=E^\chi_{s}(\bm{k})\ket{u^{\alpha\chi}_s(\bm{k})} .
\end{equation}
For the eigenenergies, we obtain
\begin{equation}\label{eq:dispersion_WPs}
	E^\chi_{s}(\bm{k})=s v_F \hbar \left|\bm{k}-\chi\frac{\bm{Q}}{2}\right|
\end{equation}
for all three model systems, where $s$ labels the conduction band ($s=+1$) and the valence band ($s=-1$), respectively. On the other hand, the eigenstates $\ket{u^{\alpha\chi}_s(\bm{k})}$ depend on the model system, i.e., they depend on $R_{ij}^{\alpha\chi}$.

\section{Polarizability}\label{sec:polarizability}

Within the linear-response regime, the static polarizability or charge-density response function is given by the Lindhard function
\begin{align}\label{eq:charge_response_function}
  \pi^\alpha(\bm{q}) &= -\frac{1}{\mathcal{V}}\sum_{\chi\chi's s'}\sum_{\bm{k}}
  \frac{n_F(E^\chi_{s}(\bm{k}))-n_F(E^{\chi'}_{s'}(\bm{k}'))}
  {E^\chi_{s}(\bm{k})-E^{\chi'}_{s'}(\bm{k}')+i \eta}\nonumber\\
&\quad{}\times F_{ss'}^{\alpha;\chi\chi'}(\bm{k},\bm{k}'),
\end{align}
where $\mathcal{V}$ is the volume of the system,
$\chi$ and $\chi'$ denote chiralities (valleys), $s$ and $s'$ are band indices, $n_F(E)$ is the Fermi function, and here and for the rest of the paper we set~\cite{Lv_Zhang_2013, Zhou_Chang_2015, Ghosh_Timm_2019}
\begin{equation}
\bm{k}'=\bm{k}+\bm{q} .
\end{equation}
The spinor overlap of the eigenstates is described by
\begin{equation}
	F^{\alpha ;\chi\chi '}_{s s'}(\bm{k},\bm{k}')
	= \big|\braket{u^{\alpha\chi'}_{s'}\!(\bm{k}')}{u^{\alpha\chi}_s(\bm{k})}\big|^2.
\end{equation}
The intravalley ($\chi'=\chi$) spinor overlap is independent of the spin structures of the Weyl points,
\begin{equation}\label{eq:spinor_overlap_intra}
	F^{\alpha ;\chi\chi}_{s s'}(\bm{k},\bm{k}')=\frac{1}{2}\left(1+ss'\,
	\frac{\left(\bm{k}-\chi\frac{\bm{Q}}{2}\right)\cdot\left(\bm{k}'-\chi\frac{\bm{Q}}{2}\right)}
	{\left|\bm{k}-\chi\frac{\bm{Q}}{2}\right|\left|\bm{k}'-\chi\frac{\bm{Q}}{2}\right|}\right),
\end{equation}
since the relative spin structure of each Weyl point with respect to itself is always the same.
On the other hand, the intervalley ($\chi'=-\chi\equiv\bar{\chi}$) spinor overlap does depend on the relative spin structure,
\begin{align}
&F^{\alpha ;\chi\bar{\chi}}_{s s'}(\bm{k},\bm{k}') \nonumber\\
	&\quad =\frac{1}{2}\left(1+ss'\,
	\frac{\left(\bm{k}-\chi\frac{\bm{Q}}{2}\right)^T
	(R^{\alpha -}_{ij})
	\left(\bm{k}'-\bar{\chi}\frac{\bm{Q}}{2}\right)}{\left|\bm{k}-\chi\frac{\bm{Q}}{2}\right|
	\left|\bm{k}'-\bar{\chi}\frac{\bm{Q}}{2}\right|}\right).
	\label{eq:spinor_overlap_inter}
\end{align}

The charge response function can be decomposed into intravalley and intervalley contributions as
\begin{equation}\label{eq:crf_total}
	\pi^\alpha(\bm{q})=\pi^\alpha_{\mathrm{intra}}(\bm{q})+\pi^\alpha_{\mathrm{inter}}(\bm{q}),
\end{equation}
where
\begin{align}
\pi^\alpha_{\mathrm{intra}}(\bm{q}) &= -\frac{1}{\mathcal{V}}\sum_{\chi s s'}\sum_{\bm{k}}
  \frac{n_F(E^\chi_{s}(\bm{k}))-n_F(E^{\chi}_{s'}(\bm{k}'))}
  {E^\chi_{s}(\bm{k})-E^{\chi}_{s'}(\bm{k}')+i \eta}\nonumber\\
&\quad{}\times F_{ss'}^{\alpha;\chi\chi}(\bm{k},\bm{k}'),\label{eq:crf_intra_gen}\\
\pi^\alpha_{\mathrm{inter}}(\bm{q}) &= -\frac{1}{\mathcal{V}}\sum_{\chi s s'}\sum_{\bm{k}}
  \frac{n_F(E^\chi_{s}(\bm{k}))-n_F(E^{\bar\chi}_{s'}(\bm{k}'))}
  {E^\chi_{s}(\bm{k})-E^{\bar\chi}_{s'}(\bm{k}')+i \eta}\nonumber\\
&\quad{}\times F_{ss'}^{\alpha;\chi\bar\chi}(\bm{k},\bm{k}'). \label{eq:crf_inter_gen}
\end{align}
The intravalley polarizability describes processes occurring within a single Weyl cone, whereas the intervalley polarizability describes processes occurring between Weyl cones of opposite chirality. In the following, we evaluate these two contributions separately, in the zero-tem\-pe\-ra\-ture limit.

\subsection{Intravalley polarizability}\label{sec:intra_pola}

As discussed above, the intravalley spinor overlap, Eq.\ \eqref{eq:spinor_overlap_intra}, does not depend on the spin structure of the Weyl points. Consequently, the intravalley polarizability also is independent of the spin structure and is the same for all three models. The intravalley polarizability can be split further into an extrinsic and an intrinsic contribution~\cite{Lv_Zhang_2013, Burkov_2014, Hwang_Sarma_2007, Zhou_Chang_2015},
\begin{equation}
	\pi^\alpha_{\mathrm{intra}}(\bm{q})=\pi^{\mathrm{ext},\alpha}_{\mathrm{intra}}(\bm{q})+\pi^{\mathrm{int},\alpha}_{\mathrm{intra}}(\bm{q}),\label{eq:crf_intra}
\end{equation}
which originate from the partially filled conduction band and from the fully occupied valence band, respectively. At zero temperature, the two contributions read as
\begin{align}
&\pi^{\mathrm{ext},\alpha}_{\mathrm{intra}}(\bm{q}) \nonumber\\
&=-\sum_{\chi s}\left[\int \frac{d^3k}{(2\pi)^3}\:
	\frac{\Theta(E_F-E^\chi_{+}(\bm{k}))}{E^\chi_{+}(\bm{k})-E^\chi_{s}(\bm{k}')+i\eta}\,
	F^{\alpha ;\chi\chi}_{+ s}(\bm{k},\bm{k}')\right.\nonumber\\
&\quad{} - \left.\int \frac{d^3k}{(2\pi)^3}\:
	\frac{\Theta(E_F-E^\chi_{+}(\bm{k}'))}{E^\chi_{s}(\bm{k})-E^\chi_{+}(\bm{k}')+i\eta}\,
	F^{\alpha ;\chi\chi}_{s +}(\bm{k},\bm{k}') \right],\label{eq:crf_intra_ext}
\end{align}
and
\begin{align}
&\pi^{\mathrm{int},\alpha}_{\mathrm{intra}}(\bm{q}) \nonumber\\
&=-\sum_{\chi s}\left[\int \frac{d^3k}{(2\pi)^3}\:
	\frac{\Theta(E_F-E^\chi_{-}(\bm{k}))}{E^\chi_{-}(\bm{k})-E^\chi_{s}(\bm{k}')+i\eta}\,
	F^{\alpha ;\chi\chi}_{- s}(\bm{k},\bm{k}')\right.\nonumber\\
&\quad{} - \left. \int \frac{d^3k}{(2\pi)^3}\:
	\frac{\Theta(E_F-E^\chi_{-}(\bm{k}'))}{E^\chi_{s}(\bm{k})-E^\chi_{-}(\bm{k}')+i\eta}\,
	F^{\alpha ;\chi\chi}_{s -}(\bm{k},\bm{k}') \right],\label{eq:crf_intra_int}
\end{align}
where $E_F\geq0$ is the Fermi energy and the integrals are over the entire momentum space. In the following, we evaluate these two contributions separately.

\begin{figure}
		\includegraphics[scale=0.8]{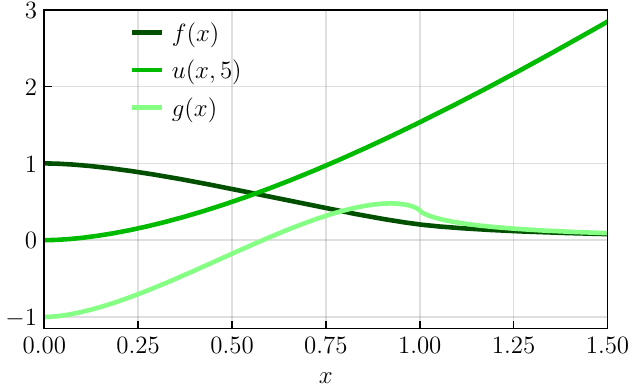}
		\caption{Functions $f(x)$, $u(x, 5)$, and $g(x)$
			of $x = q/2k_F$, given in Eqs.~\eqref{eq:func_f}, \eqref{eq:func_u}, and \eqref{eq:func_g}, respectively.}\label{fig_2}
\end{figure}

\subsubsection{Extrinsic polarizability}

The extrinsic intravalley polarizability, $\pi^{\mathrm{ext},\alpha}_{\mathrm{intra}}(\bm{q})$, corresponds to the static polarizability of two isotropic Weyl points of opposite chirality that do not interact with each other. The extrinsic contribution originates from the partially filled conduction band. In particular, only states with energies in the range of the Weyl nodes and the Fermi energy contribute. This is reflected by the Heaviside step functions in Eq.~\eqref{eq:crf_intra_ext}, which restrict the integration volume to spheres the radius of which is the Fermi wave number. The calculation has been performed in Refs.~\cite{Kapusta_Toimela_1988} and \cite{Lv_Zhang_2013}. It yields
\begin{equation}
	\pi^{\mathrm{ext},\alpha}_{\mathrm{intra}}(\bm{q})
	= 2\,\frac{2}{(2\pi)^2}\,
	\frac{k_F^2}{v_F\hbar}\, f\bigg( \frac{q}{2k_F} \bigg) ,
	\label{eq:crf_intra_ext_ana}
\end{equation}
where $k_F=E_F/v_F\hbar$ is the Fermi wave number and
\begin{equation}
	f(x) = \frac{2}{3}\left(1+\frac{1-3x^2}{4x}\, \ln \left|\frac{1+x}{1-x}\right|
	- \frac{x^2}{2}\,  \ln \left|\frac{1-x^2}{x^2}\right|\right),
	\label{eq:func_f}
\end{equation}
which is illustrated in Fig.~\ref{fig_2}. Note that $2k_F$ provides a natural momentum scale. $\pi^{\mathrm{ext},\alpha}_{\mathrm{intra}}(\bm{q})$ is a radial function the second derivative of which is logarithmically divergent at momenta with modulus $q=0$ or $q=2k_F$. The latter case defines a sphere
\begin{equation}
	S_{\mathrm{intra}}=\big\{\bm{q}\: \big|\:  |\bm{q}| = 2k_F \big\}
	\label{eq:Sintra}
\end{equation}
on which the extrinsic intravalley polarizability is singular.

Singularities in the polarizability lead to Friedel oscillations in the charge density induced by a point charge. For an isotropic metal, the first derivative of the polarizability shows a logarithmic divergence at $q=2k_F$, the origin of which lies in the Lindhard function, see Eq.\ \eqref{eq:charge_response_function}: Contributions  by momenta $q<2k_F$ permit terms with $E(\bm{k}+\bm{q})\approx E(\bm{k})$, while for $q>2k_F$ such terms cannot arise \cite{Grosso_Parravicini_solid_state_theory}. Due to the divergence in the \emph{first} derivative, the Friedel oscillations in an isotropic metal are given by $\rho\sim\cos(2k_Fr)/r^3$, where $r$ is the distance from the point charge and $2k_F r\gg 1$.

For a Weyl point, processes with $q\approx2k_F$ are suppressed by the spinor overlap since the eigenstates at opposite points on the Fermi surface are orthogonal (see Fig.~\ref{fig_1}). For this reason, the polarizability possesses a logarithmic singularity in its \emph{second} derivative at $q=2k_F$. The resulting Friedel oscillations for isolated Weyl points therefore read as
\begin{equation}\label{eq:ind_char_dens_isolated_WPs}
	\rho(\bm{r})\sim\frac{\sin 2k_F r}{r^4}
\end{equation}
for large $r$ \cite{Lv_Zhang_2013}. Since these Friedel oscillations originate in the universal intravalley polarizability, they occur for all Weyl semimetals but do not allow to distinguish between different spin structures. Note that the singularity at $q=0$ in the extrinsic polarizability does not cause any Friedel oscillations since the spatial frequency of such oscillations would be zero.

\subsubsection{Intrinsic polarizability}\label{subsubsec:intra_int}

The intrinsic intravalley polarizability in Eq.~\eqref{eq:crf_intra_int} stems from the fully occupied valence band. Within this band, no transitions are possible so that the summand with $s=-1$ vanishes. Furthermore, there are no natural momentum or length scales for two reasons. First, only transitions within a single Weyl cone are included so that the separation between the Weyl points is irrelevant. Second, we have $E_F-E^\chi_{-}(\bm{k})\geq0$ for all $\bm{k}$. Consequently, the integration volume is unlimited. This is in contrast to $\pi^{\mathrm{ext},\alpha}_{\mathrm{intra}}(\bm{q})$, where the integration volume is constrained by $k_F$.

The infinite integration volume causes the integral in Eq.\ \eqref{eq:crf_intra_int} to formally diverge for large momenta since the integrand does not fall off sufficiently rapidly. This requires a momentum cutoff $\Lambda$. For the linearized model in Eq.\ \eqref{eq:model_Hamiltonian}, the cutoff is understood as a momentum scale beyond which the linear approximation for the dispersion around the Weyl points fails \cite{Zyuzin_Tiwari_2016}. Due to the lack of any natural momentum scale, the cutoff is only limited by the size of the Brillouin zone.

The cutoff can be implemented in various ways. We follow Ref.~\cite{Lv_Zhang_2013} and introduce two spheres with radius $\Lambda$, where each sphere is centered around one Weyl point. Mathematically, this is achieved by
\begin{equation}
	\int \frac{d^3k}{(2\pi)^3} \rightarrow \int \frac{d^3k}{(2\pi)^3}\: \Theta\left(\Lambda-\left|\bm{k}-\chi \frac{\bm{Q}}{2}\right|\right).\label{eq:cutoff_intra}
\end{equation}
By applying Eq.~\eqref{eq:cutoff_intra} to Eq.~\eqref{eq:crf_intra_int} and performing a change of coordinates for each integral, we obtain an integral that has been evaluated in Ref.~\cite{Lv_Zhang_2013}. The result reads as
\begin{equation}
	\pi^{\mathrm{int},\alpha}_{\mathrm{intra}}(\bm{q}) = 2\, \frac{2}{(2\pi)^2}\,
	\frac{k_F^2}{v_F\hbar}\,
	 u\bigg(\frac{q}{2k_F},\frac{\Lambda}{2k_F}\bigg),
	\label{eq:crf_intra_int_ana}
\end{equation}
with
\begin{equation}
	u(x,\lambda)
	= \frac{2}{3}\, x^2\, \ln \frac{2\lambda}{x} .
	\label{eq:func_u}
\end{equation}
Note that the momentum scale of the extrinsic intravalley polarizability, $2k_F$, has been artificially introduced in Eq.~\eqref{eq:crf_intra_int_ana} to achieve a form similar to Eq.~\eqref{eq:crf_intra_ext_ana}; this scale cancels out. The function $u(x,5)$ is depicted in Fig.~\ref{fig_2}. The dependence of the intrinsic intravalley polarizability on the momentum cutoff differs from the one for the two-dimensional Dirac semimetal graphene, where it is independent of the cutoff~\cite{Hwang_Sarma_2007}. For $q\rightarrow 0$, $\pi^{\mathrm{int},\alpha}_{\mathrm{intra}}(\bm{q})$ is logarithmically divergent in its second derivative.

To conclude, the intravalley polarizability for a pair of isotropic Weyl points is universal. In particular, it is independent of the relative spin structure of the Weyl points. Furthermore, it depends only quantitatively but not qualitatively on the details of the band structure through $k_F$ and $\Lambda$.

\subsection{Intervalley polarizability}\label{sec:intervalley_pola}

In analogy to Eq.\ \eqref{eq:crf_intra}, the intervalley polarizability is also decomposed into extrinsic and intrinsic components,
\begin{equation}
	\pi^\alpha_{\mathrm{inter}}(\bm{q}) = \pi^{\mathrm{ext},\alpha}_{\mathrm{inter}}(\bm{q})
	+ \pi^{\mathrm{int},\alpha}_{\mathrm{inter}}(\bm{q}), \label{eq:crf_inter}
\end{equation}
where
\begin{align}
&\pi^{\mathrm{ext},\alpha}_{\mathrm{inter}}(\bm{q}) \nonumber\\
&=-\sum_{\chi s}\left[\int \frac{d^3k}{(2\pi)^3}\:
  \frac{\Theta(E_F-E^\chi_{+}(\bm{k}))}{E^\chi_{+}(\bm{k})-E^{\bar\chi}_{s}(\bm{k}')+i\eta}\,
  F^{\alpha ;\chi{\bar\chi}}_{+ s}(\bm{k},\bm{k}')\right.\nonumber\\
&\quad{} - \left. \int \frac{d^3k}{(2\pi)^3}\:
  \frac{\Theta(E_F-E^{\bar\chi}_{+}(\bm{k}'))}
  {E^\chi_{s}(\bm{k})-E^{\bar\chi}_{+}(\bm{k}')+i\eta}\,
  F^{\alpha ;\chi{\bar\chi}}_{s +}(\bm{k},\bm{k}') \right],
  \label{eq:crf_inter_ext}\\
&\pi^{\mathrm{int},\alpha}_{\mathrm{inter}}(\bm{q}) \nonumber\\
&=-\sum_{\chi s}\left[\int \frac{d^3k}{(2\pi)^3}\:
  \frac{\Theta(E_F-E^\chi_{-}(\bm{k}))}{E^\chi_{-}(\bm{k})-E^{\bar\chi}_{s}(\bm{k}')+i\eta}\,
  F^{\alpha ;\chi{\bar\chi}}_{- s}(\bm{k},\bm{k}')\right.\nonumber\\
&\quad{} - \left. \int \frac{d^3k}{(2\pi)^3}\:
  \frac{\Theta(E_F-E^{\bar\chi}_{-}(\bm{k}'))}
  {E^\chi_{s}(\bm{k})-E^{\bar\chi}_{-}(\bm{k}')+i\eta}\,
  F^{\alpha ;\chi{\bar\chi}}_{s -}(\bm{k},\bm{k}') \right].\label{eq:crf_inter_int}
\end{align}
For both $\pi^{\mathrm{ext},\alpha}_{\mathrm{inter}}(\bm{q})$ and $\pi^{\mathrm{int},\alpha}_{\mathrm{inter}}(\bm{q})$, the spinor overlaps depend on the relative spin structure of the two Weyl points, unlike for their intravalley counterparts. As a result, the intervalley polarizability in Weyl semimetals is not universal but rather contains signatures of distinct spin structures, as we will see.

\subsubsection{Extrinsic polarizability}

\begin{table}
	\caption{Coefficient functions for the extrinsic intervalley polarizability in Eq.~\eqref{eq:inter_ext_crf_ana} for each model system.}
		\begin{tabular}{cc@{\quad}c}
			\hline\hline
			Model system & $c^{\chi,\alpha}_f$  & $c^{\chi,\alpha}_g$ \\\hline
			\rule{0pt}{5ex}
			$a$ & $0$ & $\displaystyle\frac{1}{2}$ \\
			\rule{0pt}{5ex}
			$b$
			& $\displaystyle\frac{1}{2}\, \bigg[1+\bigg(\frac{y_z^\chi}{y^\chi}\bigg)^{\!2}\,\bigg]$
			& $\displaystyle\frac{1}{2}\, \bigg(\frac{y_z^\chi}{y^\chi}\bigg)^{\!2}$ \\
			\rule{0pt}{5ex}
			$c$
			& $\displaystyle\frac{1}{2}\, \bigg[1+\bigg(\frac{y_x^\chi}{y^\chi}\bigg)^{\!2}\,\bigg]$
			& $\displaystyle\frac{1}{2}\, \bigg(\frac{y_x^\chi}{y^\chi}\bigg)^{\!2}$ \\
			\hline\hline
		\end{tabular}
		\label{table:coeffs_cf_cg}
\end{table}

First, we consider the extrinsic contribution in Eq.\ \eqref{eq:crf_inter_ext}. In contrast to Eq.~\eqref{eq:crf_intra_ext}, where the integrand becomes isotropic after a suitable change of coordinates, the integrand in Eq.~\eqref{eq:crf_inter_ext} is anisotropic due to the separation of the Weyl points and the spin structure. For models $a$ and $b$, the integrands are rotationally symmetric around the $k_z$ axis, while there is a preferred direction orthogonal to the separation direction for model $c$. In Appendix \ref{app:ext_inter_crf}, we present the evaluation of Eq.~\eqref{eq:crf_inter_ext} for model $c$. The calculations for the other two models are performed analogously. For each model system, the extrinsic intervalley polarizability is of the form
\begin{equation}
	\pi^{\mathrm{ext},\alpha}_{\mathrm{inter}}(\bm{q}) = \frac{2}{(2\pi)^2}\,
	\frac{k_F^2}{v_F\hbar} \sum_\chi \big[ c_f^{\chi,\alpha} f(y^\chi)
	+ c_g^{\chi,\alpha} g(y^\chi) \big] ,
	\label{eq:inter_ext_crf_ana}
\end{equation}
where $\bm{y}^\chi=(y^\chi_x,y^\chi_y,y^\chi_z)=(\bm{q}+\chi \bm{Q})/2 k_F$, $y^\chi = |\bm{y}^\chi|$,
\begin{equation}\label{eq:func_g}
	g(x) = x^2 \ln \left|\frac{1-x^2}{x^2}\right|
	+ x \ln \left|\frac{1+x}{1-x}\right| - 1,
\end{equation}
and the coefficient functions $c^{\chi,\alpha}_f$ and $c^{\chi,\alpha}_g$, which depend on the model system, are given in Table \ref{table:coeffs_cf_cg}. The function $g(x)$ is plotted in Fig.~\ref{fig_2}.

The first point to notice is that the extrinsic intervalley polarizability contains the new function $g(y^\chi)$ for every model system, while the function $f(y^\chi)$, which was introduced for the extrinsic intravalley polarizability, appears only for models $b$ and $c$. The origin of $g(y^\chi)$ is best understood for model $a$. For this model, the two Weyl points of opposite chirality have inverted spin structures (see Fig.~\ref{fig_1}). This means that transitions between states at opposing points on Fermi surfaces surrounding the two Weyl points are enhanced. For instance, transitions with momentum transfers $\bm{q}=(Q+2k_F)\, \hat{\bm{e}}_z$ [$\bm{q}=(Q-2k_F)\, \hat{\bm{e}}_z$] are enhanced between the south (north) pole of the negative-chirality Weyl point's Fermi surface and the north (south) pole of the positive-chi\-ra\-li\-ty Weyl point's Fermi surface. This is similar to the transitions in an isotropic metal (see Sec.~\ref{sec:intra_pola}). Hence, $g(y^\chi)$ has to possess a logarithmic divergence in its \emph{first} derivative at $y^\chi=1$. Thus, there is a logarithmic divergence in the first derivative of $\pi^{\mathrm{ext},a}_{\mathrm{inter}}(\bm{q})$, and the extrinsic intervalley polarizability is singular on two spheres with radius $2k_F$ centered around $\mp\bm{Q}$:
\begin{equation}
	S^\pm_{\mathrm{inter}} = \big\{\bm{q}\: \big|\: |\bm{q}\pm Q\hat{\bm{e}}_z| = 2k_F \big\}.
	\label{eq:Sinter}
\end{equation}

For models $b$ and $c$, the spin structures of the two Weyl points are only partially inverted. For this reason, certain transitions between states at opposing points on the Fermi surfaces surrounding the two Weyl points are either enhanced or suppressed depending on the specific spin structures. Similar to model $a$, the extrinsic polarizabilities of models $b$ and $c$ show singularities on $S^\pm_{\mathrm{inter}}$. However, in contrast to model $a$, the character of the singularities on $S^\pm_{\mathrm{inter}}$ is anisotropic: In different directions logarithmic divergences in the derivatives may occur in different orders. This is reflected by the momentum-dependent coefficient functions $c_f^{\chi,\alpha}$ and $c_g^{\chi,\alpha}$. In Sec.~\ref{sec:FO}, the consequences of this anisotropy for the Friedel oscillations are discussed in more detail.

Note that the extrinsic intervalley polarizability in Eq.~\eqref{eq:inter_ext_crf_ana} has further logarithmic divergences in its second derivative at $\bm{q}=\pm\bm{Q}$ for each model system. These singularities are of essentially the same nature as the singularity of the extrinsic intravalley polarizability in Eq.~\eqref{eq:crf_intra_ext_ana} at $q=0$. This becomes apparent from the comparison of the last and first terms in $f(x)$ and $g(x)$, respectively.

\subsubsection{Intrinsic polarizability and tight-binding model}

Our next task is to evaluate the intrinsic intervalley polarizability in Eq.~\eqref{eq:crf_inter_int}. In analogy to Eq.~\eqref{eq:crf_intra_int}, the summands for $s=-1$ vanish and the integration volume is unrestricted. Therefore, we introduce the same cutoff as for the intrinsic intravalley polarizability, i.e., we apply Eq.~\eqref{eq:cutoff_intra} to \eqref{eq:crf_inter_int}. In contrast to the intrinsic intravalley polarizability, the intrinsic intervalley polarizability depends on the momentum scale $Q$. To prevent the Weyl cones from overlapping, the cutoff should satisfy $\Lambda<Q/2$.

The evaluation of Eq.\ \eqref{eq:crf_inter_int} up to the leading order in $\Lambda$ yields
\begin{equation}\label{eq:crf_inter_int_ana}
	\pi^{\mathrm{int},\alpha}_{\mathrm{inter}}(\bm{q})
	\approx \frac{2}{(2\pi)^2}\, \frac{1}{v_F\hbar}\, \Lambda^2,
\end{equation}
for all three model systems. In contrast to the weak, logarithmic divergence of $\pi^{\mathrm{int},\alpha}_{\mathrm{intra}}(\bm{q})$, $\pi^{\mathrm{int},\alpha}_{\mathrm{inter}}(\bm{q})$ shows a strong, quadratic divergence. Such strong dependencies of physical quantities on cutoffs for Weyl semimetals have been encountered before \cite{Carbotte_2016, Rodriguez_Lopez_Woods_2020}. However, Eq.\ \eqref{eq:crf_inter_int_ana} is surprising since we would not expect a large contribution of the intervalley polarizability for small $\bm{q}$. Rather, large contributions should only appear for momenta $\bm{q}$ comparable to $\bm{Q}$. Additionally, Eq.\ \eqref{eq:crf_inter_int_ana} causes unphysical long-range screening of the Coulomb interaction in the absence of free charge carriers, as shown in Appendix \ref{app:screening_intrinsic}.

The intrinsic intervalley polarizability has been obtained based on the Weyl Hamiltonian, which is an approximation for the dispersion of a realistic system in the vicinity of a Weyl point. For any reasonable approximation, the deviations between its predictions and the exact quantities should be negligible. For the  continuum Weyl approximation, this means that the contributions of the realistic lattice model that originate from the nonlinear regions away from the Weyl points must be small. These contributions are in general system specific and contain no Weyl physics. As discussed in Appendix \ref{app:int_pola}, the integrand of the intrinsic intravalley polarizability in Eq.~\eqref{eq:crf_intra_int} is small at the boundary of the linear regime, which is described by the cutoff. This suggests that contributions that stem from regions beyond the linear regime are small as well. Consequently, the intrinsic intravalley polarizability in Eq.~\eqref{eq:crf_intra_int_ana} should describe a realistic system qualitatively correctly. On the other hand, the integrand of the intrinsic intervalley polarizability is large at the boundary of the integration volume. Hence, non-linear regions may contribute significantly to the full polarizability. For this reason, Eq.~\eqref{eq:crf_inter_int_ana} is at least questionable.

Lattice models avoid the convergence problems introduced by continuum models. In the following, we study the full intrinsic polarizability for a tight-binding model, focusing on the following points. First, we need to verify whether Eq.~\eqref{eq:crf_inter_int_ana} is indeed incapable of describing a realistic system. In other words, we need to check if a constant term in the intrinsic polarizability for $q=0$ exists. Second, if the cutoff approximation fails for the intrinsic intervalley polarizability, this also raises doubts concerning its validity for the intravalley one. For this reason, we need to verify if Eq.~\eqref{eq:crf_intra_int_ana} is describing the full polarizability for small $\bm{q}$ qualitatively correctly. Third, for a full picture of the Friedel oscillations, it is necessary to investigate whether the full intrinsic polarizability of the tight-binding model contains singularities. If this is the case, we are interested in strengths and locations of these singularities.

\begin{figure}[h!]
	\hspace{-7.6cm}(a)

	\includegraphics[scale=0.75]{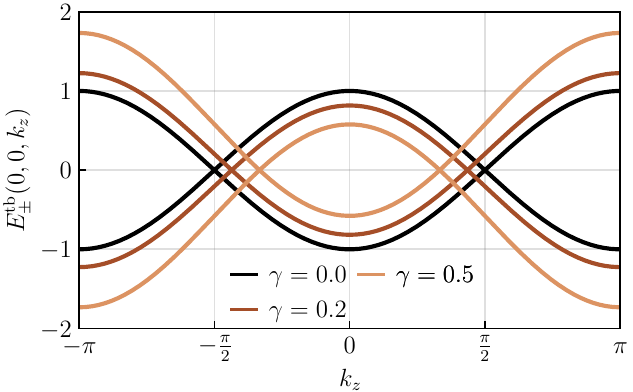}

	\hspace{-7.6cm}(b)

	\includegraphics[scale=0.75]{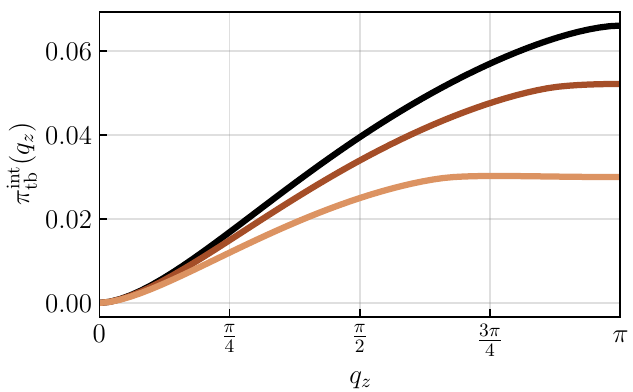}

	\hspace{-7.6cm}(c)

	\includegraphics[scale=0.75]{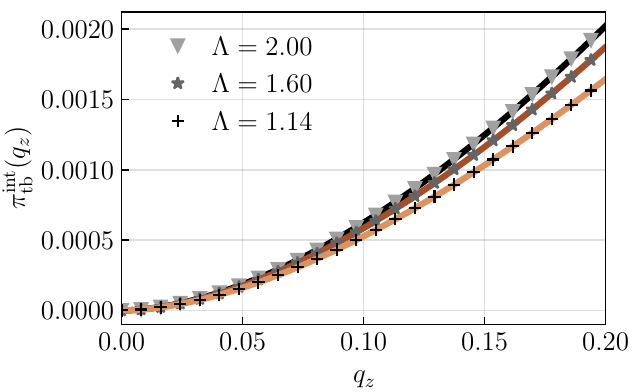}
	\caption{Band structure and intrinsic polarizability for the tight-binding model given in Eq.~\eqref{eq:tb_Ham}  with $\gamma=0$ (black), $\gamma=0.2$ (dark brown), and $\gamma=0.5$ (light brown). (a) Band structure along the $k_z$ axis. (b) Intrinsic polarizability as a function of $q_z$ for $q_x=q_y=0$. (c) Comparison of  $\pi^{\mathrm{int}}_{\mathrm{tb}}(\bm{q})$ (solid lines) and $\pi^{\mathrm{int},\alpha}_{\mathrm{intra}}(\bm{q})$ for $\Lambda=2$ (triangles), $\Lambda=1.6$ (stars), and $\Lambda=1.14$ (plus signs). The remaining parameters for the tight-binding model are $\eta=3 \times 10^{-4}$, $t_x=t_y=1$, and $t_z=1/\sqrt{1-\gamma^2}$. For $\pi^{\mathrm{int},\alpha}_{\mathrm{intra}}(\bm{q})$, $\hbar v_F=1$ was chosen.}\label{fig_3}
\end{figure}

We employ the tight-binding Hamiltonian~\cite{Yang_Lu_2011, Trescher_Bergholtz_2017, Armitage_Mele_2018}
\begin{align}
	&H_{\mathrm{tb}}(\bm{k}) = t_x \sin k_x\: \tau_x + t_y \sin k_y\: \tau_y
	\nonumber\\
	&\quad{} + t_z\, (2-\cos k_x - \cos k_y + \gamma - \cos k_z)\, \tau_z ,
	\label{eq:tb_Ham}
\end{align}
where $t_x$, $t_y$, and $t_z$ are hopping amplitudes and $\gamma$ controls the Weyl phase. This model breaks time-reversal symmetry and can be constructed employing two orbitals of opposite parity, namely, an $s$ and a $p_z$ orbital, per lattice site and a strong magnetic field that shifts states of opposite spin to high energies \cite{Armitage_Mele_2018}. The local degree of freedom described by the Pauli matrices $\tau_x$, $\tau_y$, and $\tau_z$ is a combination of spin and orbital contributions due to strong spin-orbit coupling but for simplicity we continue to call it the ``spin.'' Then, inversion symmetry is described by the matrix $P=\tau_z$ (see Appendix~\ref{app:tight_binding} for further discussions of symmetries). The dispersion relation is given by
\begin{align}
	&E_s^{\mathrm{tb}}(\bm{k}) = s\, \big[ t_x^2 \sin^2 k_x
	+ t_y^2 \sin^2 k_y \nonumber \\
	&\quad{} + t_z^2\, (2-\cos k_x - \cos k_y + \gamma - \cos k_z)^2 \big]^{1/2} ,	\label{eq:dispersion_tb}
\end{align}
where $s=\pm 1$ is the band index. In the Weyl phase, which is characterized by $-1<\gamma<1$, a pair of Weyl points exists on the $k_z$ axis at $k_z = k^\pm_{z0}=\pm\arccos\gamma$, as shown in Fig.\ \ref{fig_3}(a) for  three values of $\gamma$. For $t_x=t_y=t_F$ and $t_z=t_F/\sqrt{1-\gamma^2}$, the dispersion in the vicinity of the Weyl points is isotropic, which we assume in the following. By expanding $H_{\mathrm{tb}}(\bm{k})$ around the Weyl points up to first order in momentum, one can verify that the relative spin structure of the tight-binding model corresponds to the one of model $b$ in Eq.~\eqref{eq:pseudo_b}.

For the tight-binding model, the intrinsic polarizability is calculated using
\begin{align}
&\pi^{\mathrm{int}}_{\mathrm{tb}}(\bm{q}) \nonumber\\
&=-\left[\int_{\mathrm{BZ}} \frac{d^3k}{(2\pi)^3}\:
  \frac{1}{E^{\mathrm{tb}}_{-}(\bm{k})-E^{\mathrm{tb}}_{+}(\bm{k}')+i\eta}\,
  F^{\mathrm{tb}}_{- +}(\bm{k},\bm{k}')\right.\nonumber\\
&\quad{} - \left. \int_{\mathrm{BZ}} \frac{d^3k}{(2\pi)^3}\:
\frac{1}{E^{\mathrm{tb}}_{+}(\bm{k})-E^{\mathrm{tb}}_{-}(\bm{k}')+i\eta}\,
  F^{\mathrm{tb}}_{+ -}(\bm{k},\bm{k}') \right],\label{eq:crf_int_tb}
\end{align}
where $F^{\mathrm{tb}}_{+ -}(\bm{k},\bm{k}')$ is the absolute value squared of the spinor overlap for states with momenta $\bm{k}$ and $\bm{k}'$ in the conduction and valence bands, respectively, and the integration volume is the first Brillouin zone. In contrast to the low-energy continuum model, we cannot naturally separate intravalley and intervalley contributions for the tight-binding model.

Equation \eqref{eq:crf_int_tb} can be evaluated numerically. In Fig.\ \ref{fig_3}(b), we plot $\pi^{\mathrm{int}}_{\mathrm{tb}}(\bm{q})$ along the separation direction of the Weyl points, i.e., on the $q_z$ axis. We compare this figure with Eqs.~\eqref{eq:crf_intra_int_ana} and \eqref{eq:crf_inter_int_ana}. For $|\bm{q}|\rightarrow 0$, $\pi^{\mathrm{int}}_{\mathrm{tb}}(\bm{q})$ vanishes. This is consistent with $\pi^{\mathrm{int},\alpha}_{\mathrm{intra}}(\bm{q})$, but in contradiction to $\pi^{\mathrm{int},\alpha}_{\mathrm{inter}}(\bm{q})$. Noting the discussion in Appendix \ref{app:int_pola}, we conclude that the intrinsic intervalley polarizability cannot be calculated within the Weyl approximation.

Next, we compare $\pi^{\mathrm{int}}_{\mathrm{tb}}(\bm{q})$ and $\pi^{\mathrm{int},\alpha}_{\mathrm{intra}}(\bm{q})$ for small $q_z$. As illustrated in Fig.~\ref{fig_3}(c), good agreement between the numerical and analytical results is achieved by choosing the cutoff $\Lambda$ for the continuum model appropriately. In addition, this comparison emphasizes that for small $\bm{q}$, the intrinsic charge response function $\pi^{\mathrm{int}}_{\mathrm{tb}}(\bm{q})$ of the tight-binding model, the spin structure of which corresponds to model $b$ in Eq.\ \eqref{eq:pseudo_b}, is indeed independent of the spin structure of the Weyl points.

To summarize, there is no universal expression for the total polarizability of a centrosymmetric Weyl semimetal within the continuum Weyl approximation due to an unphysical divergence of the intrinsic intervalley polarizability. Ultimately, this divergence can be traced back to a breakdown of the low-energy continuum approximation for a realistic model system. On the other hand, our analysis suggests that the intrinsic intravalley polarizability in Eq.~\eqref{eq:crf_intra_int_ana}, which was obtained within the continuum Weyl approximation, well describes the intrinsic polarizability of a realistic system for small $\bm{q}$.

\begin{figure}[h!]
	\hspace{-7.6cm}(a)

	\includegraphics[scale=0.75]{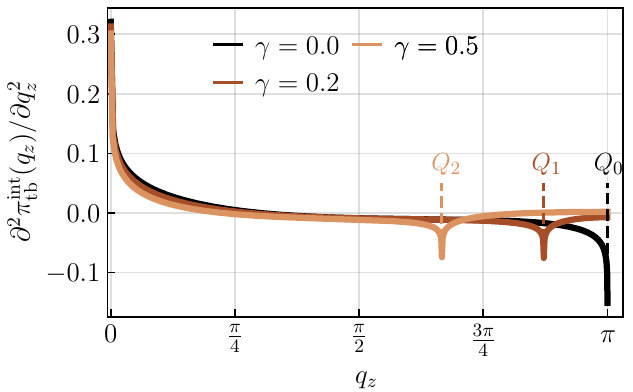}

	\hspace{-7.6cm}(b)

	\includegraphics[scale=0.75]{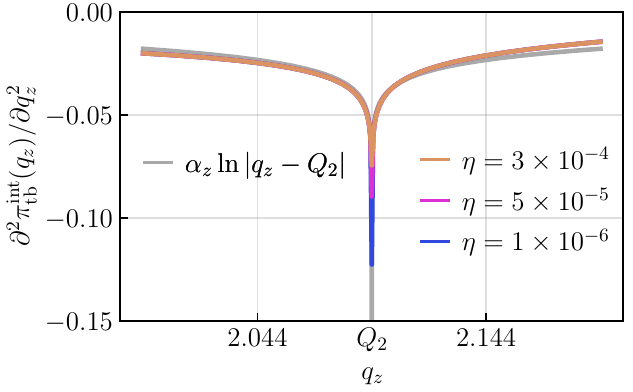}

	\hspace{-7.6cm}(c)

	\includegraphics[scale=0.75]{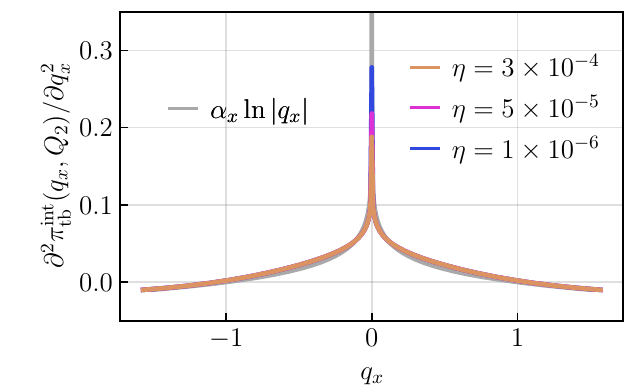}
	\caption{Second derivatives of the intrinsic polarizability of the tight-binding model with respect to different momentum components for various $\gamma$ and broadenings $\eta$.
		(a) Second derivative in $q_z$ for $q_x=q_y=0$ and various $\gamma$. The separations between the Weyl points are given by $Q_0=\pi$, $Q_1=2.74$, and $Q_2=2.09$ for the three values of $\gamma$.
		(b) For $\gamma=0.5$, comparison of the second derivative in $q_z$ in the vicinity of $Q_2$ along the $q_z$ axis for various $\eta$ with $\alpha_z \ln|q_z-Q_2|$, where $\alpha_z=0.0078$.
		(c) For $\gamma=0.5$, comparison of the second derivative in $q_x$ for $q_y=0$ and $q_z=Q_2$ for various $\eta$ with $\alpha_x \ln|q_x|$, where $\alpha_x=-0.024$.
		Unless stated otherwise, the used parameters are the same as in Fig.~\ref{fig_3}.}\label{fig_4}
\end{figure}

Although the Weyl approximation fails to describe the full intrinsic polarizability of a centrosymmetric Weyl semimetal, the full intrinsic polarizability might still contain potential singularities, which we investigate for the tight-binding model. To get further insight, we plot the second derivative of $\pi^{\mathrm{int}}_{\mathrm{tb}}(\bm{q})$ with respect to $q_z$ along the separation direction of the Weyl points in Fig.~\ref{fig_4}(a). For each value of $\gamma$, we observe a singularity at $q_z=0$. According to Eqs.~\eqref{eq:crf_intra_int_ana} and \eqref{eq:func_u}, this singularity is logarithmic. Furthermore, we find singularities at $Q=2\arccos\gamma$, the separation between the Weyl points. Note that a small but nonzero broadening $\eta$ used in the numerical evaluations cuts off the divergences. In Fig.~\ref{fig_4}(b), the singularity for $\gamma=0.5$ is shown for several $\eta$. As we can see, the singularity fits a logarithmic divergence centered at $Q_2$. Similarly, the second derivative of $\pi^{\mathrm{int}}_{\mathrm{tb}}(q_x,0,Q_2)$ with respect to $q_x$ for the same $\eta$ fits a logarithmic divergence centered around zero, see Fig.~\ref{fig_4}(c). We conclude that the intrinsic polarizability has logarithmic divergences in its second derivative at $\bm{q}=\pm Q\hat{\bm{e}}_z$.

As mentioned before, the Weyl points of our tight-binding model possess the spin structure of model $b$. Thus, the spin structures of the Weyl points are different for directions parallel and orthogonal to the separation direction of the Weyl points. However, at $\bm{q}=\pm Q\hat{\bm{e}}_z$, logarithmic divergences in the second derivatives occur along both these directions, as we have just discussed. For this reason, the strength of these singularities is independent of the spin structure. We conjecture that the intrinsic polarizability for other centrosymmetric Weyl semimetals with different spin structures shows analogous behavior near the points $\bm{q}=0$ and $\pm Q\hat{\bm{e}}_z$.

\section{Friedel oscillations}\label{sec:FO}

In this section, we calculate the charge-density Friedel oscillations induced by a point charge $Q_c>0$. For this purpose, we consider the dielectric function within the random-phase approximation, which reads as
\begin{equation}\label{eq:RPA}
	\epsilon^\alpha(\bm{q}) = 1 + V_C(\bm{q})\,\pi^\alpha(\bm{q}),
\end{equation}
where $V_C(\bm{q})={e^2}/{\epsilon_0 q^2}$ is the Coulomb interaction, $e>0$ is the elementary charge, and $\epsilon_0$ is the dielectric constant. The induced charge density is then expressed as~\cite{Grosso_Parravicini_solid_state_theory}
\begin{equation}\label{eq:ind_charge_dens_tot}
	\rho^\alpha(\bm{r})=Q_c\int \frac{d^3q}{(2\pi)^3}\,\left[\frac{1}{\epsilon^\alpha(\bm{q})}-1\right]e^{i\bm{q}\cdot\bm{r}}.
\end{equation}
As discussed above, Friedel oscillations in the induced charge density arise from singularities in the derivatives of the polarizability. For Weyl semimetals, a variety of singularities appear in the different parts of the total polarizability. To summarize, both the extrinsic and the intrinsic polarizabilities contain singularities at $\bm{q}=0$ and $\pm\bm{Q}$. However, only the former contribution contains additional singularities that occur on spheres with radius $2k_F$ centered around the aforementioned momenta $\bm{q}$. In the following, we first focus on the point singularities at $\bm{q}=0$ and $\pm\bm{Q}$.

For the intrinsic polarizability of a Weyl semimetal, there is no universal expression. From the tight-binding model, we have found three properties. First, singularities in the intrinsic polarizability occur at $\bm{q}=0$ and $\pm Q\hat{\bm{e}}_z$. Second, these singularities are logarithmic divergences in the second derivative. Third, they are independent of the spin structure of the Weyl points. All three properties also apply to the singularities of the extrinsic polarizability at the same momenta for each model system. We therefore conjecture that all centrosymmetric Weyl semimetals that host one pair of Weyl points give rise to singularities with the same properties as those of the present tight-binding model.

In the total polarizability, the singularities contributed by the extrinsic and intrinsic polarizabilities are added together. They do not cancel in general since the coefficients of the singular terms in the extrinsic and intrinsic polarizabilities depend and do not depend, respectively, on the Fermi energy [see Eqs.~\eqref{eq:crf_intra_ext_ana}, \eqref{eq:inter_ext_crf_ana}, and \eqref{eq:crf_int_tb}]. For this reason, the total polarizability of a Weyl semimetal contains logarithmic singularities in its second derivative at $\bm{q}=0$ and $\pm\bm{Q}$.

The relevant features of Friedel oscillations are the scaling of their amplitude vs.\ distance and their oscillation frequency. The former depends on the strengths (orders) of the singularities in the polarizability, while the latter is governed by the locations of the singularities in momentum space. For Weyl semimetals, the locations and strengths of all singularities of the total polarizability are identical to those of the extrinsic contribution, whereas the intrinsic polarizability has only the point singularities in common with them. For this reason, it is sufficient to calculate the Friedel oscillations in the induced charge density in Eq.~\eqref{eq:ind_charge_dens_tot} by including only the extrinsic polarizability. The singularities of the intrinsic polarizability cannot give rise to Friedel oscillations with different scaling and frequencies.

\subsection{Model system \boldmath{$a$}}

\begin{figure*}
	\begin{center}
		\raisebox{42ex}[0ex][5ex]{(a)}\includegraphics[scale=0.70]{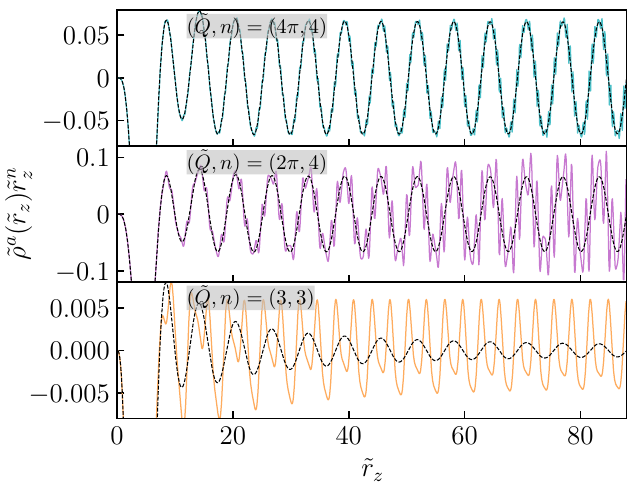}\hspace{0.5cm}
		\raisebox{42ex}[0ex][5ex]{(b)}\includegraphics[scale=0.68]{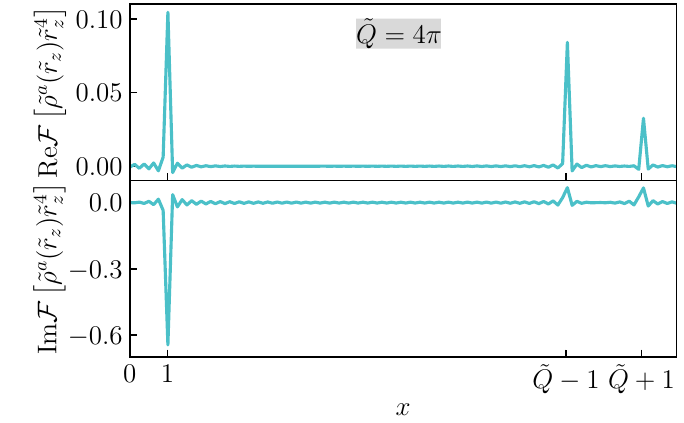}
		\raisebox{37ex}[0ex][0ex]{(c)}\includegraphics[scale=0.68]{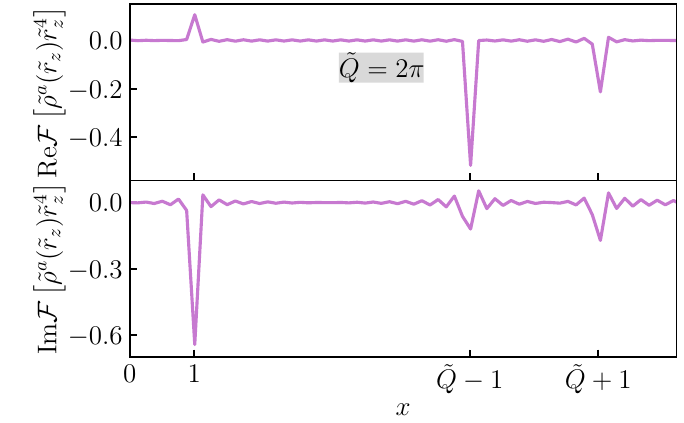}\hspace{0.5cm}
		\raisebox{37ex}[0ex][0ex]{(d)}\includegraphics[scale=0.68]{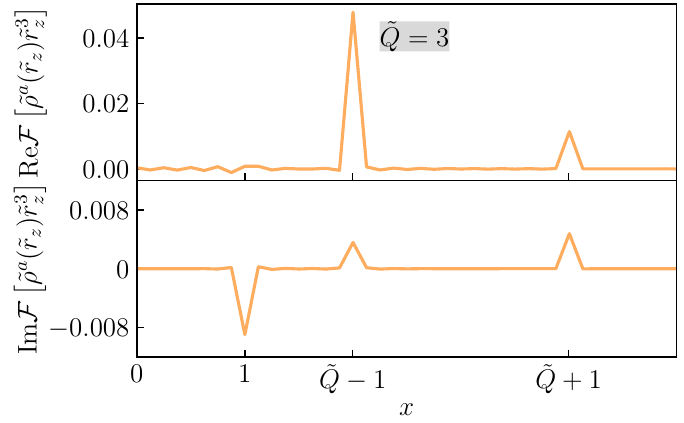}
		\caption{Induced charge density and frequency spectrum of the Friedel oscillations for model system $a$. (a) Induced charge density for various $\tilde{Q}=Q/2k_F$ along the $\tilde{r}_z$ axis ($\tilde{r}_x=\tilde{r}_y=0$), scaled with $\tilde{r}_z^n$, where $n=4$ for the upper two plots and $n=3$ for the lower plot. The black dashed curve shows the induced charge density including only the extrinsic intravalley polarizability. The effective fine-structure constant is taken to be $r_s=1$. (b)--(d) Fourier components of the discrete Fourier transform of the Friedel oscillations from panel~(a).}\label{fig_5}
	\end{center}
\end{figure*}

For the calculation of the Friedel oscillations, where we take into account only the extrinsic polarizability, it is convenient to introduce $2k_F$ as a natural momentum scale. Then, the dimensionless induced charge density due to the extrinsic polarizability reads as
\begin{equation}\label{eq:dim_less_charge_dens_ext}
	\tilde{\rho}^\alpha(\tilde{\bm{r}})=\frac{\rho^{\mathrm{ext},\alpha}_{\mathrm{ind}}(\tilde{\bm{r}})}{Q_c (2k_F)^3}=\int \frac{d^3x}{(2\pi)^3}\,\left[\frac{1}{\epsilon^{\mathrm{ext},\alpha}(\bm{x})}-1\right]e^{i\bm{x}\cdot\tilde{\bm{r}}},
\end{equation}
with $\tilde{\bm{r}}=2k_F\bm{r}$, $\bm{x}=\bm{q}/2k_F$,
\begin{align}
	\epsilon^{\mathrm{ext},\alpha}(\bm{x})
	&= 1 + \frac{r_s}{4x^2}\, \tilde{\pi}^{\mathrm{ext},\alpha}(\bm{x})
	\label{eq:external_dielec_func} , \\
	\tilde{\pi}^{\mathrm{ext},\alpha}(\bm{x})
	&= 2f(x)
	+ \sum_\chi \big[ c_f^{\chi,\alpha} f(y^\chi) + c_g^{\chi,\alpha} g(y^\chi) \big] ,
	\label{eq:induced_charge_dens_ext}
\end{align}
where $r_s=e^2/2\pi^2\epsilon_0 v_F\hbar$ is the effective fine-structure constant of the Weyl semimetal. Here, $\tilde{\pi}^{\mathrm{ext},\alpha}(\bm{x})$ is the dimensionless extrinsic polarizability consisting of intravalley and intervalley contributions. Aside from the point singularities at $\bm{q}=0$ and $\pm Q\hat{\bm{e}}_z$, $\tilde{\pi}^{\mathrm{ext},\alpha}(\bm{x})$ is singular on the surfaces $S_{\mathrm{intra}}$ and $S_{\mathrm{inter}}^\pm$ [see Eqs.\ \eqref{eq:Sintra} and \eqref{eq:Sinter}, respectively]. For each model, the interplay of these singularities causes unique Friedel oscillation patterns, as we will see in the following.

For model $a$, we have already explored the singularities on $S_{\mathrm{intra}}$ and $S^\pm_{\mathrm{inter}}$. While there are logarithmic divergences in the second derivative on $S_{\mathrm{intra}}$, logarithmic divergences appear in the first derivative of the dielectric function on $S^\pm_{\mathrm{inter}}$. A logarithmic divergence in the first derivative causes Friedel oscillations of the form $\cos\tilde r/\tilde r^3$. Since the singularities on $S^\pm_{\mathrm{inter}}$ are stronger than those on $S_{\mathrm{intra}}$ they determine the Friedel oscillations at larger separations. For directions parallel and orthogonal to the separation direction of the Weyl points, the Friedel oscillations therefore behave as
\begin{align}
	\tilde{\rho}^a(\tilde{r}_z) &\sim \frac{
	\cos\big[(\tilde{Q}-1)\tilde{r}_z\big] +
	A\cos\big[(\tilde{Q}+1)\tilde{r}_z\big]
	}{\tilde{r}_z^3},
	\label{eq:ind_char_dens_inter_parallel}\\
	\tilde{\rho}^a(\tilde{r}_\perp) &\sim \frac{\cos\tilde{r}_\perp}{\tilde{r}_\perp^3}
	\label{eq:ind_char_dens_inter_perp},
\end{align}
for large $\tilde{r}_z$ and large $\tilde{r}_\perp$, respectively. $A$ is the relative amplitude between the oscillations with spatial frequencies $\tilde{Q}-1$ and $\tilde{Q}+1$, where $\tilde{Q}=Q/2k_F$.

To understand how the distinct Friedel oscillations resulting from the singularities on $S_{\mathrm{intra}}$ and $S_{\mathrm{inter}}^\pm$ merge, we plot the dimensionless charge density from Eq.~\eqref{eq:dim_less_charge_dens_ext} along the $\tilde{r}_z$ axis for model system $a$ in Fig.~\ref{fig_5}(a). For comparison, we also show the induced charge density of systems consisting of two isolated Weyl cones, i.e., including only the intravalley contribution. The long-distance behavior is then described by Eq.~\eqref{eq:ind_char_dens_isolated_WPs}.

For Weyl points separated by $\tilde{Q}=Q/2k_F=4\pi$ [upper plot in Fig.\ \ref{fig_5}(a)], the induced charge density for small $\tilde{r}_z$ follows the charge density obtained for two isolated Weyl points. For larger $\tilde{r}_z$, small deviations become noticeable. Moving the Weyl points closer together (middle plot) reduces the small-$\tilde{r}_z$ region in which the induced charge densities for model $a$ and for isolated Weyl points agree. As one can see, the intervalley contribution generates additional oscillations, the amplitudes of which fall off more slowly than $1/\tilde{r}^4_z$. By moving the Weyl points even closer together (lower plot), we see that the induced charge density for model $a$ and for isolated Weyl points agrees only in a narrow region. Most importantly, we observe that $\tilde{\rho}^\alpha(\tilde{r}_z)$ falls off as $1/\tilde{r}^3_z$, as expected from Eq.\ \eqref{eq:ind_char_dens_inter_parallel}, in contrast to the induced charge density of isolated Weyl points, which falls off as $1/\tilde{r}_z^4$.

We analyze the spatial oscillation frequencies of the induced charge densities by calculating the discrete Fourier transforms $\mathcal{F}$ for the oscillations of $\tilde{\rho}^a(\tilde{r}_z)\, \tilde{r}^n_z$ for intervals that comprise approximately eight periods, see Appendix \ref{app:dft_fos} for details. Within the investigated intervals, the dominant oscillation frequency for $\tilde{Q}=4\pi$ is $x= q/2k_F=1$, as shown in Fig.~\ref{fig_5}(b). As $\tilde{Q}$ decreases to $\tilde{Q}=2\pi$, see Fig.~\ref{fig_5}(c), the oscillations with frequencies $x=\tilde{Q}\pm1$ increase and eventually become dominant for $\tilde{Q}=3$, as shown in Fig.~\ref{fig_5}(d). We conclude that the intravalley singularity on $S_{\mathrm{intra}}$ dominates for small $\tilde{r}_z$, while the intervalley singularities on $S_{\mathrm{inter}}^\pm$ dominate for large $\tilde{r}_z$, where the length scale of the crossover between these two regimes depends on the separation $\tilde{Q}$ between the Weyl points.

\begin{figure}
		\includegraphics[scale=0.70]{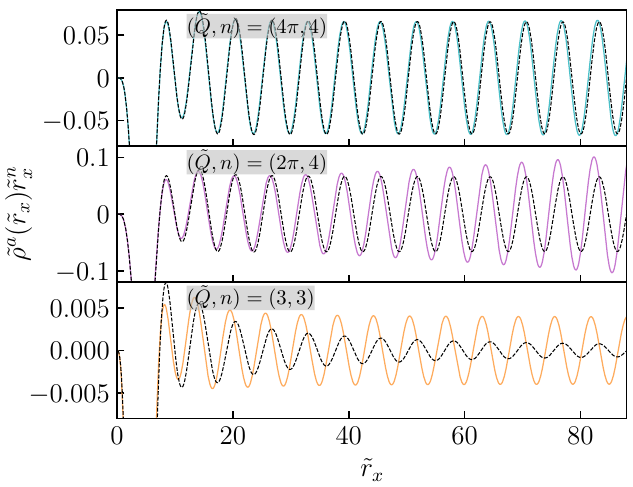}
		\caption{Induced charge density for various $\tilde{Q}=Q/2k_F$ along the $\tilde{r}_x$ axis for model system $a$. The black dashed curve shows the induced charge density including only the extrinsic intravalley polarizability. The effective fine-structure constant is taken to be $r_s=1$. }\label{fig_6}
\end{figure}

Interestingly, we do not observe Friedel oscillations with spatial frequency $x=\tilde{Q}$, which could be expected due to the logarithmic divergences in the second derivative of the dimensionless extrinsic polarizability at $\bm{x}=\pm\tilde{Q}\hat{\bm{e}}_z$. The absence of such oscillations is likely due to the fact that point singularities form a null set in momentum space compared to the Friedel oscillations originating from the surfaces $S_{\mathrm{intra}}$ and $S_{\mathrm{inter}}^\pm$. Our results are in disagreement with a conjecture in Ref.~\cite{Lv_Zhang_2013}, where cosinusoidal Friedel oscillations with spatial frequency $q=Q$ that fall off as $1/r_z^3$ have been suggested.

So far, we have focused on the direction parallel to the separation vector between the Weyl points. We now turn to the orthogonal direction. Figure \ref{fig_6} shows the Friedel oscillations along the $\tilde{r}_x$ axis. The results are similar to the previous case. Specifically, we observe a crossover from $1/\tilde{r}_x^4$ scaling for small $\tilde{r}_x$ to $1/\tilde{r}_x^3$ scaling for large $\tilde{r}_x$. For small $\tilde{r}_x$, the intravalley singularity dominates so that the induced charge density is close to the one for two isolated Weyl points. For larger $\tilde{r}_x$, the intervalley singularities become more and more pronounced so that the induced charge density approaches Eq.~\eqref{eq:ind_char_dens_inter_perp}. Note that no oscillations with frequencies on the order of $\tilde{Q}$ appear. This is reasonable because $\tilde{Q}$ refers to a direction that is orthogonal to $\tilde{r}_x$. However, the scale $\tilde{Q}$ still determines the crossover between the regimes of small and large $\tilde{r}_x$.

\subsection{Model system \boldmath{$b$}}\label{sec:FOs_b}

\begin{figure}
		\raisebox{41ex}[0ex][3ex]{(a)}\includegraphics[scale=0.70]{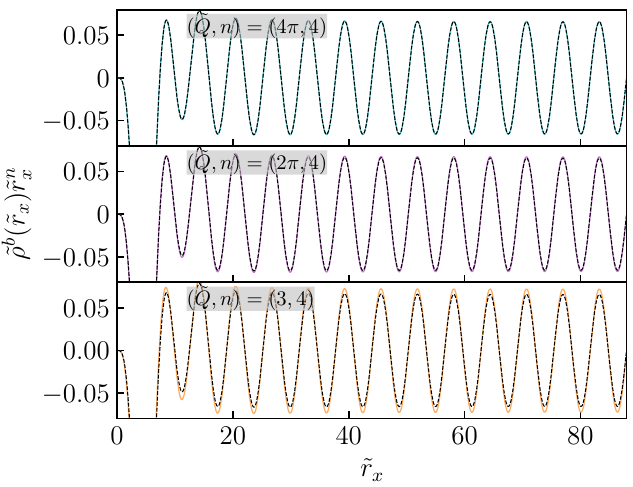}
		\raisebox{41ex}[0ex][3ex]{(b)}\includegraphics[scale=0.70]{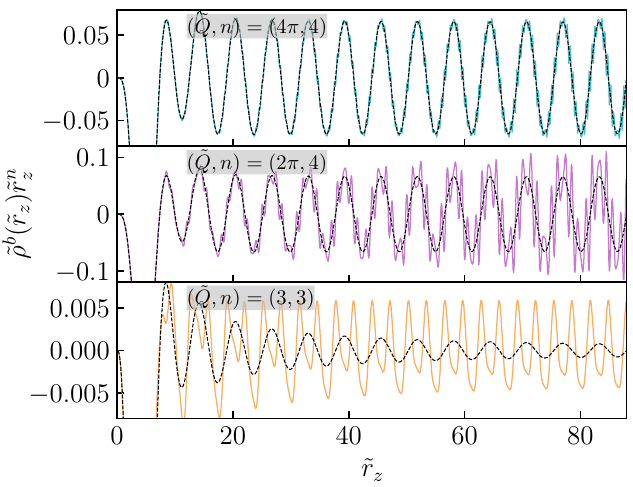}
		\caption{Induced charge density for different $\tilde{Q}=Q/(2k_F)$ for model system $b$ along (a) the $\tilde{r}_x$ axis, and (b) the $\tilde{r}_z$ axis. The black dashed curve shows the induced charge density including only the extrinsic intravalley polarizability. The effective fine-structure constant is taken to be $r_s=1$. }\label{fig_7}
\end{figure}

We now turn to model system $b$, starting with the Friedel oscillations along the $\tilde{r}_x$ axis, which are shown in Fig.~\ref{fig_7}(a). There are two major differences in comparison to model $a$ (cf.\ Fig.\ \ref{fig_6}). First, there is no crossover in the scaling of the Friedel oscillations, which fall off as $1/\tilde{r}_x^4$ for all $\tilde{Q}$. Second, only for small $\tilde{Q}$ there are deviations from the induced charge density of isolated Weyl points, which are observable near the minima and maxima.

\begin{figure}
	\begin{center}
		\includegraphics[scale=1.0]{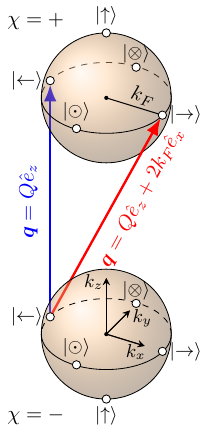}
		\caption{Illustration of transitions between states at the equators of the Fermi surfaces for model $b$. The transition shown in blue is enhanced due to the spin structure, while the one in red is suppressed.}\label{fig_8}
	\end{center}
\end{figure}

In order to explain these features, we recall that the scaling of the Friedel oscillations depends on the order of the singularities of the polarizability. To be more precise, it is also important that the singularities are found in the direction in which the integrand in Eq.~\eqref{eq:dim_less_charge_dens_ext} oscillates, which is the $x$ direction in this case. The decay of the Friedel oscillations with $1/\tilde{r}_x^4$ signifies that the logarithmic singularity in the polarizability is found in the second derivative in the $x$ direction. The Friedel oscillations  along the $\tilde{r}_x$ axis are thus in disagreement with Ref.~\cite{Lv_Zhang_2013}, where cosinusoidal Friedel oscillations with spatial frequency $x=1$ that decay as $1/\tilde{r}_x^3$  were conjectured.

For model $b$, the spin structures of both Weyl cones are the same at the equators of their Fermi surfaces, as illustrated in Fig.~\ref{fig_8}. Hence, a transition from a state at the equator of one Fermi surface to the equator of the other Fermi surface is enhanced if the momentum transfer is {$\bm{q}=\pm Q\hat{\bm{e}}_z$}, as indicated by the transition shown in blue. If the momentum transfer is such that the initial and final momenta are opposing points on the Fermi surfaces, the transition is suppressed due to the spin structure (red transition). Note that only for the latter transition the momentum transfer carries a component in the $x$ direction. Hence, we conclude that the mechanism that occurs for transitions between states at the equators of the two Fermi surfaces is essentially the same mechanism that occurs for intravalley transitions with a momentum transfer of $2k_F$ for states on the same Fermi surface and that has been discussed above. For this reason, the Friedel oscillations along the $\tilde{r}_x$ axis stem from a logarithmic singularity in the second derivative of the polarizability. This is also reflected by  Eq.~\eqref{eq:induced_charge_dens_ext}. For $\bm{q}=Q\hat{\bm{e}}_z + 2k_F\hat{\bm{e}}_x$, $\bm{y}^\chi=(\bm{q}+\chi \bm{Q})/2 k_F$, and $y^\chi = |\bm{y}^\chi|$, we have $c_g^{-,b}=0$, $c_f^{-,b}=1/2$, and $y^-=1$, i.e., we are on $S_{\mathrm{inter}}^-$. Note that for the same $\bm{q}$ the second intervalley summand with $\chi=+1$ is not evaluated on $S_{\mathrm{inter}}^+$ since $y^+\neq 1$ for this $\bm{q}$.

Since the intravalley and intervalley singularities are of the same strength, both singularities give rise to Friedel oscillations that behave like those for isolated Weyl points [see Eq.~\eqref{eq:ind_char_dens_isolated_WPs}]. However, the amplitude of the oscillations caused by intravalley singularities is larger than the amplitude of the oscillations resulting from intervalley singularities. In particular, the latter decreases for increasing $\tilde{Q}$. Hence, deviations from the induced charge density for isolated Weyl points are only significant for small $\tilde{Q}$ [see the black dashed curve in Fig.~\ref{fig_7}(a)].

At the poles, the spin structures of the Weyl points are inverted for both models $a$ and $b$. Hence, we expect the Friedel oscillations along the $\tilde{r}_z$ axis to behave in the same way for models $a$ and $b$. As depicted in Fig.~\ref{fig_7}(b), the induced charge density for model $b$ along the $\tilde{r}_z$ axis shows the same crossover as for model $a$, where the induced charge density  corresponds to the one for isolated Weyl points for small $\tilde{r}_z$ and to Eq.~\eqref{eq:ind_char_dens_inter_parallel} for large $\tilde{r}_z$ [see also Fig.~\ref{fig_5}(a) for comparison].

\subsection{Model system \boldmath{$c$}}\label{sec:FOs_c}

\begin{figure}
	\begin{center}
		\raisebox{21ex}[0ex][3ex]{(a)}
		\includegraphics[scale=0.70]{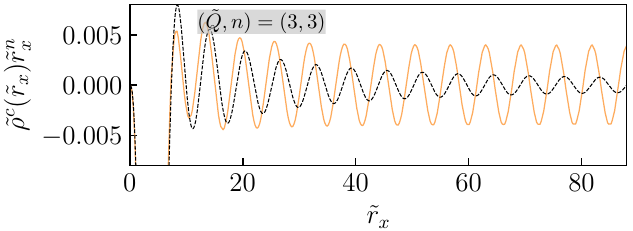}

		\raisebox{41ex}[0ex][3ex]{(b)}
		\includegraphics[scale=0.70]{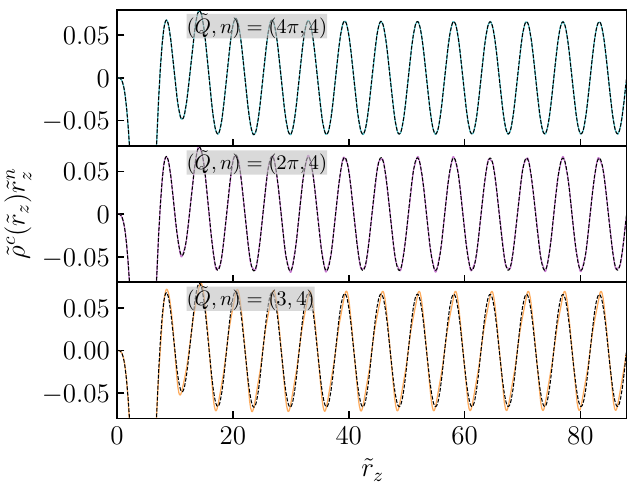}
		\caption{Induced charge density for model system $c$ along (a) the $\tilde{r}_x$ axis for $\tilde{Q}=Q/(2k_F)=3$, and (b) the $\tilde{r}_z$ axis for different $\tilde{Q}$. The black dashed curve shows the induced charge density including only the extrinsic intravalley polarizability. The effective fine-structure constant is taken to be $r_s=1$.}\label{fig_9}
	\end{center}
\end{figure}

For model system $c$, the spin structures of the two Weyl points are inverted along the $x$ direction.
Hence, the Friedel oscillations along the $\tilde{r}_x$ axis behave like those for model $a$ in the same direction, i.e., as described by Eq.~\eqref{eq:ind_char_dens_inter_perp} and plotted in Fig.\ \ref{fig_6}. In Fig.~\ref{fig_9}(a), the induced charge density along the $\tilde{r}_x$ axis is depicted for model system $c$ and $\tilde{Q}=3$~\cite{footnote_model_c}.

Along the $z$ direction, the spin structures of the Weyl points are identical. Thus, the Friedel oscillations fall off as $1/\tilde{r}_z^4$. While the Friedel oscillations caused by the intravalley singularity oscillate with spatial frequency $2k_F$, the ones due to the intervalley singularities oscillate with frequencies $Q\pm 2k_F$. However, the amplitude of the latter is smaller than the amplitude of the former. This is similar to the Friedel oscillations along the $\tilde{r}_x$ axis for model $b$. Therefore, deviations from the induced charge density of isolated Weyl points are only observable for small $\tilde{Q}$. In a weakly doped Weyl semimetal, the separation between the Weyl points is much larger than the extensions of the Fermi surfaces, i.e., $\tilde{Q}=Q/2k_F$ is large. Thus, we expect Friedel oscillations with spatial frequency $x=1$. Our results for the charge-density oscillations of model $c$ along the $\tilde{r}_z$ axis contradict Ref.~\cite{Lv_Zhang_2013}, where Friedel oscillations with spatial frequencies $\tilde{Q}$ and $\tilde Q\pm 1$ and decaying as $1/\tilde r_z^3$ have been conjectured. In Fig.~\ref{fig_9}(b), the Friedel oscillations for model system $c$ along the $\tilde{r}_z$ axis are illustrated.

\section{Beyond centrosymmetric isotropic Weyl points}
\label{sec:beyond_cWSM}

In this section, we discuss extensions of our work beyond the minimal centrosymmetric models considered so far. The Weyl points in real Weyl semimetals are frequently anisotropic. An anisotropic Weyl point can be described by an anisotropic velocity tensor in Eq.~\eqref{eq:eff_Ham_iso_WP}. Consequently, the spin structure and the shape of the Fermi surface change. For anisotropic Fermi surfaces, different momentum scales $k_F$ are expected in different directions. Therefore, the spatial oscillation frequencies are direction dependent. Like for the isotropic case, two Weyl points with opposite chirality cannot have the same spin structure. In the directions with identical (inverted) spin structure, the Friedel oscillations are sinusoidal (cosinusoidal) and their amplitude falls off as $1/r^4$ ($1/r^3$). By investigating the Friedel oscillations along different directions, the relative spin structure can thus be deduced from the exponent of the inverse power law, while the spatial frequencies provide information about the shapes of the Fermi surfaces.

A distinct form of anisotropy is described by adding a term of the form $\bm{v}_0\cdot\bm{k}\,\sigma_0$ to the Weyl Hamiltonian in Eq.~\eqref{eq:eff_Ham_iso_WP}. While such a term tilts the Weyl cone in the direction parallel to $\bm{v}_0$, it does not affect the eigenstates since the term is proportional to the identity matrix. Hence, the spin structure is left invariant but the Fermi surface becomes anisotropic. We expect that our analysis essentially remains valid for type-I Weyl semimetals, which have closed Fermi surfaces, but not for over-tilted type-II Weyl semimetals.


Most Weyl semimetals contain more than two Weyl points. For such systems, intervalley processes between all possible pairs contribute to the Friedel oscillations. Every individual contribution conforms to the analysis presented in this work. Note that a pair of Weyl points with the same chirality in general has a nontrivial relative spin structure, which is reflected by the Friedel oscillations. By investigating the direction dependence of the Friedel oscillations and noting relations between the spin structures of Weyl points imposed by crystal symmetries, it should generally be possible to determine all relative spin structures.

Although we have focused on centrosymmetric Weyl semimetals, our results are also applicable to Weyl semimetals with broken inversion symmetry. We first discuss a minimal model of a time-reversal-symmetric, i.e., nonmagnetic, Weyl semimetal with isotropic Weyl points. Due to time-reversal symmetry, two Weyl points at momenta $\bm{k}_0$ and $-\bm{k}_0$ form a pair, where the two Weyl points possess identical spin structures since time reversal inverts both spin and momentum. Thus, both Weyl points have the same chirality. Based on the results of Secs.~\ref{sec:FOs_b} and \ref{sec:FOs_c}, the Friedel oscillations due to this pair are of the form $\frac{\sin 2k_Fr}{r^4}$ in directions parallel and orthogonal to the separation vector of the Weyl points. Note that the amplitudes of oscillations with spatial frequencies $q=Q\pm2k_F$ are suppressed for Weyl points possessing identical spin structures along their separation direction. Time-reversal-symmetric Weyl semimetals host at least a second pair of Weyl points with opposite chirality. The relative spin structure between all Weyl points can again be analyzed by investigating the direction dependence of the Friedel oscillations.

A different situation occurs in nonmagnetic chiral crystals, where so-called Kramers--Weyl points are enforced at all time-reversal-invariant momenta, e.g., the $\Gamma$ point, as shown by symmetry-based classifications \cite{Chang_Wieder_2018,Knoll_Timm_2022}. These Kramers--Weyl points do not possess a time-reversal partner since time-reversal symmetry maps them onto themselves. In nonmagnetic chiral crystals, the fermion-doubling theorem is satisfied by Weyl points or higher-dimensional band touchings that are not related by symmetry and are thus not energetically degenerate.


So far, we have been concerned with Weyl semimetals, which are characterized by twofold-degenerate linearly dispersing Weyl points. However, there exist also semimetals with higher-order band-touching points. For example, cubic systems such as Pr$_2$Ir$_2$O$_7$ host quadratic band-touching points \cite{Cheng_Ohtsuki_2017}, while ferromagnetic HgCr$_2$Se$_4$ hosts ``double Weyl points'' with quadratic dispersion in two directions and linear dispersion in the remaining one \cite{Fang_Gilbert_2012}. Friedel oscillations are charaterized by the scaling of their amplitude vs.\ distance and their spatial oscillation frequency. Neither of these two features depends on the leading power law in the dispersion. Consequently, we expect that our results, which have their origins in the relative spin structure, are essentially also applicable to higher-order semimetals. In particular, we expect sinusoidal (cosinusoidal) Friedel oscillations decaying as $1/r^4$ ($1/r^3$) along directions in which $2k_F$ transitions between opposing sites of the different Fermi surfaces are suppressed (enhanced).

\section{Summary and conclusions}\label{sec:summary_conclusion}

In this work, we have shown that the Friedel oscillations of the induced charge density qualitatively depend on the relative spin structure of the Weyl points in Weyl semimetals. We have employed minimal centrosymmetric and time-reversal-symmetry-breaking models with two Weyl points of opposite chirality. For three model systems with different spin structures, we have analytically calculated the extrinsic polarizability. This allowed us to compute the induced charge densities due to a test charge for each model along various directions. The results are summarized in Table \ref{table:summary}.

\begin{table}[tb!]
	\caption{Behavior of the Friedel oscillations for large distances from the origin for each model system along the $\tilde{r}_z$ and $\tilde{r}_x$ axes. The oscillatory behavior in Eq.~\eqref{eq:ind_char_dens_inter_parallel} is abbreviated as $\tilde{\rho}^\alpha(\tilde{r}_z)\sim\frac{\cos(\tilde{Q}\pm 1)\tilde{r}_z}{\tilde{r}_z^3}$. }
	\begin{threeparttable}
		\begin{tabular*}{0.8\linewidth}{c@{\quad}c@{\qquad}c}
			\hline\hline
			Model system & $\tilde{\rho}^\alpha(\tilde{r}_z)$  & $\tilde{\rho}^\alpha(\tilde{r}_x)$ \\\hline
			\rule{0pt}{5ex}
			$a$
			& $\displaystyle {}\sim\frac{\cos(\tilde{Q}\pm 1)\tilde{r}_z}{\tilde{r}_z^3}$
			& $\displaystyle {}\sim\frac{\cos \tilde{r}_x}{\tilde{r}_x^3}$ \\
			\rule{0pt}{5ex}
			$b$
			& $\displaystyle {}\sim\frac{\cos(\tilde{Q}\pm 1)\tilde{r}_z}{\tilde{r}_z^3}$
			& $\displaystyle {}\sim\frac{\sin \tilde{r}_x}{\tilde{r}_x^4}$ \\
			\rule{0pt}{5ex}
			$c$\vspace{1mm}
			& $\displaystyle {}\sim\left.\frac{\sin\tilde{r}_z}{\tilde{r}_z^4}\right.^*$
			& $\displaystyle {}\sim\frac{\cos \tilde{r}_x}{\tilde{r}_x^3}$ \\
			\hline\hline
		\end{tabular*}
		\begin{tablenotes}
			\item[*] There are additional oscillations $\tilde{\rho}^a(\tilde{r}_z)\sim\frac{\sin(\tilde{Q}\pm 1)\tilde{r}_z}{\tilde{r}_z^4}$ with a much smaller amplitude.
		\end{tablenotes}
	\end{threeparttable}
	\label{table:summary}
\end{table}

Our work shows that Weyl semimetals possess spin-structure-dependent Friedel oscillations due to an interplay of intravalley and intervalley singularities on the surfaces $S_{\mathrm{intra}}$ and $S_{\mathrm{inter}}^\pm$, which are spheres with radius $2k_F$ centered around the points $\bm{q}=0$ and $\bm{q}=\mp\bm{Q}$, respectively. The Friedel oscillations show the following properties: First, for small distances $r$, the Friedel oscillations due to the intravalley singularity, which are sinusoidal and decay as $1/r^4$ \cite{Lv_Zhang_2013}, are dominant for all Weyl semimetals. The length scale that divides small from large distances, at which intravalley and intervalley singularities dominate, respectively, depends on the separation $Q$ between the Weyl points in momentum space.
Second, for large distances $r$, intervalley processes dominate and the Friedel oscillations depend on the relative spin structure of the two Weyl points. If the spin structures of the Weyl points are identical along a certain direction the Friedel oscillations are sinusoidal and fall off as $1/r^4$, while for inverted spin structures they are phase shifted to a cosine and fall off more slowly as $1/r^3$. Third, along directions that are orthogonal to the separation vector of the Weyl points, the spatial oscillation frequency is in general $2k_F$. Fourth, for large distances $r$, the spatial oscillation frequencies along directions that are parallel to the separation vector of the Weyl points depend on the relative spin structure. If the spin structures are identical along these directions, the oscillation frequency equals $2k_F$ since the amplitude of the oscillation due to the intravalley singularity is much larger than the ones from the intervalley singularities. In contrast, the intervalley singularities dominate the Friedel oscillations for large distances for inverted spin structures, where the corresponding spatial frequencies are $Q\pm2k_F$. Note that we do not find oscillations with spatial frequency $Q$, which have been conjectured in Ref.~\cite{Lv_Zhang_2013}.


On surfaces, the local density of states associated with surface Friedel oscillations can be observed using scanning tunneling microscopy \cite{Binnig_Rohrer_1984, Hasegawa_Avouris_1993, Crommie_Lutz_1993, Mallet_Brihuega_2016}. It is highly desirable to extend our analysis to systems with a surface, where the presence of Fermi-arc surface states is expected to lead to additional contributions \cite{Hosur_2012}. The observation of Friedel oscillations in the bulk is difficult. Their indirect observation might be possible using inelastic x-ray scattering (IXS) \cite{Abbamonte_Finkelstein_2004, Abbamonte_Wong_2010, Schuelke_2007}. In IXS experiments, the dynamic structure factor $S(\bm{q},\omega)$ is measured \cite{Ament_Veenendaal_2011}. Using the fluctuation-dissipation theorem, one can calculate the imaginary part of the polarizability as $\mathrm{Im}\, \pi(\bm{q},\omega) = -\pi S(\bm{q},\omega)$ \cite{Hoeller_Krotscheck_2015}. The real part of the polarizability is then calculated by employing the Kramers--Kronig relation. In particular, the static polarizabilities of graphene \cite{Reed_Uchoa_2010} and lithium \cite{Hagiya_Matsuda_2020} have been measured using this approach. In IXS experiments, the total polarizability is obtained, which consists of extrinsic and intrinsic contributions. Although the latter dominates over the former for $\bm{q}\approx\bm{Q}$ in a Weyl semimetal, the singularities on $S_{\mathrm{inter}}^\pm$ in the derivatives of the extrinsic polarizability still give rise to spin-structure-dependent Friedel oscillations. Recently, the material K$_2$Mn$_3$(AsO$_4$)$_3$ has been proposed to be a magnetic Weyl semimetal with only two Weyl points \cite{Nie_Hashimoto_2022}. Similar to the tight-binding model employed in Sec.~\ref{sec:intervalley_pola}, it belongs to the point group $C_{2h}$ and gives rise to a pair of Weyl points on the $k_z$ axis. Therefore, K$_2$Mn$_3$(AsO$_4$)$_3$ constitutes an ideal experimental test candidate for our theory.

Our work emphasizes the importance of intervalley processes for the response of Weyl semimetals. These processes have often been neglected or treated approximately. We expect that spin-structure-dependent intervalley processes are not only relevant for Friedel oscillations but also for other phenomena.  As we have seen, the induced charge distribution due to an impurity is anisotropic for Weyl semimetals. This suggests an anisotropic Ruderman-Kittel-Kasuya-Yosida (RKKY) interaction \cite{Ruderman_Kittel_1954, Kasuya_1956, Yosida_1957} between local moments. We hence expect that the RKKY interaction is strongly affected by the relative spin structure of the Weyl points. Furthermore, effects of scattering on magnetotransport, where the spin structure enhances or suppresses certain transitions, are another promising example for signatures for the relative spin structure.

\vspace*{3ex}
\begin{acknowledgments}

We are grateful to L. M. Woods, J. Bouaziz, and C. Koschenz for helpful discussions. Financial support by the Deut\-sche For\-schungs\-ge\-mein\-schaft through Collaborative Research Center SFB 1143, project A04, project id 247310070, and the W\"urzburg-Dresden Cluster of Excellence ct.qmat, EXC 2147, project id 390858490, is gratefully acknowledged.

\end{acknowledgments}

\appendix

\begin{widetext}

	\section{Calculation of the extrinsic intervalley polarizability}\label{app:ext_inter_crf}

	In this appendix, we calculate the extrinsic intervalley polarizability in Eq.~\eqref{eq:crf_inter_ext} for model system $c$. By performing the coordinate transformation $\tilde{\bm{k}}=\bm{k}-\chi\frac{\bm{Q}}{2}$, we obtain
	\begin{align}
		\pi^{\mathrm{ext},c}_{\mathrm{inter}}(\bm{q})
		&= - \frac{1}{v_F\hbar}\sum_{\chi s}
		\left[\int \frac{d^3\tilde{k}}{(2\pi)^3}\:
		\frac{\Theta(k_F-\tilde{k})}{\tilde{k}-s|\tilde{\bm{k}}+\bm{q}+\chi\bm{Q}|+i \tilde{\eta}}\,
		F^{c ;\chi{\bar\chi}}_{+ s}\left(\tilde{\bm{k}}+\chi\frac{\bm{Q}}{2},\tilde{\bm{k}}+\bm{q}+\chi\frac{\bm{Q}}{2}\right)
		\right.\nonumber\\
		&\qquad \left. {}-\int
		\frac{d^3\tilde{k}}{(2\pi)^3}\: \frac{\Theta(k_F-|\tilde{\bm{k}}+\bm{q}+\chi\bm{Q}|)}
		{s\tilde{k}-|\tilde{\bm{k}}+\bm{q}+\chi\bm{Q}|+i \tilde{\eta}}\,
		F^{c ;\chi{\bar\chi}}_{s+}\left(\tilde{\bm{k}}
		+\chi\frac{\bm{Q}}{2},\tilde{\bm{k}}+\bm{q}+\chi\frac{\bm{Q}}{2}\right)	\right],
		\label{eq:ext_inter_c_eq1}
	\end{align}
	where $\tilde{\eta}=\eta/v_F\hbar$. After substituting $\bm{k}=-(\tilde{\bm{k}}+\bm{q}^\chi)$ with $\bm{q}^\chi=\bm{q}+\chi\bm{Q}$ in the second integral and relabeling $\tilde{\bm{k}}=\bm{k}$ in the first one, we obtain
	\begin{align}
		\pi^{\mathrm{ext},c}_{\mathrm{inter}}(\bm{q})
		&= - \frac{1}{v_F\hbar} \sum_{\chi s}
		\int \frac{d^3k}{(2\pi)^3}\, \Theta(k_F-k) \left(\frac{1}{k-s|\bm{k}+\bm{q}^\chi|+i\tilde{\eta}}
		+\frac{1}{k-s|\bm{k}+\bm{q}^\chi|-i\tilde{\eta}}\right) \nonumber\\
		&\quad{} \times
		\frac{1}{2} \left[1+s\, \frac{-k_x(k_x+q_x^\chi)+k_y(k_y+q_y^\chi)+k_z(k_z+q_z^\chi)}{k|\bm{k}+\bm{q}^\chi|}\right] .
		\label{eq:ext_inter_c_eq2}
	\end{align}
	Next, we rewrite Eq.~\eqref{eq:ext_inter_c_eq2} as
	\begin{align}
		\pi^{\mathrm{ext},c}_{\mathrm{inter}}(\bm{q})=\pi^{\mathrm{ext},c}_{\mathrm{inter},1}(\bm{q})+\pi^{\mathrm{ext},c}_{\mathrm{inter},2}(\bm{q}) ,
		\label{eq:ext_inter_c_eq4}
	\end{align}
	where
	\begin{align}
		\pi^{\mathrm{ext},c}_{\mathrm{inter},1}(\bm{q})
		&= -  \frac{1}{v_F\hbar}\sum_{\chi s}
		\int \frac{d^3k}{(2\pi)^3}\,
		\Theta(k_F-k) \left(\frac{1}{k-s|\bm{k}+\bm{q}^\chi|+i\tilde{\eta}}
		+ \frac{1}{k-s|\bm{k}+\bm{q}^\chi|-i\tilde{\eta}}\right)
		\frac{1}{2}\left[1+s\,\frac{\bm{k}\cdot(\bm{k}+\bm{q}^\chi)}{k|\bm{k}+\bm{q}^\chi|}\right],
		\label{eq:ext_inter_c_eq4a}\\
		\pi^{\mathrm{ext},c}_{\mathrm{inter},2}(\bm{q})
		&= \frac{1}{v_F\hbar}\sum_{\chi s}
		\int \frac{d^3k}{(2\pi)^3}\,
		\Theta(k_F-k)\left(\frac{1}{k-s|\bm{k}+\bm{q}^\chi|+i\tilde{\eta}}
		+ \frac{1}{k-s|\bm{k}+\bm{q}^\chi|-i\tilde{\eta}}\right) s\, \frac{k_x(k_x+q_x^\chi)}{k|\bm{k}+\bm{q}^\chi|}.
		\label{eq:ext_inter_c_eq4b}
	\end{align}
	We first focus on Eq.~\eqref{eq:ext_inter_c_eq4a}. The integrand only depends on scalar products involving the vectors $\bm{k}$ and $\bm{q}^\chi$. Hence, we can choose $\bm{k}$ such that $\bm{q}^\chi$ is parallel to the $z$ direction, $\bm{q}^\chi\|\hat{\bm{e}}_z$. To solve the two-center integral, it is useful to employ prolate spheroidal coordinates,
	\begin{align}
		k_x &= \frac{1}{2}\, q^\chi \sinh\mu\sin\nu\cos\theta,\label{eq:prolate_x} \\
		k_y &= \frac{1}{2}\, q^\chi \sinh\mu\sin\nu\sin\theta,\label{eq:prolate_y} \\
		k_z &= \frac{1}{2}\, q^\chi (\cosh\mu\cos\nu-1)\label{eq:prolate_z},
	\end{align}
	where $\mu\in[0,\infty)$, $\nu\in[0,\pi)$, $\theta\in[0,2\pi)$. These coordinates describe spheroids the foci of which are separated by $q^\chi$ along the $z$ direction. In these coordinates, the Jacobi determinant is written as
	\begin{equation}
		J=\frac{1}{8}(q^\chi)^3\sinh\mu\sin\nu\, (\cosh^2\!\mu-\cos^2\!\nu),
	\end{equation}
	and we have
	\begin{align}
		k &= \frac{1}{2}\, q^\chi(\cosh\mu-\cos\nu),\\
		|\bm{k}+\bm{q}^\chi| &= \frac{1}{2}\, q^\chi(\cosh\mu+\cos\nu),\\
		\bm{k}\cdot(\bm{k}+\bm{q}^\chi) &= \frac{1}{4}\, (q^\chi)^2(\cosh\!^2\mu+\cos\!^2\nu-2).
	\end{align}
	Evaluating Eq.~\eqref{eq:ext_inter_c_eq4a}, we arrive at
	\begin{equation}
		\pi^{\mathrm{ext},c}_{\mathrm{inter},1}(\bm{q})
		= \sum_\chi \frac{2}{(2\pi)^2}\,
		\frac{k_F^2}{v_F\hbar}\, f(y^\chi).
		\label{eq:ext_inter_c_eq5}
	\end{equation}
	Note that Eq.~\eqref{eq:ext_inter_c_eq5} is similar to the extrinsic intravalley polarizability in Eq.~\eqref{eq:crf_intra_ext_ana}.

	Next, we consider Eq.~\eqref{eq:ext_inter_c_eq4b}. Since we want to employ the same prolate spheroidal coordinates, we have to rotate the coordinate system such that $\bm{q}^\chi \| \hat{\bm{e}}_z$. In contrast to Eq.~\eqref{eq:ext_inter_c_eq4a}, the integral in Eq.~\eqref{eq:ext_inter_c_eq4b} contains terms that are not expressible as scalar products of the two vectors $\bm{k}$ and $\bm{q}^\chi$. As a consequence, any rotations of the coordinate system result in additional terms that stem from the rotations and have to be taken into account.

	Let $(R_{ij})\in\mathrm{SO}(3)$ be a rotation matrix such that
	\begin{equation}\label{eq:definition_rot_mat}
		q^\chi \hat{\bm{e}}_z = (R_{ij})\,\bm{q}^\chi.
	\end{equation}
	We suppress the dependence of $(R_{ij})$ on $\bm{q}^\chi$ for notational convenience. Using $(R_{ij})$, we define new coordinates via
	\begin{equation}\label{eq:defintion_rotated_coords}
		\tilde{\bm{k}}=(R_{ij})\bm{k}.
	\end{equation}
	Plugging Eq.~\eqref{eq:defintion_rotated_coords} into Eq.~\eqref{eq:ext_inter_c_eq4b}, we obtain
	\begin{equation}
		\pi^{\mathrm{ext},c}_{\mathrm{inter},2}(\bm{q})=\pi^{\mathrm{ext},c}_{\mathrm{inter},2A}(\bm{q})+\pi^{\mathrm{ext},c}_{\mathrm{inter},2B}(\bm{q}), \label{eq:ext_inter_2}
	\end{equation}
	with
	\begin{align}
		\pi^{\mathrm{ext},c}_{\mathrm{inter},2A}(\bm{q})
		&= \frac{1}{v_F\hbar} \sum_{\chi s}
		\int \frac{d^3\tilde{k}}{(2\pi)^3}\,
		\Theta(k_F-\tilde{k}) \left(\frac{1}{\tilde{k}-s|\tilde{\bm{k}}+\tilde{\bm{q}}^\chi|
			+ i\tilde{\eta}}
		+ \frac{1}{\tilde{k}-s|\tilde{\bm{k}}+\tilde{\bm{q}}^\chi|-i\tilde{\eta}}\right)
		\nonumber\\
		&\qquad{} \times s\, \frac{(R_{11})^2\tilde{k}_x^2+(R_{21})^2\tilde{k}_y^2+(R_{31})^2\tilde{k}_z^2}
		{\tilde{k}|\tilde{\bm{k}}+\tilde{\bm{q}}^\chi|},
		\label{eq:crf_ext_inter_2A}
	\end{align}
	and
	\begin{align}
		\pi^{\mathrm{ext},c}_{\mathrm{inter},2B}(\bm{q})
		= \frac{1}{v_F\hbar}\sum_{\chi s}q_x^\chi
		\int \frac{d^3\tilde{k}}{(2\pi)^3}\,
		\Theta(k_F-\tilde{k})
		\left(\frac{1}{\tilde{k}-s|\tilde{\bm{k}}+\tilde{\bm{q}}^\chi|
			+ i\tilde{\eta}}
		+ \frac{1}{\tilde{k}-s|\tilde{\bm{k}}+\tilde{\bm{q}}^\chi|-i\tilde{\eta}}\right)
		s\, \frac{ R_{31}\tilde{k}_z}{\tilde{k}|\tilde{\bm{k}}
			+ \tilde{\bm{q}}^\chi|},
		\label{eq:crf_ext_inter_2B}
	\end{align}
	where $\tilde{\bm{q}}^\chi=q^\chi \hat{\bm{e}}_z$. In Eq.~\eqref{eq:crf_ext_inter_2A}, we have omitted terms that are odd functions of $\tilde{k}_x$ or $\tilde{k}_y$ since they vanish after performing the integration.  Evaluation of Eq.~\eqref{eq:crf_ext_inter_2A} yields
	\begin{equation}
		\pi^{\mathrm{ext},c}_{\mathrm{inter},2A}(\bm{q}) = -\frac{1}{2}\, \frac{2}{(2\pi)^2}\,
		\frac{1}{v_F\hbar} \sum_\chi \left(
		\left[(R_{11})^2 + (R_{21})^2\right]f(y^\chi)+ (R_{31})^2 g(y^\chi) \right)
		\label{eq:ext_inter_2A_rot_mat}
	\end{equation}
	and from Eq.~\eqref{eq:crf_ext_inter_2B} we obtain
	\begin{equation}
		\pi^{\mathrm{ext},c}_{\mathrm{inter},2B}(\bm{q}) = \frac{2}{(2\pi)^2}\,
		\frac{1}{v_F\hbar} \sum_\chi R_{31}\, \frac{y^\chi_x}{y^\chi}\, g(y^\chi) .
		\label{eq:ext_inter_2B_rot_mat}
	\end{equation}

	In order to determine the rotation matrix in Eq.~\eqref{eq:definition_rot_mat}, we employ Rodrigues' formula
	\begin{equation}\label{eq:rodrigues_formula}
		R_{ij} = \cos\theta\,\delta_{ij}
		+ (1-\cos\theta)\, n_i n_j - \sin\theta\,\epsilon_{ijk}n_k ,
	\end{equation}
	where $\bm{n}={\bm{v}_1\times\bm{v}_2}/{|\bm{v}_1\times\bm{v}_2|}$, and $\cos\theta={\bm{v}_1\cdot\bm{v}_2}/{|\bm{v}_1| |\bm{v}_2|}$. With $\bm{v}_1=\tilde{\bm{q}}^\chi/q^\chi$ and $\bm{v}_2=\bm{q}^\chi/q^\chi$ in Eq.~\eqref{eq:rodrigues_formula}, we find
	\begin{align}
		\pi^{\mathrm{ext},c}_{\mathrm{inter},2A}(\bm{q})
		&= -\frac{1}{2}\, \frac{2}{(2\pi)^2}\, \frac{1}{v_F\hbar} \sum_\chi \left[
		\frac{(y_y^\chi)^2+(y_z^\chi)^2}{(y^\chi)^2}\, f(y^\chi)
		+ \left(\frac{y^\chi_x}{y^\chi}\right)^{\!2} g(y^\chi) \right] ,
		\label{eq:ext_int_2A_final} \\
		\pi^{\mathrm{ext},c}_{\mathrm{inter},2B}(\bm{q})
		&= \frac{2}{(2\pi)^2}\, \frac{1}{v_F\hbar} \sum_\chi
		\left(\frac{y^\chi_x}{y^\chi}\right)^{\!2} g(y^\chi).
		\label{eq:ext_inter_2B_final}
	\end{align}
	Utilizing Eqs.~\eqref{eq:ext_inter_c_eq5}, \eqref{eq:ext_inter_2}, \eqref{eq:ext_int_2A_final}, and \eqref{eq:ext_inter_2B_final}, the final result is Eq.~\eqref{eq:inter_ext_crf_ana}. For  model systems $a$ and $b$, the calculations are performed analogously.
\end{widetext}

\section{Coulomb screening due to the intrinsic polarizability}\label{app:screening_intrinsic}

The intrinsic intervalley polarizability given in Eq.~\eqref{eq:crf_inter_int_ana} leads to unphysical screening of the Coulomb interaction for the continuum models, as we show in the following. Using Eq.~\eqref{eq:RPA}, the screened Coulomb interaction is calculated as
\begin{equation}\label{eq:screened_Coulomb}
	V^\alpha_{\mathrm{sc}}(\bm{q})
	= \frac{V_C(\bm{q})}{\epsilon^\alpha(\bm{q})}
	{ = \frac{V_C(\bm{q})}{1+V_C(\bm{q})\,\pi^\alpha(\bm{q})} } .
\end{equation}
Furthermore, we assume that the Fermi energy is right at the Weyl nodes so that the extrinsic polarizability vanishes. If we only include the intravalley contribution $\pi^{\mathrm{int},\alpha}_{\mathrm{intra}}(\bm{q})$ from Eq.~\eqref{eq:crf_intra_int_ana}, we find that the dielectric function diverges logarithmically for $q \to 0$. The Coulomb potential is thus anomalously weakly screened in the long-distance limit \cite{Abrikosov_Beneslavskii_1971}. This result seems reasonable since the density of states at the Weyl nodes vanishes. On the other hand, if we also include the intervalley contribution $\pi^{\mathrm{int},\alpha}_{\mathrm{inter}}(\bm{q})$ from Eq.\ \eqref{eq:crf_inter_int_ana} the dielectric function approaches the Thomas-Fermi form for $q \to 0$. Hence, there is metallic screening, which is unexpected since there are still no free charge carriers. As shown in Appendix \ref{app:int_pola}, this unphysical result is an artifact of the continuum model.

\section{Strong cutoff dependence of the intrinsic intervalley polarizability }\label{app:int_pola}

The goal of this Appendix is to understand the origin of the constant term in the intrinsic intervalley polarizability in Eq.~\eqref{eq:crf_inter_int_ana}, whereas such a term is absent in the intrinsic intravalley polarizability in Eq.~\eqref{eq:crf_intra_int_ana}. In the following, we focus on model $a$. First, we inspect the intrinsic intravalley Lindhard function in Eq.~\eqref{eq:crf_intra_int}. Since we are interested in constant terms in the polarizability we assume $\bm{q}$ to be small. The dispersion appears in the denominator in the integrand, while the spinor overlap is found in the nominator. Let us ignore the spinor overlap for the moment. We are left with the dispersion in the denominator and the Jacobi determinant. In spherical coordinates and for large momenta, the product of the two is linear in momentum $\bm{k}$. The intrinsic intravalley spinor overlap reads as
\begin{equation}\label{eq:spinor_overlap_a_int_intra}
	F^{a ;\chi\chi}_{+-}(\bm{k},\bm{k}')=\frac{1}{2}\left(1-\frac{\left(\bm{k}-\chi\frac{\bm{Q}}{2}\right)\cdot\left(\bm{k}'-\chi\frac{\bm{Q}}{2}\right)}{\left|\bm{k}-\chi\frac{\bm{Q}}{2}\right|\left|\bm{k}'-\chi\frac{\bm{Q}}{2}\right|}\right).
\end{equation}
For large $\bm{k}$, we have $\bm{k}\approx\bm{k}'$ and the intrinsic intravalley spinor overlap vanishes. As one can verify, the full integrand in Eq.~\eqref{eq:crf_intra_int} vanishes for large momenta as well. However, the full integrand does not fall off sufficiently rapidly for the integral to be convergent. For this reason, the restriction of the integration volume by a momentum cutoff is necessary, as we have argued in Sec.~\ref{subsubsec:intra_int}. As a result, the intrinsic intravalley polarizability depends logarithmically on the cutoff [see Eq.~\eqref{eq:crf_intra_int_ana}]. Note that the integrand in Eq.~\eqref{eq:crf_intra_int} is small at the boundary of the integration volume, which is described by the cutoff. For a realistic system, this suggests that contributions that stem from regions beyond the linear regime are insignificant.

Next, we turn to the intrinsic intervalley Lindhard function in Eq.~\eqref{eq:crf_inter_int}. In the linear regime and for large momenta, the product of the Jacobi determinant and the inverse of the dispersion is again linear in momentum. The different cutoff dependencies in Eqs.~\eqref{eq:crf_intra_int_ana} and \eqref{eq:crf_inter_int_ana} must therefore result from the intrinsic intervalley spinor overlap, which is given by
\begin{equation}\label{eq:spinor_overlap_a_int_inter}
	F^{a ;\chi\bar\chi}_{+-}(\bm{k},\bm{k}')=\frac{1}{2}\left(1+\frac{\left(\bm{k}-\chi\frac{\bm{Q}}{2}\right)\cdot\left(\bm{k}'-\bar{\chi}\frac{\bm{Q}}{2}\right)}{\left|\bm{k}-\chi\frac{\bm{Q}}{2}\right|\left|\bm{k}'-\bar{\chi}\frac{\bm{Q}}{2}\right|}\right).
\end{equation}
At the boundary of the integration volume, this spinor overlap remains finite. This is best seen in the limit  $|\bm{k}|\rightarrow\infty$, where $F^{a ;\chi\bar\chi}_{+-}(\bm{k},\bm{k}')\rightarrow 1$. Hence, the integrand in Eq.~\eqref{eq:crf_inter_int} grows towards the boundary of the integration volume described by the cutoff. For this reason, Eq.~\eqref{eq:crf_inter_int_ana} diverges strongly as a function of the cutoff. In a realistic system, the integrand is large at the boundary of the linear regime so that it is also large and still increasing at the beginning of the nonlinear regime. Therefore, significant contributions from the nonlinear region are expected, which may counter the strong cutoff dependence of the intrinsic intervalley polarizability in the linear regime.

Ultimately, the nonvanishing intervalley interband spinor overlap is a consequence of the different spin structures of Weyl points with opposite chiralities. For this reason, the above arguments also apply to models $b$ and $c$. The cutoff dependence can be avoided by working with lattice models, such as the tight-binding model introduced in the main text.

\section{Additional information on the tight-binding model}\label{app:tight_binding}

Let us consider the point group $C_{2h}$, for which the symmetry operators are written as  $P=\tau_z$ (inversion), $U_{C_{2z}}=\tau_z$ (rotation about the \textit{z} axis), and $U_{\sigma_h}=\tau_0$ (mirror reflection in the horizontal plane). For the irreducible representations $A_g$ and $B_u$, the  orbital-pseudospin and momentum basis functions up to second order are given by
\begin{align}
	A_g{:} &\quad \tau_z, 1, k_x^2, k_y^2, k_x k_y, k_z^2,\\
	B_u{:} &\quad \tau_x, \tau_y, k_x, k_y.
\end{align}
By regularizing the basis functions on a simple cubic lattice and choosing the coefficients appropriately, the model Hamiltonian in Eq.~\eqref{eq:tb_Ham} is obtained. To simplify the Hamiltonian, the coefficient of the second-order term $k_x k_y$ is set to zero since this term does not affect the existence of the Weyl points on the $k_z$ axis. For the isotropic case described in the main text, the point group is extended to $C_{4h}$.

\section{Discrete Fourier transform of Friedel oscillations}\label{app:dft_fos}

\begin{figure}
		\includegraphics[scale=0.8]{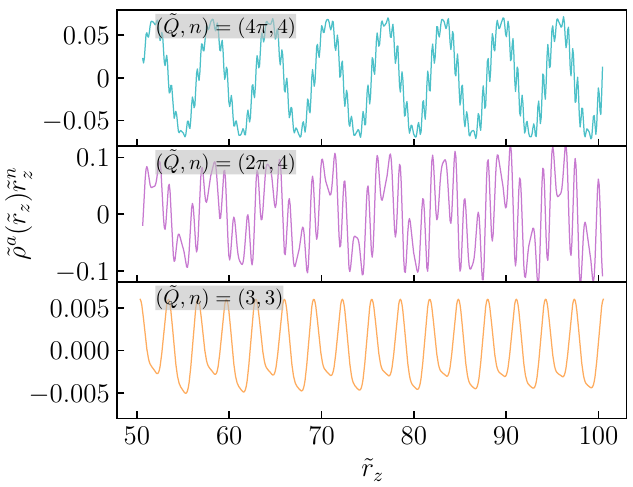}
		\caption{Intervals of the Friedel oscillations in Fig.~\ref{fig_5}(a) chosen to calculate the Fourier components in Figs.~\ref{fig_5}(b)--(d).}\label{fig_10}
\end{figure}

For the calculation of the discrete Fourier transforms in Figs.~\ref{fig_5}(b)--(d), we have used the Friedel oscillations in the intervals depicted in Fig.~\ref{fig_10}. For $\tilde{Q}=3$, the interval is $\left[50.36,100.53\right]$, while the interval is $\left[50.62,100.44\right]$ for $\tilde{Q}=2\pi$ and $\tilde{Q}=4\pi$. The intervals were chosen in such a way that they comprise eight full periods of the Friedel oscillations as good as possible. However, since the Friedel oscillations in the investigated intervals are not perfectly periodic functions, additional nonzero Fourier components apart from $x=1$ and $\tilde{Q}\pm 1$ occur.

\newpage

\bibliography{Knoll_literature}

\begin{thebibliography}{76}%
\makeatletter
\providecommand \@ifxundefined [1]{%
 \@ifx{#1\undefined}
}%
\providecommand \@ifnum [1]{%
 \ifnum #1\expandafter \@firstoftwo
 \else \expandafter \@secondoftwo
 \fi
}%
\providecommand \@ifx [1]{%
 \ifx #1\expandafter \@firstoftwo
 \else \expandafter \@secondoftwo
 \fi
}%
\providecommand \natexlab [1]{#1}%
\providecommand \enquote  [1]{``#1''}%
\providecommand \bibnamefont  [1]{#1}%
\providecommand \bibfnamefont [1]{#1}%
\providecommand \citenamefont [1]{#1}%
\providecommand \href@noop [0]{\@secondoftwo}%
\providecommand \href [0]{\begingroup \@sanitize@url \@href}%
\providecommand \@href[1]{\@@startlink{#1}\@@href}%
\providecommand \@@href[1]{\endgroup#1\@@endlink}%
\providecommand \@sanitize@url [0]{\catcode `\\12\catcode `\$12\catcode
  `\&12\catcode `\#12\catcode `\^12\catcode `\_12\catcode `\%12\relax}%
\providecommand \@@startlink[1]{}%
\providecommand \@@endlink[0]{}%
\providecommand \url  [0]{\begingroup\@sanitize@url \@url }%
\providecommand \@url [1]{\endgroup\@href {#1}{\urlprefix }}%
\providecommand \urlprefix  [0]{URL }%
\providecommand \Eprint [0]{\href }%
\providecommand \doibase [0]{https://doi.org/}%
\providecommand \selectlanguage [0]{\@gobble}%
\providecommand \bibinfo  [0]{\@secondoftwo}%
\providecommand \bibfield  [0]{\@secondoftwo}%
\providecommand \translation [1]{[#1]}%
\providecommand \BibitemOpen [0]{}%
\providecommand \bibitemStop [0]{}%
\providecommand \bibitemNoStop [0]{.\EOS\space}%
\providecommand \EOS [0]{\spacefactor3000\relax}%
\providecommand \BibitemShut  [1]{\csname bibitem#1\endcsname}%
\let\auto@bib@innerbib\@empty
\bibitem [{\citenamefont {Dirac}\ and\ \citenamefont
  {Fowler}(1928)}]{Dirac_1928}%
  \BibitemOpen
  \bibfield  {author} {\bibinfo {author} {\bibfnamefont {P.~A.~M.}\
  \bibnamefont {Dirac}}\ and\ \bibinfo {author} {\bibfnamefont {R.~H.}\
  \bibnamefont {Fowler}},\ }\bibfield  {title} {\bibinfo {title} {{The quantum
  theory of the electron}},\ }\href {https://doi.org/10.1098/rspa.1928.0023}
  {\bibfield  {journal} {\bibinfo  {journal} {Proc. R. Soc. A}\ }\textbf
  {\bibinfo {volume} {117}},\ \bibinfo {pages} {610} (\bibinfo {year}
  {1928})}\BibitemShut {NoStop}%
\bibitem [{\citenamefont {Weyl}(1929)}]{Weyl_1929}%
  \BibitemOpen
  \bibfield  {author} {\bibinfo {author} {\bibfnamefont {H.}~\bibnamefont
  {Weyl}},\ }\bibfield  {title} {\bibinfo {title} {{Gravitation and the
  Electron}},\ }\href {https://doi.org/10.1073/pnas.15.4.323} {\bibfield
  {journal} {\bibinfo  {journal} {Proc. Natl. Acad. Sci. (U.S.A.)}\ }\textbf
  {\bibinfo {volume} {15}},\ \bibinfo {pages} {323} (\bibinfo {year}
  {1929})}\BibitemShut {NoStop}%
\bibitem [{\citenamefont {Peskin}\ and\ \citenamefont
  {Schroeder}(1995)}]{Peskin_Schroeder_book}%
  \BibitemOpen
  \bibfield  {author} {\bibinfo {author} {\bibfnamefont {M.~E.}\ \bibnamefont
  {Peskin}}\ and\ \bibinfo {author} {\bibfnamefont {D.~V.}\ \bibnamefont
  {Schroeder}},\ }\href@noop {} {\emph {\bibinfo {title} {{An Introduction to
  Quantum Field Theory}}}}\ (\bibinfo  {publisher} {Westview Press},\ \bibinfo
  {year} {1995})\BibitemShut {NoStop}%
\bibitem [{\citenamefont {Xu}\ \emph {et~al.}(2015)\citenamefont {Xu},
  \citenamefont {Alidoust}, \citenamefont {Belopolski}, \citenamefont {Yuan},
  \citenamefont {Bian}, \citenamefont {Chang}, \citenamefont {Zheng},
  \citenamefont {Strocov}, \citenamefont {Sanchez}, \citenamefont {Chang},
  \citenamefont {Zhang}, \citenamefont {Mou}, \citenamefont {Wu}, \citenamefont
  {Huang}, \citenamefont {Lee}, \citenamefont {Huang}, \citenamefont {Wang},
  \citenamefont {Bansil}, \citenamefont {Jeng}, \citenamefont {Neupert},
  \citenamefont {Kaminski}, \citenamefont {Lin}, \citenamefont {Jia},\ and\
  \citenamefont {Hasan}}]{Xu_Nasser_Belopolski_2015}%
  \BibitemOpen
  \bibfield  {author} {\bibinfo {author} {\bibfnamefont {S.-Y.}\ \bibnamefont
  {Xu}}, \bibinfo {author} {\bibfnamefont {N.}~\bibnamefont {Alidoust}},
  \bibinfo {author} {\bibfnamefont {I.}~\bibnamefont {Belopolski}}, \bibinfo
  {author} {\bibfnamefont {Z.}~\bibnamefont {Yuan}}, \bibinfo {author}
  {\bibfnamefont {G.}~\bibnamefont {Bian}}, \bibinfo {author} {\bibfnamefont
  {T.-R.}\ \bibnamefont {Chang}}, \bibinfo {author} {\bibfnamefont
  {H.}~\bibnamefont {Zheng}}, \bibinfo {author} {\bibfnamefont {V.~N.}\
  \bibnamefont {Strocov}}, \bibinfo {author} {\bibfnamefont {D.~S.}\
  \bibnamefont {Sanchez}}, \bibinfo {author} {\bibfnamefont {G.}~\bibnamefont
  {Chang}}, \bibinfo {author} {\bibfnamefont {C.}~\bibnamefont {Zhang}},
  \bibinfo {author} {\bibfnamefont {D.}~\bibnamefont {Mou}}, \bibinfo {author}
  {\bibfnamefont {Y.}~\bibnamefont {Wu}}, \bibinfo {author} {\bibfnamefont
  {L.}~\bibnamefont {Huang}}, \bibinfo {author} {\bibfnamefont {C.-C.}\
  \bibnamefont {Lee}}, \bibinfo {author} {\bibfnamefont {S.-M.}\ \bibnamefont
  {Huang}}, \bibinfo {author} {\bibfnamefont {B.}~\bibnamefont {Wang}},
  \bibinfo {author} {\bibfnamefont {A.}~\bibnamefont {Bansil}}, \bibinfo
  {author} {\bibfnamefont {H.-T.}\ \bibnamefont {Jeng}}, \bibinfo {author}
  {\bibfnamefont {T.}~\bibnamefont {Neupert}}, \bibinfo {author} {\bibfnamefont
  {A.}~\bibnamefont {Kaminski}}, \bibinfo {author} {\bibfnamefont
  {H.}~\bibnamefont {Lin}}, \bibinfo {author} {\bibfnamefont {S.}~\bibnamefont
  {Jia}},\ and\ \bibinfo {author} {\bibfnamefont {M.~Z.}\ \bibnamefont
  {Hasan}},\ }\bibfield  {title} {\bibinfo {title} {{Discovery of a Weyl
  fermion state with Fermi arcs in niobium arsenide}},\ }\href
  {https://doi.org/10.1038/nphys3437} {\bibfield  {journal} {\bibinfo
  {journal} {Nature Phys.}\ }\textbf {\bibinfo {volume} {11}},\ \bibinfo
  {pages} {748} (\bibinfo {year} {2015})}\BibitemShut {NoStop}%
\bibitem [{\citenamefont {Yang}\ \emph {et~al.}(2015)\citenamefont {Yang},
  \citenamefont {Liu}, \citenamefont {Sun}, \citenamefont {Peng}, \citenamefont
  {Yang}, \citenamefont {Zhang}, \citenamefont {Zhou}, \citenamefont {Zhang},
  \citenamefont {Guo}, \citenamefont {Rahn}, \citenamefont {Prabhakaran},
  \citenamefont {Hussain}, \citenamefont {Mo}, \citenamefont {Felser},
  \citenamefont {Yan},\ and\ \citenamefont {Chen}}]{Yang_Liu_Sun_2015}%
  \BibitemOpen
  \bibfield  {author} {\bibinfo {author} {\bibfnamefont {L.~X.}\ \bibnamefont
  {Yang}}, \bibinfo {author} {\bibfnamefont {Z.~K.}\ \bibnamefont {Liu}},
  \bibinfo {author} {\bibfnamefont {Y.}~\bibnamefont {Sun}}, \bibinfo {author}
  {\bibfnamefont {H.}~\bibnamefont {Peng}}, \bibinfo {author} {\bibfnamefont
  {H.~F.}\ \bibnamefont {Yang}}, \bibinfo {author} {\bibfnamefont
  {T.}~\bibnamefont {Zhang}}, \bibinfo {author} {\bibfnamefont
  {B.}~\bibnamefont {Zhou}}, \bibinfo {author} {\bibfnamefont {Y.}~\bibnamefont
  {Zhang}}, \bibinfo {author} {\bibfnamefont {Y.~F.}\ \bibnamefont {Guo}},
  \bibinfo {author} {\bibfnamefont {M.}~\bibnamefont {Rahn}}, \bibinfo {author}
  {\bibfnamefont {D.}~\bibnamefont {Prabhakaran}}, \bibinfo {author}
  {\bibfnamefont {Z.}~\bibnamefont {Hussain}}, \bibinfo {author} {\bibfnamefont
  {S.-K.}\ \bibnamefont {Mo}}, \bibinfo {author} {\bibfnamefont
  {C.}~\bibnamefont {Felser}}, \bibinfo {author} {\bibfnamefont
  {B.}~\bibnamefont {Yan}},\ and\ \bibinfo {author} {\bibfnamefont {Y.~L.}\
  \bibnamefont {Chen}},\ }\bibfield  {title} {\bibinfo {title} {{Weyl semimetal
  phase in the non-centrosymmetric compound {TaAs}}},\ }\href
  {https://doi.org/10.1038/nphys3425} {\bibfield  {journal} {\bibinfo
  {journal} {Nature Phys.}\ }\textbf {\bibinfo {volume} {11}},\ \bibinfo
  {pages} {728} (\bibinfo {year} {2015})}\BibitemShut {NoStop}%
\bibitem [{\citenamefont {Lv}\ \emph {et~al.}(2015{\natexlab{a}})\citenamefont
  {Lv}, \citenamefont {Xu}, \citenamefont {Weng}, \citenamefont {Ma},
  \citenamefont {Richard}, \citenamefont {Huang}, \citenamefont {Zhao},
  \citenamefont {Chen}, \citenamefont {Matt}, \citenamefont {Bisti},
  \citenamefont {Strocov}, \citenamefont {Mesot}, \citenamefont {Fang},
  \citenamefont {Dai}, \citenamefont {Qian}, \citenamefont {Shi},\ and\
  \citenamefont {Ding}}]{Lv_Xu_Weng_2015}%
  \BibitemOpen
  \bibfield  {author} {\bibinfo {author} {\bibfnamefont {B.~Q.}\ \bibnamefont
  {Lv}}, \bibinfo {author} {\bibfnamefont {N.}~\bibnamefont {Xu}}, \bibinfo
  {author} {\bibfnamefont {H.~M.}\ \bibnamefont {Weng}}, \bibinfo {author}
  {\bibfnamefont {J.~Z.}\ \bibnamefont {Ma}}, \bibinfo {author} {\bibfnamefont
  {P.}~\bibnamefont {Richard}}, \bibinfo {author} {\bibfnamefont {X.~C.}\
  \bibnamefont {Huang}}, \bibinfo {author} {\bibfnamefont {L.~X.}\ \bibnamefont
  {Zhao}}, \bibinfo {author} {\bibfnamefont {G.~F.}\ \bibnamefont {Chen}},
  \bibinfo {author} {\bibfnamefont {C.~E.}\ \bibnamefont {Matt}}, \bibinfo
  {author} {\bibfnamefont {F.}~\bibnamefont {Bisti}}, \bibinfo {author}
  {\bibfnamefont {V.~N.}\ \bibnamefont {Strocov}}, \bibinfo {author}
  {\bibfnamefont {J.}~\bibnamefont {Mesot}}, \bibinfo {author} {\bibfnamefont
  {Z.}~\bibnamefont {Fang}}, \bibinfo {author} {\bibfnamefont {X.}~\bibnamefont
  {Dai}}, \bibinfo {author} {\bibfnamefont {T.}~\bibnamefont {Qian}}, \bibinfo
  {author} {\bibfnamefont {M.}~\bibnamefont {Shi}},\ and\ \bibinfo {author}
  {\bibfnamefont {H.}~\bibnamefont {Ding}},\ }\bibfield  {title} {\bibinfo
  {title} {{Observation of Weyl nodes in {TaAs}}},\ }\href
  {https://doi.org/10.1038/nphys3426} {\bibfield  {journal} {\bibinfo
  {journal} {Nature Phys.}\ }\textbf {\bibinfo {volume} {11}},\ \bibinfo
  {pages} {724} (\bibinfo {year} {2015}{\natexlab{a}})}\BibitemShut {NoStop}%
\bibitem [{\citenamefont {Wan}\ \emph {et~al.}(2011)\citenamefont {Wan},
  \citenamefont {Turner}, \citenamefont {Vishwanath},\ and\ \citenamefont
  {Savrasov}}]{Wan_Turner_2011}%
  \BibitemOpen
  \bibfield  {author} {\bibinfo {author} {\bibfnamefont {X.}~\bibnamefont
  {Wan}}, \bibinfo {author} {\bibfnamefont {A.~M.}\ \bibnamefont {Turner}},
  \bibinfo {author} {\bibfnamefont {A.}~\bibnamefont {Vishwanath}},\ and\
  \bibinfo {author} {\bibfnamefont {S.~Y.}\ \bibnamefont {Savrasov}},\
  }\bibfield  {title} {\bibinfo {title} {{Topological semimetal and Fermi-arc
  surface states in the electronic structure of pyrochlore iridates}},\ }\href
  {https://doi.org/10.1103/PhysRevB.83.205101} {\bibfield  {journal} {\bibinfo
  {journal} {Phys. Rev. B}\ }\textbf {\bibinfo {volume} {83}},\ \bibinfo
  {pages} {205101} (\bibinfo {year} {2011})}\BibitemShut {NoStop}%
\bibitem [{\citenamefont {Armitage}\ \emph {et~al.}(2018)\citenamefont
  {Armitage}, \citenamefont {Mele},\ and\ \citenamefont
  {Vishwanath}}]{Armitage_Mele_2018}%
  \BibitemOpen
  \bibfield  {author} {\bibinfo {author} {\bibfnamefont {N.~P.}\ \bibnamefont
  {Armitage}}, \bibinfo {author} {\bibfnamefont {E.~J.}\ \bibnamefont {Mele}},\
  and\ \bibinfo {author} {\bibfnamefont {A.}~\bibnamefont {Vishwanath}},\
  }\bibfield  {title} {\bibinfo {title} {{Weyl and Dirac semimetals in
  three-dimensional solids}},\ }\href
  {https://doi.org/10.1103/RevModPhys.90.015001} {\bibfield  {journal}
  {\bibinfo  {journal} {Rev. Mod. Phys.}\ }\textbf {\bibinfo {volume} {90}},\
  \bibinfo {pages} {015001} (\bibinfo {year} {2018})}\BibitemShut {NoStop}%
\bibitem [{\citenamefont {Berry}(1985)}]{Berry_1985}%
  \BibitemOpen
  \bibfield  {author} {\bibinfo {author} {\bibfnamefont {M.~V.}\ \bibnamefont
  {Berry}},\ }\bibinfo {title} {{Aspects of Degeneracy}},\ in\ \href@noop {}
  {\emph {\bibinfo {booktitle} {{Chaotic Behavior in Quantum Systems: Theory
  and Applications}}}},\ \bibinfo {editor} {edited by\ \bibinfo {editor}
  {\bibfnamefont {G.}~\bibnamefont {Casati}}}\ (\bibinfo  {publisher} {Springer
  US},\ \bibinfo {year} {1985})\BibitemShut {NoStop}%
\bibitem [{\citenamefont {Volovik}(2009)}]{Volovik_helium_drop_2009}%
  \BibitemOpen
  \bibfield  {author} {\bibinfo {author} {\bibfnamefont {G.~E.}\ \bibnamefont
  {Volovik}},\ }\href
  {https://doi.org/10.1093/acprof:oso/9780199564842.001.0001} {\emph {\bibinfo
  {title} {{The Universe in a Helium Droplet}}}}\ (\bibinfo  {publisher}
  {Oxford University Press},\ \bibinfo {year} {2009})\BibitemShut {NoStop}%
\bibitem [{\citenamefont {Simon}(1983)}]{Simon_1983}%
  \BibitemOpen
  \bibfield  {author} {\bibinfo {author} {\bibfnamefont {B.}~\bibnamefont
  {Simon}},\ }\bibfield  {title} {\bibinfo {title} {{Holonomy, the Quantum
  Adiabatic Theorem, and Berry's Phase}},\ }\href
  {https://doi.org/10.1103/PhysRevLett.51.2167} {\bibfield  {journal} {\bibinfo
   {journal} {Phys. Rev. Lett.}\ }\textbf {\bibinfo {volume} {51}},\ \bibinfo
  {pages} {2167} (\bibinfo {year} {1983})}\BibitemShut {NoStop}%
\bibitem [{\citenamefont {Berry}(1984)}]{Berry_1984}%
  \BibitemOpen
  \bibfield  {author} {\bibinfo {author} {\bibfnamefont {M.~V.}\ \bibnamefont
  {Berry}},\ }\bibfield  {title} {\bibinfo {title} {{Quantal phase factors
  accompanying adiabatic changes}},\ }\href
  {https://doi.org/10.1098/rspa.1984.0023} {\bibfield  {journal} {\bibinfo
  {journal} {Proc. R. Soc. A}\ }\textbf {\bibinfo {volume} {392}},\ \bibinfo
  {pages} {45} (\bibinfo {year} {1984})}\BibitemShut {NoStop}%
\bibitem [{\citenamefont {Nielsen}\ and\ \citenamefont
  {Ninomiya}(1981{\natexlab{a}})}]{Nielsen_Ninomiya_1981_a}%
  \BibitemOpen
  \bibfield  {author} {\bibinfo {author} {\bibfnamefont {H.}~\bibnamefont
  {Nielsen}}\ and\ \bibinfo {author} {\bibfnamefont {M.}~\bibnamefont
  {Ninomiya}},\ }\bibfield  {title} {\bibinfo {title} {{Absence of neutrinos on
  a lattice: (I). Proof by homotopy theory}},\ }\href
  {https://doi.org/https://doi.org/10.1016/0550-3213(81)90361-8} {\bibfield
  {journal} {\bibinfo  {journal} {Nucl. Phys. B}\ }\textbf {\bibinfo {volume}
  {185}},\ \bibinfo {pages} {20} (\bibinfo {year}
  {1981}{\natexlab{a}})}\BibitemShut {NoStop}%
\bibitem [{\citenamefont {Nielsen}\ and\ \citenamefont
  {Ninomiya}(1981{\natexlab{b}})}]{Nielsen_Ninomiya_1981_b}%
  \BibitemOpen
  \bibfield  {author} {\bibinfo {author} {\bibfnamefont {H.}~\bibnamefont
  {Nielsen}}\ and\ \bibinfo {author} {\bibfnamefont {M.}~\bibnamefont
  {Ninomiya}},\ }\bibfield  {title} {\bibinfo {title} {{Absence of neutrinos on
  a lattice: (II). Intuitive topological proof}},\ }\href
  {https://doi.org/https://doi.org/10.1016/0550-3213(81)90524-1} {\bibfield
  {journal} {\bibinfo  {journal} {Nucl. Phys. B}\ }\textbf {\bibinfo {volume}
  {193}},\ \bibinfo {pages} {173} (\bibinfo {year}
  {1981}{\natexlab{b}})}\BibitemShut {NoStop}%
\bibitem [{\citenamefont {Lv}\ \emph {et~al.}(2021)\citenamefont {Lv},
  \citenamefont {Qian},\ and\ \citenamefont {Ding}}]{Lv_Qian_Ding_2021}%
  \BibitemOpen
  \bibfield  {author} {\bibinfo {author} {\bibfnamefont {B.~Q.}\ \bibnamefont
  {Lv}}, \bibinfo {author} {\bibfnamefont {T.}~\bibnamefont {Qian}},\ and\
  \bibinfo {author} {\bibfnamefont {H.}~\bibnamefont {Ding}},\ }\bibfield
  {title} {\bibinfo {title} {{Experimental perspective on three-dimensional
  topological semimetals}},\ }\href
  {https://doi.org/10.1103/RevModPhys.93.025002} {\bibfield  {journal}
  {\bibinfo  {journal} {Rev. Mod. Phys.}\ }\textbf {\bibinfo {volume} {93}},\
  \bibinfo {pages} {025002} (\bibinfo {year} {2021})}\BibitemShut {NoStop}%
\bibitem [{\citenamefont {Adler}(1969)}]{Adler_1969}%
  \BibitemOpen
  \bibfield  {author} {\bibinfo {author} {\bibfnamefont {S.~L.}\ \bibnamefont
  {Adler}},\ }\bibfield  {title} {\bibinfo {title} {{Axial-Vector Vertex in
  Spinor Electrodynamics}},\ }\href {https://doi.org/10.1103/PhysRev.177.2426}
  {\bibfield  {journal} {\bibinfo  {journal} {Phys. Rev.}\ }\textbf {\bibinfo
  {volume} {177}},\ \bibinfo {pages} {2426} (\bibinfo {year}
  {1969})}\BibitemShut {NoStop}%
\bibitem [{\citenamefont {Bell}\ and\ \citenamefont
  {Jackiw}(1969)}]{Bell_Jackiw_1969}%
  \BibitemOpen
  \bibfield  {author} {\bibinfo {author} {\bibfnamefont {J.~S.}\ \bibnamefont
  {Bell}}\ and\ \bibinfo {author} {\bibfnamefont {R.}~\bibnamefont {Jackiw}},\
  }\bibfield  {title} {\bibinfo {title} {{A {PCAC} puzzle:
  $\pi^0\rightarrow\gamma\gamma$ in the $\sigma$-model}},\ }\href
  {https://doi.org/10.1007/bf02823296} {\bibfield  {journal} {\bibinfo
  {journal} {Il Nuovo Cimento A}\ }\textbf {\bibinfo {volume} {60}},\ \bibinfo
  {pages} {47} (\bibinfo {year} {1969})}\BibitemShut {NoStop}%
\bibitem [{\citenamefont {Nielsen}\ and\ \citenamefont
  {Ninomiya}(1983)}]{Nielsen_Ninomiya_1983}%
  \BibitemOpen
  \bibfield  {author} {\bibinfo {author} {\bibfnamefont {H.}~\bibnamefont
  {Nielsen}}\ and\ \bibinfo {author} {\bibfnamefont {M.}~\bibnamefont
  {Ninomiya}},\ }\bibfield  {title} {\bibinfo {title} {{The Adler-Bell-Jackiw
  anomaly and Weyl fermions in a crystal}},\ }\href
  {https://doi.org/https://doi.org/10.1016/0370-2693(83)91529-0} {\bibfield
  {journal} {\bibinfo  {journal} {Phys. Lett. B}\ }\textbf {\bibinfo {volume}
  {130}},\ \bibinfo {pages} {389} (\bibinfo {year} {1983})}\BibitemShut
  {NoStop}%
\bibitem [{\citenamefont {Son}\ and\ \citenamefont
  {Spivak}(2013)}]{Son_Spivak_2013}%
  \BibitemOpen
  \bibfield  {author} {\bibinfo {author} {\bibfnamefont {D.~T.}\ \bibnamefont
  {Son}}\ and\ \bibinfo {author} {\bibfnamefont {B.~Z.}\ \bibnamefont
  {Spivak}},\ }\bibfield  {title} {\bibinfo {title} {{Chiral anomaly and
  classical negative magnetoresistance of Weyl metals}},\ }\href
  {https://doi.org/10.1103/PhysRevB.88.104412} {\bibfield  {journal} {\bibinfo
  {journal} {Phys. Rev. B}\ }\textbf {\bibinfo {volume} {88}},\ \bibinfo
  {pages} {104412} (\bibinfo {year} {2013})}\BibitemShut {NoStop}%
\bibitem [{\citenamefont {Kim}\ \emph {et~al.}(2014)\citenamefont {Kim},
  \citenamefont {Kim},\ and\ \citenamefont {Sasaki}}]{Kim_Kim_Sasaki_2014}%
  \BibitemOpen
  \bibfield  {author} {\bibinfo {author} {\bibfnamefont {K.-S.}\ \bibnamefont
  {Kim}}, \bibinfo {author} {\bibfnamefont {H.-J.}\ \bibnamefont {Kim}},\ and\
  \bibinfo {author} {\bibfnamefont {M.}~\bibnamefont {Sasaki}},\ }\bibfield
  {title} {\bibinfo {title} {{Boltzmann equation approach to anomalous
  transport in a Weyl metal}},\ }\href
  {https://doi.org/10.1103/PhysRevB.89.195137} {\bibfield  {journal} {\bibinfo
  {journal} {Phys. Rev. B}\ }\textbf {\bibinfo {volume} {89}},\ \bibinfo
  {pages} {195137} (\bibinfo {year} {2014})}\BibitemShut {NoStop}%
\bibitem [{\citenamefont {Burkov}(2014{\natexlab{a}})}]{Burkov_chiral_a_2014}%
  \BibitemOpen
  \bibfield  {author} {\bibinfo {author} {\bibfnamefont {A.~A.}\ \bibnamefont
  {Burkov}},\ }\bibfield  {title} {\bibinfo {title} {{Chiral Anomaly and
  Diffusive Magnetotransport in Weyl Metals}},\ }\href
  {https://doi.org/10.1103/PhysRevLett.113.247203} {\bibfield  {journal}
  {\bibinfo  {journal} {Phys. Rev. Lett.}\ }\textbf {\bibinfo {volume} {113}},\
  \bibinfo {pages} {247203} (\bibinfo {year} {2014}{\natexlab{a}})}\BibitemShut
  {NoStop}%
\bibitem [{\citenamefont {Xiong}\ \emph {et~al.}(2015)\citenamefont {Xiong},
  \citenamefont {Kushwaha}, \citenamefont {Liang}, \citenamefont {Krizan},
  \citenamefont {Hirschberger}, \citenamefont {Wang}, \citenamefont {Cava},\
  and\ \citenamefont {Ong}}]{Xiong_Kushwaha_Liang_2015}%
  \BibitemOpen
  \bibfield  {author} {\bibinfo {author} {\bibfnamefont {J.}~\bibnamefont
  {Xiong}}, \bibinfo {author} {\bibfnamefont {S.~K.}\ \bibnamefont {Kushwaha}},
  \bibinfo {author} {\bibfnamefont {T.}~\bibnamefont {Liang}}, \bibinfo
  {author} {\bibfnamefont {J.~W.}\ \bibnamefont {Krizan}}, \bibinfo {author}
  {\bibfnamefont {M.}~\bibnamefont {Hirschberger}}, \bibinfo {author}
  {\bibfnamefont {W.}~\bibnamefont {Wang}}, \bibinfo {author} {\bibfnamefont
  {R.~J.}\ \bibnamefont {Cava}},\ and\ \bibinfo {author} {\bibfnamefont
  {N.~P.}\ \bibnamefont {Ong}},\ }\bibfield  {title} {\bibinfo {title}
  {{Evidence for the chiral anomaly in the Dirac semimetal Na$_3$Bi}},\ }\href
  {https://doi.org/10.1126/science.aac6089} {\bibfield  {journal} {\bibinfo
  {journal} {Science}\ }\textbf {\bibinfo {volume} {350}},\ \bibinfo {pages}
  {413} (\bibinfo {year} {2015})}\BibitemShut {NoStop}%
\bibitem [{\citenamefont {Huang}\ \emph {et~al.}(2015)\citenamefont {Huang},
  \citenamefont {Zhao}, \citenamefont {Long}, \citenamefont {Wang},
  \citenamefont {Chen}, \citenamefont {Yang}, \citenamefont {Liang},
  \citenamefont {Xue}, \citenamefont {Weng}, \citenamefont {Fang},
  \citenamefont {Dai},\ and\ \citenamefont {Chen}}]{Huang_Zhao_Long_2015}%
  \BibitemOpen
  \bibfield  {author} {\bibinfo {author} {\bibfnamefont {X.}~\bibnamefont
  {Huang}}, \bibinfo {author} {\bibfnamefont {L.}~\bibnamefont {Zhao}},
  \bibinfo {author} {\bibfnamefont {Y.}~\bibnamefont {Long}}, \bibinfo {author}
  {\bibfnamefont {P.}~\bibnamefont {Wang}}, \bibinfo {author} {\bibfnamefont
  {D.}~\bibnamefont {Chen}}, \bibinfo {author} {\bibfnamefont {Z.}~\bibnamefont
  {Yang}}, \bibinfo {author} {\bibfnamefont {H.}~\bibnamefont {Liang}},
  \bibinfo {author} {\bibfnamefont {M.}~\bibnamefont {Xue}}, \bibinfo {author}
  {\bibfnamefont {H.}~\bibnamefont {Weng}}, \bibinfo {author} {\bibfnamefont
  {Z.}~\bibnamefont {Fang}}, \bibinfo {author} {\bibfnamefont {X.}~\bibnamefont
  {Dai}},\ and\ \bibinfo {author} {\bibfnamefont {G.}~\bibnamefont {Chen}},\
  }\bibfield  {title} {\bibinfo {title} {{Observation of the
  Chiral-Anomaly-Induced Negative Magnetoresistance in 3D Weyl Semimetal
  TaAs}},\ }\href {https://doi.org/10.1103/PhysRevX.5.031023} {\bibfield
  {journal} {\bibinfo  {journal} {Phys. Rev. X}\ }\textbf {\bibinfo {volume}
  {5}},\ \bibinfo {pages} {031023} (\bibinfo {year} {2015})}\BibitemShut
  {NoStop}%
\bibitem [{\citenamefont {Zhang}\ \emph {et~al.}(2016)\citenamefont {Zhang},
  \citenamefont {Xu}, \citenamefont {Belopolski}, \citenamefont {Yuan},
  \citenamefont {Lin}, \citenamefont {Tong}, \citenamefont {Bian},
  \citenamefont {Alidoust}, \citenamefont {Lee}, \citenamefont {Huang},
  \citenamefont {Chang}, \citenamefont {Chang}, \citenamefont {Hsu},
  \citenamefont {Jeng}, \citenamefont {Neupane}, \citenamefont {Sanchez},
  \citenamefont {Zheng}, \citenamefont {Wang}, \citenamefont {Lin},
  \citenamefont {Zhang}, \citenamefont {Lu}, \citenamefont {Shen},
  \citenamefont {Neupert}, \citenamefont {Hasan},\ and\ \citenamefont
  {Jia}}]{Zhang_Xu_2016}%
  \BibitemOpen
  \bibfield  {author} {\bibinfo {author} {\bibfnamefont {C.-L.}\ \bibnamefont
  {Zhang}}, \bibinfo {author} {\bibfnamefont {S.-Y.}\ \bibnamefont {Xu}},
  \bibinfo {author} {\bibfnamefont {I.}~\bibnamefont {Belopolski}}, \bibinfo
  {author} {\bibfnamefont {Z.}~\bibnamefont {Yuan}}, \bibinfo {author}
  {\bibfnamefont {Z.}~\bibnamefont {Lin}}, \bibinfo {author} {\bibfnamefont
  {B.}~\bibnamefont {Tong}}, \bibinfo {author} {\bibfnamefont {G.}~\bibnamefont
  {Bian}}, \bibinfo {author} {\bibfnamefont {N.}~\bibnamefont {Alidoust}},
  \bibinfo {author} {\bibfnamefont {C.-C.}\ \bibnamefont {Lee}}, \bibinfo
  {author} {\bibfnamefont {S.-M.}\ \bibnamefont {Huang}}, \bibinfo {author}
  {\bibfnamefont {T.-R.}\ \bibnamefont {Chang}}, \bibinfo {author}
  {\bibfnamefont {G.}~\bibnamefont {Chang}}, \bibinfo {author} {\bibfnamefont
  {C.-H.}\ \bibnamefont {Hsu}}, \bibinfo {author} {\bibfnamefont {H.-T.}\
  \bibnamefont {Jeng}}, \bibinfo {author} {\bibfnamefont {M.}~\bibnamefont
  {Neupane}}, \bibinfo {author} {\bibfnamefont {D.~S.}\ \bibnamefont
  {Sanchez}}, \bibinfo {author} {\bibfnamefont {H.}~\bibnamefont {Zheng}},
  \bibinfo {author} {\bibfnamefont {J.}~\bibnamefont {Wang}}, \bibinfo {author}
  {\bibfnamefont {H.}~\bibnamefont {Lin}}, \bibinfo {author} {\bibfnamefont
  {C.}~\bibnamefont {Zhang}}, \bibinfo {author} {\bibfnamefont {H.-Z.}\
  \bibnamefont {Lu}}, \bibinfo {author} {\bibfnamefont {S.-Q.}\ \bibnamefont
  {Shen}}, \bibinfo {author} {\bibfnamefont {T.}~\bibnamefont {Neupert}},
  \bibinfo {author} {\bibfnamefont {M.~Z.}\ \bibnamefont {Hasan}},\ and\
  \bibinfo {author} {\bibfnamefont {S.}~\bibnamefont {Jia}},\ }\bibfield
  {title} {\bibinfo {title} {{Signatures of the
  Adler{\textendash}Bell{\textendash}Jackiw chiral anomaly in a Weyl fermion
  semimetal}},\ }\href {https://doi.org/10.1038/ncomms10735} {\bibfield
  {journal} {\bibinfo  {journal} {Nature Commun.}\ }\textbf {\bibinfo {volume}
  {7}},\ \bibinfo {pages} {10735} (\bibinfo {year} {2016})}\BibitemShut
  {NoStop}%
\bibitem [{\citenamefont {Hirschberger}\ \emph {et~al.}(2016)\citenamefont
  {Hirschberger}, \citenamefont {Kushwaha}, \citenamefont {Wang}, \citenamefont
  {Gibson}, \citenamefont {Liang}, \citenamefont {Belvin}, \citenamefont
  {Bernevig}, \citenamefont {Cava},\ and\ \citenamefont
  {Ong}}]{Hirschberger_Kushwaha_Wang_2016}%
  \BibitemOpen
  \bibfield  {author} {\bibinfo {author} {\bibfnamefont {M.}~\bibnamefont
  {Hirschberger}}, \bibinfo {author} {\bibfnamefont {S.}~\bibnamefont
  {Kushwaha}}, \bibinfo {author} {\bibfnamefont {Z.}~\bibnamefont {Wang}},
  \bibinfo {author} {\bibfnamefont {Q.}~\bibnamefont {Gibson}}, \bibinfo
  {author} {\bibfnamefont {S.}~\bibnamefont {Liang}}, \bibinfo {author}
  {\bibfnamefont {C.~A.}\ \bibnamefont {Belvin}}, \bibinfo {author}
  {\bibfnamefont {B.~A.}\ \bibnamefont {Bernevig}}, \bibinfo {author}
  {\bibfnamefont {R.~J.}\ \bibnamefont {Cava}},\ and\ \bibinfo {author}
  {\bibfnamefont {N.~P.}\ \bibnamefont {Ong}},\ }\bibfield  {title} {\bibinfo
  {title} {{The chiral anomaly and thermopower of Weyl fermions in the
  half-Heusler~{GdPtBi}}},\ }\href {https://doi.org/10.1038/nmat4684}
  {\bibfield  {journal} {\bibinfo  {journal} {Nature Mater.}\ }\textbf
  {\bibinfo {volume} {15}},\ \bibinfo {pages} {1161} (\bibinfo {year}
  {2016})}\BibitemShut {NoStop}%
\bibitem [{\citenamefont {Li}\ \emph {et~al.}(2017)\citenamefont {Li},
  \citenamefont {Wang}, \citenamefont {Li}, \citenamefont {Yang}, \citenamefont
  {Shen}, \citenamefont {Sheng}, \citenamefont {Li}, \citenamefont {Lu},
  \citenamefont {Zheng},\ and\ \citenamefont {Xu}}]{Li_Wang_Li_2017}%
  \BibitemOpen
  \bibfield  {author} {\bibinfo {author} {\bibfnamefont {Y.}~\bibnamefont
  {Li}}, \bibinfo {author} {\bibfnamefont {Z.}~\bibnamefont {Wang}}, \bibinfo
  {author} {\bibfnamefont {P.}~\bibnamefont {Li}}, \bibinfo {author}
  {\bibfnamefont {X.}~\bibnamefont {Yang}}, \bibinfo {author} {\bibfnamefont
  {Z.}~\bibnamefont {Shen}}, \bibinfo {author} {\bibfnamefont {F.}~\bibnamefont
  {Sheng}}, \bibinfo {author} {\bibfnamefont {X.}~\bibnamefont {Li}}, \bibinfo
  {author} {\bibfnamefont {Y.}~\bibnamefont {Lu}}, \bibinfo {author}
  {\bibfnamefont {Y.}~\bibnamefont {Zheng}},\ and\ \bibinfo {author}
  {\bibfnamefont {Z.-A.}\ \bibnamefont {Xu}},\ }\bibfield  {title} {\bibinfo
  {title} {{Negative magnetoresistance in Weyl semimetals {NbAs} and {NbP}:
  Intrinsic chiral anomaly and extrinsic effects}},\ }\href
  {https://doi.org/10.1007/s11467-016-0636-8} {\bibfield  {journal} {\bibinfo
  {journal} {Front. Phys.}\ }\textbf {\bibinfo {volume} {12}},\ \bibinfo
  {pages} {127205} (\bibinfo {year} {2017})}\BibitemShut {NoStop}%
\bibitem [{\citenamefont {Dantas}\ \emph {et~al.}(2018)\citenamefont {Dantas},
  \citenamefont {Pe{\~{n}}a-Benitez}, \citenamefont {Roy},\ and\ \citenamefont
  {Sur{\'{o}}wka}}]{Dantas_Benitez_Roy_2018}%
  \BibitemOpen
  \bibfield  {author} {\bibinfo {author} {\bibfnamefont {R.~M.~A.}\
  \bibnamefont {Dantas}}, \bibinfo {author} {\bibfnamefont {F.}~\bibnamefont
  {Pe{\~{n}}a-Benitez}}, \bibinfo {author} {\bibfnamefont {B.}~\bibnamefont
  {Roy}},\ and\ \bibinfo {author} {\bibfnamefont {P.}~\bibnamefont
  {Sur{\'{o}}wka}},\ }\bibfield  {title} {\bibinfo {title} {{Magnetotransport
  in multi-Weyl semimetals: a kinetic theory approach}},\ }\href
  {https://doi.org/10.1007/jhep12(2018)069} {\bibfield  {journal} {\bibinfo
  {journal} {J. High Energ. Phys.}\ }\textbf {\bibinfo {volume} {2018}}\bibinfo
   {number} { (12)},\ \bibinfo {pages} {69}}\BibitemShut {NoStop}%
\bibitem [{\citenamefont {Xiao}\ \emph {et~al.}(2010)\citenamefont {Xiao},
  \citenamefont {Chang},\ and\ \citenamefont {Niu}}]{Xiao_Chang_2010}%
  \BibitemOpen
\bibfield  {number} {  }\bibfield  {author} {\bibinfo {author} {\bibfnamefont
  {D.}~\bibnamefont {Xiao}}, \bibinfo {author} {\bibfnamefont {M.-C.}\
  \bibnamefont {Chang}},\ and\ \bibinfo {author} {\bibfnamefont
  {Q.}~\bibnamefont {Niu}},\ }\bibfield  {title} {\bibinfo {title} {{Berry
  phase effects on electronic properties}},\ }\href
  {https://doi.org/10.1103/RevModPhys.82.1959} {\bibfield  {journal} {\bibinfo
  {journal} {Rev. Mod. Phys.}\ }\textbf {\bibinfo {volume} {82}},\ \bibinfo
  {pages} {1959} (\bibinfo {year} {2010})}\BibitemShut {NoStop}%
\bibitem [{\citenamefont {Sundaram}\ and\ \citenamefont
  {Niu}(1999)}]{Sundaram_Niu_1999}%
  \BibitemOpen
  \bibfield  {author} {\bibinfo {author} {\bibfnamefont {G.}~\bibnamefont
  {Sundaram}}\ and\ \bibinfo {author} {\bibfnamefont {Q.}~\bibnamefont {Niu}},\
  }\bibfield  {title} {\bibinfo {title} {{Wave-packet dynamics in slowly
  perturbed crystals: Gradient corrections and Berry-phase effects}},\ }\href
  {https://doi.org/10.1103/PhysRevB.59.14915} {\bibfield  {journal} {\bibinfo
  {journal} {Phys. Rev. B}\ }\textbf {\bibinfo {volume} {59}},\ \bibinfo
  {pages} {14915} (\bibinfo {year} {1999})}\BibitemShut {NoStop}%
\bibitem [{\citenamefont {Knoll}\ \emph {et~al.}(2020)\citenamefont {Knoll},
  \citenamefont {Timm},\ and\ \citenamefont {Meng}}]{KTM20}%
  \BibitemOpen
  \bibfield  {author} {\bibinfo {author} {\bibfnamefont {A.}~\bibnamefont
  {Knoll}}, \bibinfo {author} {\bibfnamefont {C.}~\bibnamefont {Timm}},\ and\
  \bibinfo {author} {\bibfnamefont {T.}~\bibnamefont {Meng}},\ }\bibfield
  {title} {\bibinfo {title} {{Negative longitudinal magnetoconductance at weak
  fields in Weyl semimetals}},\ }\href
  {https://doi.org/10.1103/PhysRevB.101.201402} {\bibfield  {journal} {\bibinfo
   {journal} {Phys. Rev. B}\ }\textbf {\bibinfo {volume} {101}},\ \bibinfo
  {pages} {201402(R)} (\bibinfo {year} {2020})}\BibitemShut {NoStop}%
\bibitem [{\citenamefont {Xu}\ \emph {et~al.}(2011)\citenamefont {Xu},
  \citenamefont {Weng}, \citenamefont {Wang}, \citenamefont {Dai},\ and\
  \citenamefont {Fang}}]{Xu_Weng_Wang_2011}%
  \BibitemOpen
  \bibfield  {author} {\bibinfo {author} {\bibfnamefont {G.}~\bibnamefont
  {Xu}}, \bibinfo {author} {\bibfnamefont {H.}~\bibnamefont {Weng}}, \bibinfo
  {author} {\bibfnamefont {Z.}~\bibnamefont {Wang}}, \bibinfo {author}
  {\bibfnamefont {X.}~\bibnamefont {Dai}},\ and\ \bibinfo {author}
  {\bibfnamefont {Z.}~\bibnamefont {Fang}},\ }\bibfield  {title} {\bibinfo
  {title} {{Chern Semimetal and the Quantized Anomalous Hall Effect in
  $\mathrm{HgCr}_2\mathrm{Se}_4$}},\ }\href
  {https://doi.org/10.1103/PhysRevLett.107.186806} {\bibfield  {journal}
  {\bibinfo  {journal} {Phys. Rev. Lett.}\ }\textbf {\bibinfo {volume} {107}},\
  \bibinfo {pages} {186806} (\bibinfo {year} {2011})}\BibitemShut {NoStop}%
\bibitem [{\citenamefont {Lv}\ \emph {et~al.}(2015{\natexlab{b}})\citenamefont
  {Lv}, \citenamefont {Weng}, \citenamefont {Fu}, \citenamefont {Wang},
  \citenamefont {Miao}, \citenamefont {Ma}, \citenamefont {Richard},
  \citenamefont {Huang}, \citenamefont {Zhao}, \citenamefont {Chen},
  \citenamefont {Fang}, \citenamefont {Dai}, \citenamefont {Qian},\ and\
  \citenamefont {Ding}}]{Lv_Weng_Fu_2015}%
  \BibitemOpen
  \bibfield  {author} {\bibinfo {author} {\bibfnamefont {B.~Q.}\ \bibnamefont
  {Lv}}, \bibinfo {author} {\bibfnamefont {H.~M.}\ \bibnamefont {Weng}},
  \bibinfo {author} {\bibfnamefont {B.~B.}\ \bibnamefont {Fu}}, \bibinfo
  {author} {\bibfnamefont {X.~P.}\ \bibnamefont {Wang}}, \bibinfo {author}
  {\bibfnamefont {H.}~\bibnamefont {Miao}}, \bibinfo {author} {\bibfnamefont
  {J.}~\bibnamefont {Ma}}, \bibinfo {author} {\bibfnamefont {P.}~\bibnamefont
  {Richard}}, \bibinfo {author} {\bibfnamefont {X.~C.}\ \bibnamefont {Huang}},
  \bibinfo {author} {\bibfnamefont {L.~X.}\ \bibnamefont {Zhao}}, \bibinfo
  {author} {\bibfnamefont {G.~F.}\ \bibnamefont {Chen}}, \bibinfo {author}
  {\bibfnamefont {Z.}~\bibnamefont {Fang}}, \bibinfo {author} {\bibfnamefont
  {X.}~\bibnamefont {Dai}}, \bibinfo {author} {\bibfnamefont {T.}~\bibnamefont
  {Qian}},\ and\ \bibinfo {author} {\bibfnamefont {H.}~\bibnamefont {Ding}},\
  }\bibfield  {title} {\bibinfo {title} {{Experimental Discovery of Weyl
  Semimetal TaAs}},\ }\href {https://doi.org/10.1103/PhysRevX.5.031013}
  {\bibfield  {journal} {\bibinfo  {journal} {Phys. Rev. X}\ }\textbf {\bibinfo
  {volume} {5}},\ \bibinfo {pages} {031013} (\bibinfo {year}
  {2015}{\natexlab{b}})}\BibitemShut {NoStop}%
\bibitem [{\citenamefont {Grassano}\ \emph {et~al.}(2018)\citenamefont
  {Grassano}, \citenamefont {Pulci}, \citenamefont {Conte},\ and\ \citenamefont
  {Bechstedt}}]{Grassano_Pulci_2018}%
  \BibitemOpen
  \bibfield  {author} {\bibinfo {author} {\bibfnamefont {D.}~\bibnamefont
  {Grassano}}, \bibinfo {author} {\bibfnamefont {O.}~\bibnamefont {Pulci}},
  \bibinfo {author} {\bibfnamefont {A.~M.}\ \bibnamefont {Conte}},\ and\
  \bibinfo {author} {\bibfnamefont {F.}~\bibnamefont {Bechstedt}},\ }\bibfield
  {title} {\bibinfo {title} {{Validity of Weyl fermion picture for transition
  metals monopnictides {TaAs}, {TaP}, {NbAs}, and {NbP} from ab initio
  studies}},\ }\href {https://doi.org/10.1038/s41598-018-21465-z} {\bibfield
  {journal} {\bibinfo  {journal} {Sci. Rep.}\ }\textbf {\bibinfo {volume}
  {8}},\ \bibinfo {pages} {3534} (\bibinfo {year} {2018})}\BibitemShut
  {NoStop}%
\bibitem [{\citenamefont {Grassano}\ \emph {et~al.}(2020)\citenamefont
  {Grassano}, \citenamefont {Pulci}, \citenamefont {Cannuccia},\ and\
  \citenamefont {Bechstedt}}]{Grassano_Pulci_Cannuccia_2020}%
  \BibitemOpen
  \bibfield  {author} {\bibinfo {author} {\bibfnamefont {D.}~\bibnamefont
  {Grassano}}, \bibinfo {author} {\bibfnamefont {O.}~\bibnamefont {Pulci}},
  \bibinfo {author} {\bibfnamefont {E.}~\bibnamefont {Cannuccia}},\ and\
  \bibinfo {author} {\bibfnamefont {F.}~\bibnamefont {Bechstedt}},\ }\bibfield
  {title} {\bibinfo {title} {{Influence of anisotropy, tilt and pairing of Weyl
  nodes: the Weyl semimetals {TaAs}, {TaP}, {NbAs} and {NbP}}},\ }\href
  {https://doi.org/10.1140/epjb/e2020-10110-x} {\bibfield  {journal} {\bibinfo
  {journal} {Euro. Phys. J. B}\ }\textbf {\bibinfo {volume} {93}},\ \bibinfo
  {pages} {157} (\bibinfo {year} {2020})}\BibitemShut {NoStop}%
\bibitem [{\citenamefont {Trescher}\ \emph {et~al.}(2015)\citenamefont
  {Trescher}, \citenamefont {Sbierski}, \citenamefont {Brouwer},\ and\
  \citenamefont {Bergholtz}}]{Trescher_Sbierski_2015}%
  \BibitemOpen
  \bibfield  {author} {\bibinfo {author} {\bibfnamefont {M.}~\bibnamefont
  {Trescher}}, \bibinfo {author} {\bibfnamefont {B.}~\bibnamefont {Sbierski}},
  \bibinfo {author} {\bibfnamefont {P.~W.}\ \bibnamefont {Brouwer}},\ and\
  \bibinfo {author} {\bibfnamefont {E.~J.}\ \bibnamefont {Bergholtz}},\
  }\bibfield  {title} {\bibinfo {title} {Quantum transport in {Dirac}
  materials: Signatures of tilted and anisotropic {Dirac} and {Weyl} cones},\
  }\href {https://doi.org/10.1103/PhysRevB.91.115135} {\bibfield  {journal}
  {\bibinfo  {journal} {Phys. Rev. B}\ }\textbf {\bibinfo {volume} {91}},\
  \bibinfo {pages} {115135} (\bibinfo {year} {2015})}\BibitemShut {NoStop}%
\bibitem [{\citenamefont {Johansson}\ \emph {et~al.}(2019)\citenamefont
  {Johansson}, \citenamefont {Henk},\ and\ \citenamefont
  {Mertig}}]{Johansson_Henk_2019}%
  \BibitemOpen
  \bibfield  {author} {\bibinfo {author} {\bibfnamefont {A.}~\bibnamefont
  {Johansson}}, \bibinfo {author} {\bibfnamefont {J.}~\bibnamefont {Henk}},\
  and\ \bibinfo {author} {\bibfnamefont {I.}~\bibnamefont {Mertig}},\
  }\bibfield  {title} {\bibinfo {title} {{Chiral anomaly in type-I Weyl
  semimetals: Comprehensive analysis within a semiclassical Fermi surface
  harmonics approach}},\ }\href {https://doi.org/10.1103/PhysRevB.99.075114}
  {\bibfield  {journal} {\bibinfo  {journal} {Phys. Rev. B}\ }\textbf {\bibinfo
  {volume} {99}},\ \bibinfo {pages} {075114} (\bibinfo {year}
  {2019})}\BibitemShut {NoStop}%
\bibitem [{\citenamefont {Panfilov}\ \emph {et~al.}(2014)\citenamefont
  {Panfilov}, \citenamefont {Burkov},\ and\ \citenamefont
  {Pesin}}]{Panfilov_Burkov_2014}%
  \BibitemOpen
  \bibfield  {author} {\bibinfo {author} {\bibfnamefont {I.}~\bibnamefont
  {Panfilov}}, \bibinfo {author} {\bibfnamefont {A.~A.}\ \bibnamefont
  {Burkov}},\ and\ \bibinfo {author} {\bibfnamefont {D.~A.}\ \bibnamefont
  {Pesin}},\ }\bibfield  {title} {\bibinfo {title} {{Density response in Weyl
  metals}},\ }\href {https://doi.org/10.1103/PhysRevB.89.245103} {\bibfield
  {journal} {\bibinfo  {journal} {Phys. Rev. B}\ }\textbf {\bibinfo {volume}
  {89}},\ \bibinfo {pages} {245103} (\bibinfo {year} {2014})}\BibitemShut
  {NoStop}%
\bibitem [{\citenamefont {Kapusta}\ and\ \citenamefont
  {Toimela}(1988)}]{Kapusta_Toimela_1988}%
  \BibitemOpen
  \bibfield  {author} {\bibinfo {author} {\bibfnamefont {J.}~\bibnamefont
  {Kapusta}}\ and\ \bibinfo {author} {\bibfnamefont {T.}~\bibnamefont
  {Toimela}},\ }\bibfield  {title} {\bibinfo {title} {{Friedel oscillations in
  relativistic QED and QCD}},\ }\href
  {https://doi.org/10.1103/PhysRevD.37.3731} {\bibfield  {journal} {\bibinfo
  {journal} {Phys. Rev. D}\ }\textbf {\bibinfo {volume} {37}},\ \bibinfo
  {pages} {3731} (\bibinfo {year} {1988})}\BibitemShut {NoStop}%
\bibitem [{\citenamefont {Lv}\ and\ \citenamefont
  {Zhang}(2013)}]{Lv_Zhang_2013}%
  \BibitemOpen
  \bibfield  {author} {\bibinfo {author} {\bibfnamefont {M.}~\bibnamefont
  {Lv}}\ and\ \bibinfo {author} {\bibfnamefont {S.-C.}\ \bibnamefont {Zhang}},\
  }\bibfield  {title} {\bibinfo {title} {{Dielectric function, Friedel
  oscillation and plasmons in Weyl semimetals}},\ }\href
  {https://doi.org/10.1142/S0217979213501774} {\bibfield  {journal} {\bibinfo
  {journal} {Int. J. Mod. Phys. B}\ }\textbf {\bibinfo {volume} {27}},\
  \bibinfo {pages} {1350177} (\bibinfo {year} {2013})}\BibitemShut {NoStop}%
\bibitem [{\citenamefont {Burkov}(2014{\natexlab{b}})}]{Burkov_2014}%
  \BibitemOpen
  \bibfield  {author} {\bibinfo {author} {\bibfnamefont {A.~A.}\ \bibnamefont
  {Burkov}},\ }\bibfield  {title} {\bibinfo {title} {Topological response in
  ferromagnets},\ }\href {https://doi.org/10.1103/PhysRevB.89.155104}
  {\bibfield  {journal} {\bibinfo  {journal} {Phys. Rev. B}\ }\textbf {\bibinfo
  {volume} {89}},\ \bibinfo {pages} {155104} (\bibinfo {year}
  {2014}{\natexlab{b}})}\BibitemShut {NoStop}%
\bibitem [{\citenamefont {Zhou}\ \emph {et~al.}(2015)\citenamefont {Zhou},
  \citenamefont {Chang},\ and\ \citenamefont {Xiao}}]{Zhou_Chang_2015}%
  \BibitemOpen
  \bibfield  {author} {\bibinfo {author} {\bibfnamefont {J.}~\bibnamefont
  {Zhou}}, \bibinfo {author} {\bibfnamefont {H.-R.}\ \bibnamefont {Chang}},\
  and\ \bibinfo {author} {\bibfnamefont {D.}~\bibnamefont {Xiao}},\ }\bibfield
  {title} {\bibinfo {title} {{Plasmon mode as a detection of the chiral anomaly
  in Weyl semimetals}},\ }\href {https://doi.org/10.1103/PhysRevB.91.035114}
  {\bibfield  {journal} {\bibinfo  {journal} {Phys. Rev. B}\ }\textbf {\bibinfo
  {volume} {91}},\ \bibinfo {pages} {035114} (\bibinfo {year}
  {2015})}\BibitemShut {NoStop}%
\bibitem [{\citenamefont {Ghosh}\ and\ \citenamefont
  {Timm}(2019)}]{Ghosh_Timm_2019}%
  \BibitemOpen
  \bibfield  {author} {\bibinfo {author} {\bibfnamefont {S.}~\bibnamefont
  {Ghosh}}\ and\ \bibinfo {author} {\bibfnamefont {C.}~\bibnamefont {Timm}},\
  }\bibfield  {title} {\bibinfo {title} {{Charge-spin response and collective
  excitations in Weyl semimetals}},\ }\href
  {https://doi.org/10.1103/PhysRevB.99.075104} {\bibfield  {journal} {\bibinfo
  {journal} {Phys. Rev. B}\ }\textbf {\bibinfo {volume} {99}},\ \bibinfo
  {pages} {075104} (\bibinfo {year} {2019})}\BibitemShut {NoStop}%
\bibitem [{\citenamefont {Zheng}\ \emph {et~al.}(2016)\citenamefont {Zheng},
  \citenamefont {Wang}, \citenamefont {Zhong},\ and\ \citenamefont
  {Duan}}]{Zheng_Wang_Zhong_2016}%
  \BibitemOpen
  \bibfield  {author} {\bibinfo {author} {\bibfnamefont {S.-H.}\ \bibnamefont
  {Zheng}}, \bibinfo {author} {\bibfnamefont {R.-Q.}\ \bibnamefont {Wang}},
  \bibinfo {author} {\bibfnamefont {M.}~\bibnamefont {Zhong}},\ and\ \bibinfo
  {author} {\bibfnamefont {H.-J.}\ \bibnamefont {Duan}},\ }\bibfield  {title}
  {\bibinfo {title} {{Resonance states and beating pattern induced by quantum
  impurity scattering in Weyl/Dirac semimetals}},\ }\href
  {https://doi.org/10.1038/srep36106} {\bibfield  {journal} {\bibinfo
  {journal} {Sci. Rep.}\ }\textbf {\bibinfo {volume} {6}},\ \bibinfo {pages}
  {36106} (\bibinfo {year} {2016})}\BibitemShut {NoStop}%
\bibitem [{\citenamefont {D\'iaz-Fern\'andez}\ \emph
  {et~al.}(2021)\citenamefont {D\'iaz-Fern\'andez}, \citenamefont
  {Domínguez-Adame},\ and\ \citenamefont
  {de~Abril}}]{Diaz_Fernandez_Domingues_Adame_2021}%
  \BibitemOpen
  \bibfield  {author} {\bibinfo {author} {\bibfnamefont {A.}~\bibnamefont
  {D\'iaz-Fern\'andez}}, \bibinfo {author} {\bibfnamefont {F.}~\bibnamefont
  {Domínguez-Adame}},\ and\ \bibinfo {author} {\bibfnamefont {O.}~\bibnamefont
  {de~Abril}},\ }\bibfield  {title} {\bibinfo {title} {{Electron scattering by
  magnetic impurity in Weyl semimetals}},\ }\href
  {https://doi.org/10.1088/1367-2630/ac14ce} {\bibfield  {journal} {\bibinfo
  {journal} {New J. Phys.}\ }\textbf {\bibinfo {volume} {23}},\ \bibinfo
  {pages} {083003} (\bibinfo {year} {2021})}\BibitemShut {NoStop}%
\bibitem [{\citenamefont {Hosur}(2012)}]{Hosur_2012}%
  \BibitemOpen
  \bibfield  {author} {\bibinfo {author} {\bibfnamefont {P.}~\bibnamefont
  {Hosur}},\ }\bibfield  {title} {\bibinfo {title} {{Friedel oscillations due
  to Fermi arcs in Weyl semimetals}},\ }\href
  {https://doi.org/10.1103/PhysRevB.86.195102} {\bibfield  {journal} {\bibinfo
  {journal} {Phys. Rev. B}\ }\textbf {\bibinfo {volume} {86}},\ \bibinfo
  {pages} {195102} (\bibinfo {year} {2012})}\BibitemShut {NoStop}%
\bibitem [{\citenamefont {Wunsch}\ \emph {et~al.}(2006)\citenamefont {Wunsch},
  \citenamefont {Stauber}, \citenamefont {Sols},\ and\ \citenamefont
  {Guinea}}]{Wunsch_2006}%
  \BibitemOpen
  \bibfield  {author} {\bibinfo {author} {\bibfnamefont {B.}~\bibnamefont
  {Wunsch}}, \bibinfo {author} {\bibfnamefont {T.}~\bibnamefont {Stauber}},
  \bibinfo {author} {\bibfnamefont {F.}~\bibnamefont {Sols}},\ and\ \bibinfo
  {author} {\bibfnamefont {F.}~\bibnamefont {Guinea}},\ }\bibfield  {title}
  {\bibinfo {title} {Dynamical polarization of graphene at finite doping},\
  }\href {https://doi.org/10.1088/1367-2630/8/12/318} {\bibfield  {journal}
  {\bibinfo  {journal} {New Journal of Physics}\ }\textbf {\bibinfo {volume}
  {8}},\ \bibinfo {pages} {318} (\bibinfo {year} {2006})}\BibitemShut {NoStop}%
\bibitem [{\citenamefont {Cheianov}\ and\ \citenamefont
  {Fal'ko}(2006)}]{ChF06}%
  \BibitemOpen
  \bibfield  {author} {\bibinfo {author} {\bibfnamefont {V.~V.}\ \bibnamefont
  {Cheianov}}\ and\ \bibinfo {author} {\bibfnamefont {V.~I.}\ \bibnamefont
  {Fal'ko}},\ }\bibfield  {title} {\bibinfo {title} {Friedel oscillations,
  impurity scattering, and temperature dependence of resistivity in graphene},\
  }\href {https://doi.org/10.1103/PhysRevLett.97.226801} {\bibfield  {journal}
  {\bibinfo  {journal} {Phys. Rev. Lett.}\ }\textbf {\bibinfo {volume} {97}},\
  \bibinfo {pages} {226801} (\bibinfo {year} {2006})}\BibitemShut {NoStop}%
\bibitem [{\citenamefont {Wang}\ \emph {et~al.}(2021)\citenamefont {Wang},
  \citenamefont {Raikh},\ and\ \citenamefont {Sedrakyan}}]{Wang_Raikh_2021}%
  \BibitemOpen
  \bibfield  {author} {\bibinfo {author} {\bibfnamefont {K.}~\bibnamefont
  {Wang}}, \bibinfo {author} {\bibfnamefont {M.~E.}\ \bibnamefont {Raikh}},\
  and\ \bibinfo {author} {\bibfnamefont {T.~A.}\ \bibnamefont {Sedrakyan}},\
  }\bibfield  {title} {\bibinfo {title} {{Persistent Friedel oscillations in
  graphene due to a weak magnetic field}},\ }\href
  {https://doi.org/10.1103/PhysRevB.103.085418} {\bibfield  {journal} {\bibinfo
   {journal} {Phys. Rev. B}\ }\textbf {\bibinfo {volume} {103}},\ \bibinfo
  {pages} {085418} (\bibinfo {year} {2021})}\BibitemShut {NoStop}%
\bibitem [{\citenamefont {Hwang}\ and\ \citenamefont
  {Das~Sarma}(2007)}]{Hwang_Sarma_2007}%
  \BibitemOpen
  \bibfield  {author} {\bibinfo {author} {\bibfnamefont {E.~H.}\ \bibnamefont
  {Hwang}}\ and\ \bibinfo {author} {\bibfnamefont {S.}~\bibnamefont
  {Das~Sarma}},\ }\bibfield  {title} {\bibinfo {title} {{Dielectric function,
  screening, and plasmons in two-dimensional graphene}},\ }\href
  {https://doi.org/10.1103/PhysRevB.75.205418} {\bibfield  {journal} {\bibinfo
  {journal} {Phys. Rev. B}\ }\textbf {\bibinfo {volume} {75}},\ \bibinfo
  {pages} {205418} (\bibinfo {year} {2007})}\BibitemShut {NoStop}%
\bibitem [{\citenamefont {Grosso}\ and\ \citenamefont
  {Parravicini}(2000)}]{Grosso_Parravicini_solid_state_theory}%
  \BibitemOpen
  \bibfield  {author} {\bibinfo {author} {\bibfnamefont {G.}~\bibnamefont
  {Grosso}}\ and\ \bibinfo {author} {\bibfnamefont {G.~P.}\ \bibnamefont
  {Parravicini}},\ }\href {https://doi.org/10.1016/b978-0-12-304460-0.x5000-2}
  {\emph {\bibinfo {title} {Solid State Physics}}}\ (\bibinfo  {publisher}
  {Elsevier},\ \bibinfo {year} {2000})\BibitemShut {NoStop}%
\bibitem [{\citenamefont {Zyuzin}\ and\ \citenamefont
  {Tiwari}(2016)}]{Zyuzin_Tiwari_2016}%
  \BibitemOpen
  \bibfield  {author} {\bibinfo {author} {\bibfnamefont {A.~A.}\ \bibnamefont
  {Zyuzin}}\ and\ \bibinfo {author} {\bibfnamefont {R.~P.}\ \bibnamefont
  {Tiwari}},\ }\bibfield  {title} {\bibinfo {title} {{Intrinsic anomalous Hall
  effect in type-{II} Weyl semimetals}},\ }\href
  {https://doi.org/10.1134/s002136401611014x} {\bibfield  {journal} {\bibinfo
  {journal} {{JETP} Lett.}\ }\textbf {\bibinfo {volume} {103}},\ \bibinfo
  {pages} {717} (\bibinfo {year} {2016})}\BibitemShut {NoStop}%
\bibitem [{\citenamefont {Carbotte}(2016)}]{Carbotte_2016}%
  \BibitemOpen
  \bibfield  {author} {\bibinfo {author} {\bibfnamefont {J.~P.}\ \bibnamefont
  {Carbotte}},\ }\bibfield  {title} {\bibinfo {title} {{Dirac cone tilt on
  interband optical background of type-I and type-II Weyl semimetals}},\ }\href
  {https://doi.org/10.1103/PhysRevB.94.165111} {\bibfield  {journal} {\bibinfo
  {journal} {Phys. Rev. B}\ }\textbf {\bibinfo {volume} {94}},\ \bibinfo
  {pages} {165111} (\bibinfo {year} {2016})}\BibitemShut {NoStop}%
\bibitem [{\citenamefont {Rodriguez-Lopez}\ \emph {et~al.}(2020)\citenamefont
  {Rodriguez-Lopez}, \citenamefont {Popescu}, \citenamefont {Fialkovsky},
  \citenamefont {Khusnutdinov},\ and\ \citenamefont
  {Woods}}]{Rodriguez_Lopez_Woods_2020}%
  \BibitemOpen
  \bibfield  {author} {\bibinfo {author} {\bibfnamefont {P.}~\bibnamefont
  {Rodriguez-Lopez}}, \bibinfo {author} {\bibfnamefont {A.}~\bibnamefont
  {Popescu}}, \bibinfo {author} {\bibfnamefont {I.}~\bibnamefont {Fialkovsky}},
  \bibinfo {author} {\bibfnamefont {N.}~\bibnamefont {Khusnutdinov}},\ and\
  \bibinfo {author} {\bibfnamefont {L.~M.}\ \bibnamefont {Woods}},\ }\bibfield
  {title} {\bibinfo {title} {{Signatures of complex optical response in Casimir
  interactions of type I and {II} Weyl semimetals}},\ }\href
  {https://doi.org/10.1038/s43246-020-0015-4} {\bibfield  {journal} {\bibinfo
  {journal} {Commun. Mat.}\ }\textbf {\bibinfo {volume} {1}},\ \bibinfo {pages}
  {14} (\bibinfo {year} {2020})}\BibitemShut {NoStop}%
\bibitem [{\citenamefont {Yang}\ \emph {et~al.}(2011)\citenamefont {Yang},
  \citenamefont {Lu},\ and\ \citenamefont {Ran}}]{Yang_Lu_2011}%
  \BibitemOpen
  \bibfield  {author} {\bibinfo {author} {\bibfnamefont {K.-Y.}\ \bibnamefont
  {Yang}}, \bibinfo {author} {\bibfnamefont {Y.-M.}\ \bibnamefont {Lu}},\ and\
  \bibinfo {author} {\bibfnamefont {Y.}~\bibnamefont {Ran}},\ }\bibfield
  {title} {\bibinfo {title} {Quantum {Hall} effects in a {Weyl} semimetal:
  Possible application in pyrochlore iridates},\ }\href
  {https://doi.org/10.1103/PhysRevB.84.075129} {\bibfield  {journal} {\bibinfo
  {journal} {Phys. Rev. B}\ }\textbf {\bibinfo {volume} {84}},\ \bibinfo
  {pages} {075129} (\bibinfo {year} {2011})}\BibitemShut {NoStop}%
\bibitem [{\citenamefont {Trescher}\ \emph {et~al.}(2017)\citenamefont
  {Trescher}, \citenamefont {Bergholtz}, \citenamefont {Udagawa},\ and\
  \citenamefont {Knolle}}]{Trescher_Bergholtz_2017}%
  \BibitemOpen
  \bibfield  {author} {\bibinfo {author} {\bibfnamefont {M.}~\bibnamefont
  {Trescher}}, \bibinfo {author} {\bibfnamefont {E.~J.}\ \bibnamefont
  {Bergholtz}}, \bibinfo {author} {\bibfnamefont {M.}~\bibnamefont {Udagawa}},\
  and\ \bibinfo {author} {\bibfnamefont {J.}~\bibnamefont {Knolle}},\
  }\bibfield  {title} {\bibinfo {title} {Charge density wave instabilities of
  type-{II} {Weyl} semimetals in a strong magnetic field},\ }\href
  {https://doi.org/10.1103/PhysRevB.96.201101} {\bibfield  {journal} {\bibinfo
  {journal} {Phys. Rev. B}\ }\textbf {\bibinfo {volume} {96}},\ \bibinfo
  {pages} {201101} (\bibinfo {year} {2017})}\BibitemShut {NoStop}%
\bibitem [{foo()}]{footnote_model_c}%
  \BibitemOpen
  \href@noop {} {}\bibinfo {note} {{In cylindrical coordinates, the
  azimuthal-angle integration in Eq.~\eqref{eq:dim_less_charge_dens_ext} for
  $\tilde{r}_y=\tilde{r}_z=0$ for model \textit{c} cannot be performed
  analytically. Hence, the full three-dimensional integration has to be
  performed numerically. In contrast, for all other charge-density plots in
  this work, one integral was calculated analytically. Hence, the calculations
  for the induced charge density for model system \textit{c} along the
  $\tilde{r}_x$ axis are more costly, for which reason we have calculated the
  induced charge density only for the value $\tilde{Q}=3$.}}\BibitemShut
  {Stop}%
\bibitem [{\citenamefont {Chang}\ \emph {et~al.}(2018)\citenamefont {Chang},
  \citenamefont {Wieder}, \citenamefont {Schindler}, \citenamefont {Sanchez},
  \citenamefont {Belopolski}, \citenamefont {Huang}, \citenamefont {Singh},
  \citenamefont {Wu}, \citenamefont {Chang}, \citenamefont {Neupert},
  \citenamefont {Xu}, \citenamefont {Lin},\ and\ \citenamefont
  {Hasan}}]{Chang_Wieder_2018}%
  \BibitemOpen
  \bibfield  {author} {\bibinfo {author} {\bibfnamefont {G.}~\bibnamefont
  {Chang}}, \bibinfo {author} {\bibfnamefont {B.~J.}\ \bibnamefont {Wieder}},
  \bibinfo {author} {\bibfnamefont {F.}~\bibnamefont {Schindler}}, \bibinfo
  {author} {\bibfnamefont {D.~S.}\ \bibnamefont {Sanchez}}, \bibinfo {author}
  {\bibfnamefont {I.}~\bibnamefont {Belopolski}}, \bibinfo {author}
  {\bibfnamefont {S.-M.}\ \bibnamefont {Huang}}, \bibinfo {author}
  {\bibfnamefont {B.}~\bibnamefont {Singh}}, \bibinfo {author} {\bibfnamefont
  {D.}~\bibnamefont {Wu}}, \bibinfo {author} {\bibfnamefont {T.-R.}\
  \bibnamefont {Chang}}, \bibinfo {author} {\bibfnamefont {T.}~\bibnamefont
  {Neupert}}, \bibinfo {author} {\bibfnamefont {S.-Y.}\ \bibnamefont {Xu}},
  \bibinfo {author} {\bibfnamefont {H.}~\bibnamefont {Lin}},\ and\ \bibinfo
  {author} {\bibfnamefont {M.~Z.}\ \bibnamefont {Hasan}},\ }\bibfield  {title}
  {\bibinfo {title} {{Topological quantum properties of chiral crystals}},\
  }\href {https://doi.org/10.1038/s41563-018-0169-3} {\bibfield  {journal}
  {\bibinfo  {journal} {Nat. Mater.}\ }\textbf {\bibinfo {volume} {17}},\
  \bibinfo {pages} {978} (\bibinfo {year} {2018})}\BibitemShut {NoStop}%
\bibitem [{\citenamefont {Knoll}\ and\ \citenamefont
  {Timm}(2022)}]{Knoll_Timm_2022}%
  \BibitemOpen
  \bibfield  {author} {\bibinfo {author} {\bibfnamefont {A.}~\bibnamefont
  {Knoll}}\ and\ \bibinfo {author} {\bibfnamefont {C.}~\bibnamefont {Timm}},\
  }\bibfield  {title} {\bibinfo {title} {{Classification of Weyl points and
  nodal lines based on magnetic point groups for spin-$\frac{1}{2}$
  quasiparticles}},\ }\href {https://doi.org/10.1103/PhysRevB.105.115109}
  {\bibfield  {journal} {\bibinfo  {journal} {Phys. Rev. B}\ }\textbf {\bibinfo
  {volume} {105}},\ \bibinfo {pages} {115109} (\bibinfo {year}
  {2022})}\BibitemShut {NoStop}%
\bibitem [{\citenamefont {Cheng}\ \emph {et~al.}(2017)\citenamefont {Cheng},
  \citenamefont {Ohtsuki}, \citenamefont {Chaudhuri}, \citenamefont
  {Nakatsuji}, \citenamefont {Lippmaa},\ and\ \citenamefont
  {Armitage}}]{Cheng_Ohtsuki_2017}%
  \BibitemOpen
  \bibfield  {author} {\bibinfo {author} {\bibfnamefont {B.}~\bibnamefont
  {Cheng}}, \bibinfo {author} {\bibfnamefont {T.}~\bibnamefont {Ohtsuki}},
  \bibinfo {author} {\bibfnamefont {D.}~\bibnamefont {Chaudhuri}}, \bibinfo
  {author} {\bibfnamefont {S.}~\bibnamefont {Nakatsuji}}, \bibinfo {author}
  {\bibfnamefont {M.}~\bibnamefont {Lippmaa}},\ and\ \bibinfo {author}
  {\bibfnamefont {N.~P.}\ \bibnamefont {Armitage}},\ }\bibfield  {title}
  {\bibinfo {title} {{Dielectric anomalies and interactions in the
  three-dimensional quadratic band touching Luttinger semimetal
  Pr$_2$Ir$_2$O$_7$}},\ }\href {https://doi.org/10.1038/s41467-017-02121-y}
  {\bibfield  {journal} {\bibinfo  {journal} {Nat. Commun.}\ }\textbf {\bibinfo
  {volume} {8}},\ \bibinfo {pages} {2097} (\bibinfo {year} {2017})}\BibitemShut
  {NoStop}%
\bibitem [{\citenamefont {Fang}\ \emph {et~al.}(2012)\citenamefont {Fang},
  \citenamefont {Gilbert}, \citenamefont {Dai},\ and\ \citenamefont
  {Bernevig}}]{Fang_Gilbert_2012}%
  \BibitemOpen
  \bibfield  {author} {\bibinfo {author} {\bibfnamefont {C.}~\bibnamefont
  {Fang}}, \bibinfo {author} {\bibfnamefont {M.~J.}\ \bibnamefont {Gilbert}},
  \bibinfo {author} {\bibfnamefont {X.}~\bibnamefont {Dai}},\ and\ \bibinfo
  {author} {\bibfnamefont {B.~A.}\ \bibnamefont {Bernevig}},\ }\bibfield
  {title} {\bibinfo {title} {{Multi-Weyl Topological Semimetals Stabilized by
  Point Group Symmetry}},\ }\href
  {https://doi.org/10.1103/PhysRevLett.108.266802} {\bibfield  {journal}
  {\bibinfo  {journal} {Phys. Rev. Lett.}\ }\textbf {\bibinfo {volume} {108}},\
  \bibinfo {pages} {266802} (\bibinfo {year} {2012})}\BibitemShut {NoStop}%
\bibitem [{\citenamefont {Binnig}\ and\ \citenamefont
  {Rohrer}(1984)}]{Binnig_Rohrer_1984}%
  \BibitemOpen
  \bibfield  {author} {\bibinfo {author} {\bibfnamefont {G.}~\bibnamefont
  {Binnig}}\ and\ \bibinfo {author} {\bibfnamefont {H.}~\bibnamefont
  {Rohrer}},\ }\bibfield  {title} {\bibinfo {title} {{Scanning tunneling
  microscopy}},\ }\href
  {https://doi.org/https://doi.org/10.1016/S0378-4363(84)80008-X} {\bibfield
  {journal} {\bibinfo  {journal} {Physica B+C}\ }\textbf {\bibinfo {volume}
  {127}},\ \bibinfo {pages} {37} (\bibinfo {year} {1984})}\BibitemShut
  {NoStop}%
\bibitem [{\citenamefont {Hasegawa}\ and\ \citenamefont
  {Avouris}(1993)}]{Hasegawa_Avouris_1993}%
  \BibitemOpen
  \bibfield  {author} {\bibinfo {author} {\bibfnamefont {Y.}~\bibnamefont
  {Hasegawa}}\ and\ \bibinfo {author} {\bibfnamefont {P.}~\bibnamefont
  {Avouris}},\ }\bibfield  {title} {\bibinfo {title} {{Direct observation of
  standing wave formation at surface steps using scanning tunneling
  spectroscopy}},\ }\href {https://doi.org/10.1103/PhysRevLett.71.1071}
  {\bibfield  {journal} {\bibinfo  {journal} {Phys. Rev. Lett.}\ }\textbf
  {\bibinfo {volume} {71}},\ \bibinfo {pages} {1071} (\bibinfo {year}
  {1993})}\BibitemShut {NoStop}%
\bibitem [{\citenamefont {Crommie}\ \emph {et~al.}(1993)\citenamefont
  {Crommie}, \citenamefont {Lutz},\ and\ \citenamefont
  {Eigler}}]{Crommie_Lutz_1993}%
  \BibitemOpen
  \bibfield  {author} {\bibinfo {author} {\bibfnamefont {M.~F.}\ \bibnamefont
  {Crommie}}, \bibinfo {author} {\bibfnamefont {C.~P.}\ \bibnamefont {Lutz}},\
  and\ \bibinfo {author} {\bibfnamefont {D.~M.}\ \bibnamefont {Eigler}},\
  }\bibfield  {title} {\bibinfo {title} {Imaging standing waves in a
  two-dimensional electron gas},\ }\href {https://doi.org/10.1038/363524a0}
  {\bibfield  {journal} {\bibinfo  {journal} {Nature}\ }\textbf {\bibinfo
  {volume} {363}},\ \bibinfo {pages} {524} (\bibinfo {year}
  {1993})}\BibitemShut {NoStop}%
\bibitem [{\citenamefont {Mallet}\ \emph {et~al.}(2016)\citenamefont {Mallet},
  \citenamefont {Brihuega}, \citenamefont {Cherkez}, \citenamefont
  {Gómez-Rodríguez},\ and\ \citenamefont {Veuillen}}]{Mallet_Brihuega_2016}%
  \BibitemOpen
  \bibfield  {author} {\bibinfo {author} {\bibfnamefont {P.}~\bibnamefont
  {Mallet}}, \bibinfo {author} {\bibfnamefont {I.}~\bibnamefont {Brihuega}},
  \bibinfo {author} {\bibfnamefont {V.}~\bibnamefont {Cherkez}}, \bibinfo
  {author} {\bibfnamefont {J.~M.}\ \bibnamefont {Gómez-Rodríguez}},\ and\
  \bibinfo {author} {\bibfnamefont {J.-Y.}\ \bibnamefont {Veuillen}},\
  }\bibfield  {title} {\bibinfo {title} {{Friedel oscillations in
  graphene-based systems probed by Scanning Tunneling Microscopy}},\ }\href
  {https://doi.org/https://doi.org/10.1016/j.crhy.2015.12.013} {\bibfield
  {journal} {\bibinfo  {journal} {Comptes Rendus Physique}\ }\textbf {\bibinfo
  {volume} {17}},\ \bibinfo {pages} {294} (\bibinfo {year} {2016})}\BibitemShut
  {NoStop}%
\bibitem [{\citenamefont {Abbamonte}\ \emph {et~al.}(2004)\citenamefont
  {Abbamonte}, \citenamefont {Finkelstein}, \citenamefont {Collins},\ and\
  \citenamefont {Gruner}}]{Abbamonte_Finkelstein_2004}%
  \BibitemOpen
  \bibfield  {author} {\bibinfo {author} {\bibfnamefont {P.}~\bibnamefont
  {Abbamonte}}, \bibinfo {author} {\bibfnamefont {K.~D.}\ \bibnamefont
  {Finkelstein}}, \bibinfo {author} {\bibfnamefont {M.~D.}\ \bibnamefont
  {Collins}},\ and\ \bibinfo {author} {\bibfnamefont {S.~M.}\ \bibnamefont
  {Gruner}},\ }\bibfield  {title} {\bibinfo {title} {{Imaging Density
  Disturbances in Water with a 41.3-Attosecond Time Resolution}},\ }\href
  {https://doi.org/10.1103/PhysRevLett.92.237401} {\bibfield  {journal}
  {\bibinfo  {journal} {Phys. Rev. Lett.}\ }\textbf {\bibinfo {volume} {92}},\
  \bibinfo {pages} {237401} (\bibinfo {year} {2004})}\BibitemShut {NoStop}%
\bibitem [{\citenamefont {Abbamonte}\ \emph {et~al.}(2010)\citenamefont
  {Abbamonte}, \citenamefont {Wong}, \citenamefont {Cahill}, \citenamefont
  {Reed}, \citenamefont {Coridan}, \citenamefont {Schmidt}, \citenamefont
  {Lai}, \citenamefont {Joe},\ and\ \citenamefont
  {Casa}}]{Abbamonte_Wong_2010}%
  \BibitemOpen
  \bibfield  {author} {\bibinfo {author} {\bibfnamefont {P.}~\bibnamefont
  {Abbamonte}}, \bibinfo {author} {\bibfnamefont {G.~C.~L.}\ \bibnamefont
  {Wong}}, \bibinfo {author} {\bibfnamefont {D.~G.}\ \bibnamefont {Cahill}},
  \bibinfo {author} {\bibfnamefont {J.~P.}\ \bibnamefont {Reed}}, \bibinfo
  {author} {\bibfnamefont {R.~H.}\ \bibnamefont {Coridan}}, \bibinfo {author}
  {\bibfnamefont {N.~W.}\ \bibnamefont {Schmidt}}, \bibinfo {author}
  {\bibfnamefont {G.~H.}\ \bibnamefont {Lai}}, \bibinfo {author} {\bibfnamefont
  {Y.~I.}\ \bibnamefont {Joe}},\ and\ \bibinfo {author} {\bibfnamefont
  {D.}~\bibnamefont {Casa}},\ }\bibfield  {title} {\bibinfo {title} {{Ultrafast
  Imaging and the Phase Problem for Inelastic X-Ray Scattering}},\ }\href
  {https://doi.org/https://doi.org/10.1002/adma.200904098} {\bibfield
  {journal} {\bibinfo  {journal} {Adv. Mat.}\ }\textbf {\bibinfo {volume}
  {22}},\ \bibinfo {pages} {1141} (\bibinfo {year} {2010})}\BibitemShut
  {NoStop}%
\bibitem [{\citenamefont {Schuelke}(2007)}]{Schuelke_2007}%
  \BibitemOpen
  \bibfield  {author} {\bibinfo {author} {\bibfnamefont {W.}~\bibnamefont
  {Schuelke}},\ }\href@noop {} {\emph {\bibinfo {title} {{Electron dynamics by
  inelastic X-ray scattering}}}}\ (\bibinfo  {publisher} {Oxford University
  Press},\ \bibinfo {year} {2007})\BibitemShut {NoStop}%
\bibitem [{\citenamefont {Ament}\ \emph {et~al.}(2011)\citenamefont {Ament},
  \citenamefont {van Veenendaal}, \citenamefont {Devereaux}, \citenamefont
  {Hill},\ and\ \citenamefont {van~den Brink}}]{Ament_Veenendaal_2011}%
  \BibitemOpen
  \bibfield  {author} {\bibinfo {author} {\bibfnamefont {L.~J.~P.}\
  \bibnamefont {Ament}}, \bibinfo {author} {\bibfnamefont {M.}~\bibnamefont
  {van Veenendaal}}, \bibinfo {author} {\bibfnamefont {T.~P.}\ \bibnamefont
  {Devereaux}}, \bibinfo {author} {\bibfnamefont {J.~P.}\ \bibnamefont
  {Hill}},\ and\ \bibinfo {author} {\bibfnamefont {J.}~\bibnamefont {van~den
  Brink}},\ }\bibfield  {title} {\bibinfo {title} {{Resonant inelastic x-ray
  scattering studies of elementary excitations}},\ }\href
  {https://doi.org/10.1103/RevModPhys.83.705} {\bibfield  {journal} {\bibinfo
  {journal} {Rev. Mod. Phys.}\ }\textbf {\bibinfo {volume} {83}},\ \bibinfo
  {pages} {705} (\bibinfo {year} {2011})}\BibitemShut {NoStop}%
\bibitem [{\citenamefont {H\"{o}ller}\ \emph {et~al.}(2015)\citenamefont
  {H\"{o}ller}, \citenamefont {Krotscheck},\ and\ \citenamefont
  {Suraud}}]{Hoeller_Krotscheck_2015}%
  \BibitemOpen
  \bibfield  {author} {\bibinfo {author} {\bibfnamefont {J.}~\bibnamefont
  {H\"{o}ller}}, \bibinfo {author} {\bibfnamefont {E.}~\bibnamefont
  {Krotscheck}},\ and\ \bibinfo {author} {\bibfnamefont {{\'{E}}.}~\bibnamefont
  {Suraud}},\ }\bibfield  {title} {\bibinfo {title} {{On the response function
  of simple metal clusters}},\ }\href
  {https://doi.org/10.1140/epjd/e2015-60017-8} {\bibfield  {journal} {\bibinfo
  {journal} {Euro. Phys. J. D}\ }\textbf {\bibinfo {volume} {69}},\ \bibinfo
  {pages} {141} (\bibinfo {year} {2015})}\BibitemShut {NoStop}%
\bibitem [{\citenamefont {Reed}\ \emph {et~al.}(2010)\citenamefont {Reed},
  \citenamefont {Uchoa}, \citenamefont {Joe}, \citenamefont {Gan},
  \citenamefont {Casa}, \citenamefont {Fradkin},\ and\ \citenamefont
  {Abbamonte}}]{Reed_Uchoa_2010}%
  \BibitemOpen
  \bibfield  {author} {\bibinfo {author} {\bibfnamefont {J.~P.}\ \bibnamefont
  {Reed}}, \bibinfo {author} {\bibfnamefont {B.}~\bibnamefont {Uchoa}},
  \bibinfo {author} {\bibfnamefont {Y.~I.}\ \bibnamefont {Joe}}, \bibinfo
  {author} {\bibfnamefont {Y.}~\bibnamefont {Gan}}, \bibinfo {author}
  {\bibfnamefont {D.}~\bibnamefont {Casa}}, \bibinfo {author} {\bibfnamefont
  {E.}~\bibnamefont {Fradkin}},\ and\ \bibinfo {author} {\bibfnamefont
  {P.}~\bibnamefont {Abbamonte}},\ }\bibfield  {title} {\bibinfo {title} {{The
  Effective Fine-Structure Constant of Freestanding Graphene Measured in
  Graphite}},\ }\href {https://doi.org/10.1126/science.1190920} {\bibfield
  {journal} {\bibinfo  {journal} {Science}\ }\textbf {\bibinfo {volume}
  {330}},\ \bibinfo {pages} {805} (\bibinfo {year} {2010})}\BibitemShut
  {NoStop}%
\bibitem [{\citenamefont {Hagiya}\ \emph {et~al.}(2020)\citenamefont {Hagiya},
  \citenamefont {Matsuda}, \citenamefont {Hiraoka}, \citenamefont {Kajihara},
  \citenamefont {Kimura},\ and\ \citenamefont {Inui}}]{Hagiya_Matsuda_2020}%
  \BibitemOpen
  \bibfield  {author} {\bibinfo {author} {\bibfnamefont {T.}~\bibnamefont
  {Hagiya}}, \bibinfo {author} {\bibfnamefont {K.}~\bibnamefont {Matsuda}},
  \bibinfo {author} {\bibfnamefont {N.}~\bibnamefont {Hiraoka}}, \bibinfo
  {author} {\bibfnamefont {Y.}~\bibnamefont {Kajihara}}, \bibinfo {author}
  {\bibfnamefont {K.}~\bibnamefont {Kimura}},\ and\ \bibinfo {author}
  {\bibfnamefont {M.}~\bibnamefont {Inui}},\ }\bibfield  {title} {\bibinfo
  {title} {{Static density response function studied by inelastic x-ray
  scattering: Friedel oscillations in solid and liquid Li}},\ }\href
  {https://doi.org/10.1103/PhysRevB.102.054208} {\bibfield  {journal} {\bibinfo
   {journal} {Phys. Rev. B}\ }\textbf {\bibinfo {volume} {102}},\ \bibinfo
  {pages} {054208} (\bibinfo {year} {2020})}\BibitemShut {NoStop}%
\bibitem [{\citenamefont {Nie}\ \emph {et~al.}(2022)\citenamefont {Nie},
  \citenamefont {Hashimoto},\ and\ \citenamefont {Prinz}}]{Nie_Hashimoto_2022}%
  \BibitemOpen
  \bibfield  {author} {\bibinfo {author} {\bibfnamefont {S.}~\bibnamefont
  {Nie}}, \bibinfo {author} {\bibfnamefont {T.}~\bibnamefont {Hashimoto}},\
  and\ \bibinfo {author} {\bibfnamefont {F.~B.}\ \bibnamefont {Prinz}},\
  }\bibfield  {title} {\bibinfo {title} {{Magnetic Weyl Semimetal in
  ${\mathrm{K}}_{2}{\mathrm{Mn}}_{3}({\mathrm{AsO}}_{4}{)}_{3}$ with the
  Minimum Number of Weyl Points}},\ }\href
  {https://doi.org/10.1103/PhysRevLett.128.176401} {\bibfield  {journal}
  {\bibinfo  {journal} {Phys. Rev. Lett.}\ }\textbf {\bibinfo {volume} {128}},\
  \bibinfo {pages} {176401} (\bibinfo {year} {2022})}\BibitemShut {NoStop}%
\bibitem [{\citenamefont {Ruderman}\ and\ \citenamefont
  {Kittel}(1954)}]{Ruderman_Kittel_1954}%
  \BibitemOpen
  \bibfield  {author} {\bibinfo {author} {\bibfnamefont {M.~A.}\ \bibnamefont
  {Ruderman}}\ and\ \bibinfo {author} {\bibfnamefont {C.}~\bibnamefont
  {Kittel}},\ }\bibfield  {title} {\bibinfo {title} {Indirect exchange coupling
  of nuclear magnetic moments by conduction electrons},\ }\href
  {https://doi.org/10.1103/PhysRev.96.99} {\bibfield  {journal} {\bibinfo
  {journal} {Phys. Rev.}\ }\textbf {\bibinfo {volume} {96}},\ \bibinfo {pages}
  {99} (\bibinfo {year} {1954})}\BibitemShut {NoStop}%
\bibitem [{\citenamefont {Kasuya}(1956)}]{Kasuya_1956}%
  \BibitemOpen
  \bibfield  {author} {\bibinfo {author} {\bibfnamefont {T.}~\bibnamefont
  {Kasuya}},\ }\bibfield  {title} {\bibinfo {title} {{A Theory of Metallic
  Ferro- and Antiferromagnetism on Zener's Model}},\ }\href
  {https://doi.org/10.1143/PTP.16.45} {\bibfield  {journal} {\bibinfo
  {journal} {Prog. Theor. Phys.}\ }\textbf {\bibinfo {volume} {16}},\ \bibinfo
  {pages} {45} (\bibinfo {year} {1956})}\BibitemShut {NoStop}%
\bibitem [{\citenamefont {Yosida}(1957)}]{Yosida_1957}%
  \BibitemOpen
  \bibfield  {author} {\bibinfo {author} {\bibfnamefont {K.}~\bibnamefont
  {Yosida}},\ }\bibfield  {title} {\bibinfo {title} {Magnetic properties of
  {Cu}-{Mn} alloys},\ }\href {https://doi.org/10.1103/PhysRev.106.893}
  {\bibfield  {journal} {\bibinfo  {journal} {Phys. Rev.}\ }\textbf {\bibinfo
  {volume} {106}},\ \bibinfo {pages} {893} (\bibinfo {year}
  {1957})}\BibitemShut {NoStop}%
\bibitem [{\citenamefont {Abrikosov}\ and\ \citenamefont
  {Beneslavskii}()}]{Abrikosov_Beneslavskii_1971}%
  \BibitemOpen
  \bibfield  {author} {\bibinfo {author} {\bibfnamefont {A.~A.}\ \bibnamefont
  {Abrikosov}}\ and\ \bibinfo {author} {\bibfnamefont {S.}~\bibnamefont
  {Beneslavskii}},\ }\bibfield  {title} {\bibinfo {title} {{Possible Existence
  of Substances Intermediate Between Metals and Dielectrics}},\ }\href@noop {}
  {\ }\bibinfo {note} {{Zh. Eksp. Teor. Fiz. \textbf{59}, 1280 (1970)
  [\href{http://jetp.ras.ru/cgi-bin/e/index/r/59/4/p1280?a=list}{Sov.
  Phys.--JETP \textbf{32}, 699 (1971)}]}}\BibitemShut {NoStop}%
\end{thebibliography}%

\end{document}